\documentclass[superscriptaddress,secnumarabic,
amssymb,amsmath,nobibnotes,aps,prd,showkeys,showpacs,nofootinbib]{revtex4-2}%

\setcitestyle{citesep={;}}
\setcitestyle{authoryear,round}%Setting citation style. Changing braces to ()
\usepackage{graphicx}
\usepackage{epsf}
\usepackage{bm}
\usepackage{amsmath}
\usepackage{amsfonts}
\usepackage{amssymb}
\usepackage{epstopdf}
\usepackage{subfigure}
\usepackage{natbib}
\usepackage{color}%
\usepackage{hyperref}
\usepackage{caption}

\hypersetup{dvips,dvipdfm,linktoc=page,colorlinks=true,linkcolor=blue,citecolor=red,filecolor=magenta,urlcolor=magenta,bookmarks=true}
\providecommand{\U}[1]{\protect\rule{.1in}{.1in}}
\newcommand{\be}{\begin{equation}}
\newcommand{\ee}{\end{equation}}

\newcommand{\mincir}{\raise
-3.truept\hbox{\rlap{\hbox{$\sim$}}\raise4.truept\hbox{$<$}\ }}
\newcommand{\magcir}{\raise
-3.truept\hbox{\rlap{\hbox{$\sim$}}\raise4.truept\hbox{$>$}\ }}

\begin{document}

\title{Effects of particle creation rate in two-fluid interacting cosmologies }
\author{Trishit Banerjee}
\email{trishit23@gmail.com}
\affiliation{Department of mathematics, Netaji Subhash Engineering College, West Bengal, India}
\affiliation{Department of Applied Mathematics, Maulana Abul Kalam Azad University of Technology, West Bengal- 741249, India.}

\author {Goutam Mandal}
\email{gmandal243@gmail.com; rs\_goutamm@nbu.ac.in}
\affiliation{Department of Mathematics, University of North Bengal, Raja Rammohunpur, Darjeeling-734013, West Bengal, India.}

\author{Atreyee Biswas}
\email{atreyee11@gmail.com}
\affiliation{Department of Applied Mathematics, Maulana Abul Kalam Azad University of Technology, West Bengal- 741249, India.}

\author{Sujay Kr. Biswas}
\email{sujaymathju@gmail.com; 	sujay.math@nbu.ac.in}
\affiliation{Department of Mathematics, University of North Bengal, Raja Rammohunpur, Darjeeling-734013, West Bengal, India.}
%\affiliation{Department of Mathematics, Jadavpur University, Jadavpur, Kolkata - 700032, West Bengal, India}
%\keywords{Irreversible thermodynamics; GSLT; particle creation mechanism, Statefinder diagnosis, Interaction, Dynamical system, Phase space, Stability}

%%%%%%%%%%%%%%%%%%%%%%%%%%%%%%%%%%%%%%%%%%%%%%%%%%%%%%%%%%%%%%%%%%%%%%%%%%%%%%%%%%%%%%%%%%%%%%5
%%%%%%%%%%%%%%%%%%%%%%%%%%%%%%%%%%%%%%%%%%%%%%%%%%%%%%%%%%%%%%%%%%%%%%%%%%%%%%%%%%%%%%%%%%%%%%%%
\begin{abstract}
	
\end{abstract}
%%%%%%%%%%%%%%%%%%%%%%%%%%%%%%%%%%%%%%%%%%%%%%%%%%%%%%%%%%%%%%%%%%%%%%%%%%%%%%%%%%%%%%%%%%
%%%%%%%%%%%%%%%%%%%%%%%%%%%%%%%%%%%%%%%%%%%%%
\begin{abstract}
In this work, a two-fluid interacting model in a flat FLRW universe has been studied considering particle creation mechanism with a particular form of particle creation rate $\Gamma=\Gamma_0 H+\frac{\Gamma_1}{H}$ from different aspects. Statistical analysis with a combined data set of SNe Ia (Supernovae Type Ia) and Hubble data is performed to achieve the best-fit values of the model parameters, and the model is compatible with current observational data. We also perform a dynamical analysis of this model to get an overall qualitative description of the cosmological evolution by converting the governing equations into a system of ordinary differential equations considering a proper transformation of variables. We find some non-isolated sets of critical points, among which some usually are normally hyperbolic sets of points that describe the present acceleration of the universe dominated by dark energy mimicking cosmological constant or phantom fluid. Scaling solutions are also obtained from this analysis, and they can alleviate the coincidence problem successfully. Statefinder diagnosis is also carried out for this model to compare it with the $\Lambda$CDM, and any other dark energy models byfinding various statefinder parameters. Finally, the thermodynamic analysis shows that the generalized second law of thermodynamics is valid in an irreversible thermodynamic context.
\end{abstract}
%%%%%%%%%%%%%%%%%%%%%%%%%%%%%%%%%%%%%%%%%%%%%%%%%%
%%%%%%%%%%%%%%%%%%%%%%%%%%%%%%%%%%%%%%%%%%%%%%%%%%%%%%%%%%%%%%%%%%%%%%%%%%%%%%%%%%%%%%%%%%
\maketitle
%%%%%%%%%%%%%%%%%%%%%%%%%%%%%%%%%%%%%%%%%%%%%%%%%%%%%%%%%%%%%%%%%%%%%%%%%%%%%%%%%%%%%%%%%%%%%%%%%%%%%%%%%%%%%%%%%%%%%%%%%%%%%%%%%%%%%%%%%%%%%%%%%%
\section{Introduction}
Current observations like Supernovae Type Ia (SNe  Ia) (\cite{Riess2007}), Large Scale Structure (LSS) (\cite{Hawkins2003,Tegmark2006,Cole2005}), and Cosmic
Microwave Background Radiation (CMBR) (\cite{Bennet2003}) reveal that our universe is presently experiencing an
accelerating expansion. There are two distinct approaches that describe this late-time acceleration of the universe. In the first approach, an unknown component dubbed as dark energy (DE) is considered to be responsible for this acceleration. It is inferred that dark energy has negative pressure violating the strong energy condition (i.e. $\rho+3p\geq0$).

Except for the information that dark energy has negative pressure, we are totally in the dark about its nature and properties. However, from observational evidence, it is clear that almost $70\%$ of the total energy density of the universe is occupied by dark energy. The most straightforward candidate is, in this case, the cosmological constant $\Lambda$. Moreover, with Cold Dark Matter (CDM), the Cosmological Constant constitutes the $\Lambda$CDM model, which has been systematically proven consistent with many observations (\cite{Komatsu2011,Perlmutter1999,Percival2010,Sharov2014}). Despite the simplicity, it suffers from several problems like fine-tuning and coincidence problems. Several authors have shown that these problems associated with $\Lambda$CDM model can be addressed by inserting an interaction term between two dark components. As the nature of two dark components (DE and DM) is still undetermined, there is no compelling reason to exclude interaction between them. Initially, the interaction term was considered in order to reduce the enormous difference between the observed value and the theoretically predicted value of the cosmological constant (\cite{Wetterich1995}) and also to alleviate the coincidence problem (\cite{Amendola2000,Tocchini2002,Amendola2003,Campo2008,Campo2009}). However, an appropriate choice of interaction terms can influence the perturbation dynamics and also affect the lowest multipoles of the CMB spectrum (\cite{Wang2005,Wang2007}). Furthermore, coupling in the dark sector may alleviate the $H_{0}$ tension  (\cite{Valentino2017,Kumar2017,Yang2018,Pan2019,Kumar2021,Valentino2020,Akarsu2021,Lucca2020,Valentino2021,Anchordoqui2021,Renzi2022,Theodoropoulos2021,Vagnozzi2020,Visinelli2019,Valentino2020(101),Cheng2020,Vagnozzi2020(102),Pan2020,Yang2021,Gao2021,Lucca2021,Nunes2021,Guo2021,Mancini2022,Ferlito2022,Nunes2022,Pan2023}), which exists due to disagreement between the CMB measurements by the Planck satellite within $\Lambda$CDM cosmology (\cite{Aghanim2020}) and SH$0$ES (\cite{Riess2007}). The coupling also solves $S_{8}$ tension (\cite{Pourtsidou2016,An2018,Kumar2019,Araujo2021,Avila2022}), which comes into existence due to discordance (i.e. many sigmas gap) between the Planck and Weak Lensing measurements (\cite{Valentino2021(131)}). Moreover, an interacting dark energy model is also capable of explaining the phantom phase without any requirement for a scalar field with negative correction (\cite{Pan2014,Bonilla2018,Rivera2019}). In such a way, recently the dynamics of the dark energy model of cosmology have attracted much attention. We would like to recommend two works (\cite{Bolotin2015,Wang2016}) in this regard for a detailed study in coupled dark energy models. \\ 

However, in explaining the late-time acceleration of the universe, there are alternative other models of $\Lambda$CDM model such as quintessence (\cite{Erickson2002}), K-essence (\cite{Armendariz2001}), and perfect fluid models like Chaplygin gas and its generalizations (\cite{Kamenschick2001,Bento2002,Pourhassan2014,Guo2007,Fabris2002,Lazkoz2019}). All of these belong to modified matter model in which the energy-momentum tensor $T_{\mu\nu}$ (the right-hand side) of Einstein’s equation $R_{\mu\nu}-\frac{1}{2}R g_{\mu\nu}=8\pi G T_{\mu\nu}$ is modified considering an exotic matter source with negative pressure. On the other hand, modified gravity models are obtained by modifying the left-hand side of Einstein’s equation. Some popular models that belong to this class are $f(R)$ gravity, see in references (\cite{Capozziello2008,Nojiri2007}). 
%%%%%%%%%%%
There exists also several more recent and popular reviews for $f(R)$ gravity. Interested readers may follow the references (\cite{Sotiriou2010,Nojiri2011,Capozziello2011,Clifton2012,Nojiri2017}) and the more recent one in reference (\cite{Odintsov2023}). Scalar-tensor theories (\cite{Bergmann1968}), and brane world models (\cite{Maartens2004}) can also explain the accelerated late phase of the universe. \\

The particle creation mechanism is another choice for explaining the present acceleration. Historically, in 1939, Schrodinger (\cite{Schrodinger1939}), introduced first a microscopic description of particle creation in an expanding universe where gravity plays a crucial role. Then, following his idea, Parker and Zeldovich with their collaborators (\cite{Parker1968,Ford1977,Zeldovich1972}), started the possible cosmological scenarios considering particle creation method. There are two general approaches that are adopted in various literatures to understand the concept of this particle production mechanism. The first one refers to the technique of adiabatic vacuum state (\cite{Parker1969,Parker1973,Fulling1989}), and the second one is the mechanism of instantaneous Hamiltonian diagonalization (\cite{Pavlov2008,Grib1994}). Also, the matter creation scenario has many other aspects which include the possibility of future deceleration, the existence of an emergent Universe and the possibility of having a phantom phase without invoking the phantom field. It is to be mentioned that the matter creation mechanism, in a logical way, is based on Quantum Field Theory(QFT) and the concept was first proposed by Prigogine (\cite{Prigogine1989}). 
The choice of particle creation rate $\Gamma$ that goes with current observations is phenomenological since complete description of quantum field theory (QFT) in FRW universe is still not developed. There are some popular choices of $\Gamma$ that can describe the evolution of universe from early phase to current accelerating era of our universe. For instance $\Gamma=$ constant (\cite{Haro2016}) explicates the evolution of universe starting from Big Bang to current accelerating phase while $\Gamma\propto H$ (\cite{Nunes2015,Pan2015,Lima2010}) and $\Gamma\propto H^2$ (\cite{Abramo1996,Lima1996,Zimdahl2000}) unfolds respectively the matter dominated decelerating phase and the early inflationary phase of universe. On the other hand the particle creation rate $\Gamma\propto \frac{1}{H}$ (\cite{Pan2015}) explains the transition from matter dominated era to the late time acceleration of universe. To get a transition from decelerating to accelerating regime in low redshifts Lima et al (\cite{Lima2008}) added an term proportional to $H$ with a constant term i.e $\Gamma=3\gamma H_0+3\beta H$. Motivated from the above works S. Pan et. al. (\cite{Pan2016b}) considered a more general particle creation rate which is a linear combination of the above three choices of particle creation rates : $\Gamma=\-\Gamma_0+mH+n/H$ and had shown that the model can describe both the early inflationary era and the late-time accelerating phase without considering the existence of Big Bang singularity and also can predict future deceleration of universe.
Recently S. Pan et al. (\cite{Pan2019(79)}) investigated the cosmological evolution of the universe, incorporating the continuous particle production by the time-varying gravitational field where they considered a homogeneous, isotropic, and spatially flat universe with two-fluid components. One was equipped with gravitationally induced “adiabatic” matter creation, and the second simply satisfies the conservation of energy. Also, the authors considered a special form of particle creation rate, namely, $\Gamma=\Gamma_0+\frac{\Gamma_1}{H}+\frac{\Gamma_2}{H^2}+\sum_{i=3}^n\Gamma_i H^{-i}$, along with some other exceptional forms. They showed that from the present two-fluid cosmological model with gravitationally induced “adiabatic” matter creation, one can obtain singular algebraic solutions to the gravitational field equations.  Motivated from all these works, we are interested in studying cosmological consequences assuming matter creation rate $\Gamma=\Gamma_0 H+\frac{\Gamma_1}{H}$ in an interacting dark energy-dark matter model. It is to be mentioned that our choice is different from (\cite{Pan2016b}) in two ways. Firstly, We have excluded the constant term in $\Gamma$ since logically particle creation rate should not be constant irrespective of any energy scale rather should be some function of some energy scale considering the fact that our universe is evolving from high energy to low energy regime.  Secondly an interaction term in the dark sector is considered in order to examine if the combined effect of interaction along with matter creation can significantly influence the future dynamics of the universe. Now, from the construction of our chosen particle creation rate, it is evident that when $H\rightarrow \infty$ i.e when $a\rightarrow 0$ the particle creation rate also become infinite directing towards Big Bang singularity. But, we see that the chosen particle creation rate again becomes infinite when $H\rightarrow 0$, i.e., when the scale factor $a$ becomes constant which renders to static universe and therefore, to avoid this, we shall exclude the case $a$=constant.   However, we have analyzed the present model from different aspects through a) statistical analysis,  b) statefinder diagnosis, c) dynamical system analysis, and d) thermodynamic analysis.
\\\\

Statefinder diagnosis is an effective tool to distinguish between different dynamical dark energy models with similar expansion history and can efficiently choose the correct model of late-time acceleration of the universe with the help of future observational value of statefinder parameters. Several dark energy models have been studied by using statefinder diagnostic tools such as the Quintessence model (\cite{Sahni2003,Alam2003}), the interacting Quintessence model (\cite{Zimdahl2004,Zhang2005a}), the phantom model (\cite{Chang2007}), the Tachyon model (\cite{Shao2008}), the generalized chaplygin gas model (\cite{Malekjani2013}), the holographic dark energy models (\cite{Zhang2005b,Zhang2008,Setare2007}), the agegraphic dark energy model (\cite{Zhang2010,Khodam2010,Wei2007,Malekjani2010}), and model by considering the variable gravitational constant $G$ (\cite{Debnath2011,Jamil2011,Setare2011}). Statefinder diagnostic is also studied for Modified Chaplygin Gas in modified gravity in the framework of particle creation mechanism (see in reference (\cite{Singh2018})).\\

Dynamical system analysis can be employed as a powerful tool to get an overall evolutionary description of the universe through a suitable choice of dynamical variables. See the references (\cite{Mandal2023,Mandal2022,Mandal2022(35),Bahamonde2018,Biswas2017,Biswas2021}) where cosmological models have been studied through dynamical analysis.
%%%%%%%%%%%
In this context, there are much more well known  references in the literature see for example (\cite{Odintsov2017,Odintsov2018,Oikonomou2018}) and references therein where dynamical system approach have been studied in cosmologcal models of modified gravity theories. 

%%%%%%%%%%%%
In our model, we construct an autonomous system of ODEs from which non-isolated sets of critical points are extracted. Linear stability analysis is then executed by finding the eigenvalues of the linearized Jacobian matrix for the critical points. We obtain some non-isolated sets of critical points with exactly one vanishing eigenvalue, called normally hyperbolic sets. The stability of the normally hyperbolic sets can be found by observing the signature of the remaining non-vanishing eigenvalues associated with the sets. From the analysis, we observe some interesting cosmological features relevant to the future evolution of the universe dominated by DE which mimics a Cosmological Constant or phantom fluid respectively. In some parameter regions, the set can predict the late-time accelerated scaling attractor solutions solving coincidence problem in the phantom era with $0<\Omega_{d}<1$ and $0<\Omega_{m}<1$. \\

Now, the study of thermodynamic properties of the universe is a familiar issue in cosmology since the revolutionary discovery of black hole thermodynamics by Hawking and Bekenstein (\cite{Bardeen1973,Hawking1975,Bekenstein1974}) in 1970. It is to be noted that though there are many works in this field based on equilibrium thermodynamics, the concept of irreversible thermodynamics in cosmology is a little old (2002), and very few papers exist where universal thermodynamics has been studied in a non-equilibrium context. In this work, we have analyzed the non-equilibrium thermodynamic behavior of the interacting two-fluid system where the particle creation mechanism plays the role of the associated internal non-equilibrium process with a specific form of particle creation rate. Here we have followed the work of S. Saha et al. (\cite{Saha2015}).\\

The construction of this paper is as follows: \\
Section II comprises the general prescription of the particle creation mechanism in cosmology and a description of our present particle creation model. Section III is dedicated to the  statistical analysis of the model where the best fit values of the model parameters have been estimated with the help of combined SNIa and Hubble data and then some contour plots have been drawn to show the inter-relations between different model parameters followed by a model comparison analysis using  the AIC-BIC method. Section IV deals with statefinder diagnosis, while section V analyzes the evolution dynamics of the universe through phase-space analysis in the purview of the present particle creation model. Section VI presents thermodynamic analysis following the theory of irreversible thermodynamics. Finally, section VII is dedicated to a summary and some concluding remarks.

\section{Particle creation mechanism in the two dark fluid model}
Suppose a closed thermodynamical system with $N$ particles has internal energy $E$. From the conservation of the internal energy the first law of thermodynamics reads as
\begin{equation}
	dE=dQ-pdV,\label{1}
\end{equation}
where $p$ and $V$ are the usual thermodynamic pressure and comoving volume, and $d$Q represents the amount of heat the system received in time $dt$. Equivalently, the Gibbs equation reads as
\begin{equation}
	Tds=dq=d\left(\frac{\rho}{n}\right)+pd\left(\frac{1}{n}\right),\label{2}
\end{equation}
%%%%%%%%%%
where `$s$' is the entropy per particle, $\rho=\frac{E}{V}$ is the energy density, $n=\frac{N}{V}$ is the particle number density, and $dq=\frac{dQ}{N}$ is the heat per unit particle. The above Gibbs equation is valid for an open thermodynamical system when the number of fluid particles are not conserved $\left(N^{\mu}_{;\mu}\neq 0\right)$. This can be expressed mathematically as
\begin{equation}
	\dot{n}+\theta n=n\Gamma.\label{3}
\end{equation}
%%%%%%%%%%%%%
Here, $N^{\mu}=nu^{\mu}$ is the particle flow vector, $u^{\mu}$ is the particle four-velocity vector, $\theta=u^{\mu}_{;\mu}$ is the fluid expansion, $\Gamma$ stands for the rate of change of the particle number in a comoving volume $V$ and $\dot{n}=n_{,\mu}u^{\mu}$ by notation. $\Gamma$ effectively behaves as a bulk viscous pressure causing the thermodynamics to be non-equilibrium in nature. $\Gamma > 0$ indicates the creation of particles, while $\Gamma <0$ corresponds to the annihilation of particles. In this work, we consider the flat FRW model of the universe to be an open thermodynamical system. Suppose that the universe consists of two dark fluids: dark matter(DM) and dark energy(DE). We also consider the two dark components to be interacting, and only the creation of DM is considered. Then the Einstein field equations are (choosing $8\pi G=1=c$)
\begin{equation}\label{Friedmann equation}
	3H^{2}=\rho_{t}=\rho_{m}+\rho_{d}
\end{equation}
and
\begin{equation}\label{acceleration equation}
	2\dot{H}+3H^2=-(p_{m}+p_{c})-p_{d},
\end{equation}
%%%%%%%
where $H=\frac{\dot{a}}{a}$ is the Hubble parameter and $a$ is the scale factor.
The energy conservation equations are given by
\begin{equation}\label{continuity DM}
	\dot{\rho}_{m}+3H(\rho_{m}+p_{m}+p_c)=-Q
\end{equation}
and
\begin{equation}\label{continuity DE}
	\dot{\rho}_{d}+3H(\rho_{d}+p_{d})=Q,
\end{equation}
%%%%%%%%%%
where $(\rho_{m},p_{m},p_c)$ are respectively energy density, thermodynamic pressure and dissipative (bulk) pressure of DM and $(\rho_{d},p_{d})$ are energy density and pressure of DE and $Q$ being the interaction term between DE and DM. In this work we have assumed $Q>0$ which indicates an energy flow from DM to DE. The explicit form of the interaction $Q$ is chosen as (\cite{Callen1960,Pavon2005})
\begin{equation}\label{interaction term}
	Q=\alpha H\rho_m
\end{equation}
where $\alpha (>0)$ is a dimansioless coupling constant determines the strength of the interaction between dark sectors.
The particle number conservation equations get modified as
\begin{equation}
	\dot{n}_{m}+3Hn_{m}=\Gamma n_{m}\label{9}
\end{equation}
and
\begin{equation}
	\dot{n}_{d}+3Hn_{d}=0,\label{10}
\end{equation}
where $(n_{m},\Gamma)$ are the number density and particle creation rate for DM while $n_{d}$ is number density for DE. We have assumed that $\Gamma >0$ so that DM particles are created only. If for simplicity, we assume the thermodynamical system to be isentropic ({\it i.e.,} entropy per particle is conserved), then the dissipative pressure and the particle creation rate are related as
\begin{equation}\label{creation pressure}
	p_{c}=-\frac{\Gamma}{3H}(\rho_{m}+p_{m})
\end{equation}
Now, we assume that DM as pressure-less dust i.e $p_m=0$ and then the modified acceleration equation in Eqn. (\ref{acceleration equation}) and the conservation equation for matter in Eqn. (\ref{continuity DM}) take the form:

\begin{equation}\label{acceleration equation modified}
	\dot{H}=-\frac{1}{2} \left\{ \rho_{m}\left(1-\frac{\Gamma}{3H}\right)+ \rho_d+p_{d} \right\}
\end{equation}
and the energy conservation equations are given by
\begin{equation}\label{continuity DM modified}
	\dot{\rho}_{m}+3H\rho_{m}\left(1-\frac{\Gamma}{3H}\right)=-Q
\end{equation}
%%%%%%%%%%%%%%%%%%%
respectively. Now we choose a particular form of  particle creation rate $\Gamma$ as
\begin{equation}\label{creation rate}
	\Gamma=\Gamma_0 H+\Gamma_1 H^{-1},
\end{equation}
%%%%%%%%%
where $\Gamma_0, \Gamma_1 (>0)$ are dimensionless constants. However, if the combined two-fluid is considered as a single fluid with energy density $\rho_{t}$, the thermodynamic pressure $p_{t}=p_{d}$ and dissipative pressure $\Pi_{t}=p_c $, then combining (\ref{continuity DM}) and (\ref{continuity DE}), we have the usual conservation equation
\begin{equation}
	\dot{\rho}_{t}+3H(\rho_{t}+p_{t})=-3H\Pi_{t}.\label{13}
\end{equation}
%%%%%%%%%%5
Also, combining (\ref{9}) and (\ref{10}) we can write
\begin{equation}
	\dot{n}_{t}+3Hn_{t}=\Gamma_{t}n_{t},\label{14}
\end{equation}
%%%%%%%%%%%%
where $n_{t}=n_{m}+n_{d}$ is the total particle number density and $\Gamma_{t}$ is the particle creation rate of the combined single fluid. Further, if we assume the resulting single fluid to be isentropic, then we have
\begin{equation}
	\Pi_{t}=-\frac{\Gamma_{t}}{\theta}(\rho_{t}+p_{t}).\label{15}
\end{equation}
%%%%%%%%%%
Now, using (\ref{creation pressure}), we obtain
\begin{equation}
	\Gamma_{t}=\frac{\Gamma(\rho_{m}+p_{m})}{(\rho_{t}+p_{t})}.\label{16}
\end{equation}
%%%%%%%%%%%
The sign of $\Gamma_{t}$ indicates whether there is a creation or an annihilation of particles is occurred in the resulting single fluid.\\
%%%%%%%%
Due to the observational shreds of evidence of the present late-time acceleration, the universe is currently dominated by a dark fluid system (DM+DE). From the energy conservation relations given by  (\ref{continuity DM}) and (\ref{continuity DE}), the energy densities of DM and DE are calculated as follows:

\begin{equation}
	\rho_m\approx \rho_{0m}a^\xi \label{17}
\end{equation}
and
\begin{equation}
	\rho_d\approx a^{\xi-\eta}\rho_{0d}\left\{\frac{\alpha r_0}{\eta}\left(a^\eta-1\right)+1\right\}\label{18}
\end{equation}
where $\rho_{0m}, \rho_{0d}$ are present value of energy densities of DM and DE respectively and
\begin{eqnarray*}
	\xi &=& \frac{\Gamma_1}{H_0^2}-3-\alpha+\Gamma_0, \\
	\eta &=& \frac{\Gamma_1}{H_0^2}+3\omega_d-\alpha+\Gamma_0,
\end{eqnarray*}
%%%%%%%%%%%
where $\omega_d=\frac{p_d}{\rho_d}$ is equation of state parameter (EoS) of DE and $r_0=\frac{\rho_{0m}}{\rho_{0d}}$.
Then using the above expressions for energy densities and equation (\ref{Friedmann equation}), we obtain analytic expression for Hubble parameter $H(z)$  as
\begin{equation}
	H=(1+z)^{-\frac{\xi}{2}}H_0\left[\Omega_{0m}+\Omega_{0d}~(1+z)^\eta\left\{\frac{\alpha r_0}{\eta}\left((1+z)^{-\eta}-1\right)+1\right\}\right]^{\frac{1}{2}}\label{19}
\end{equation}
where $z=\frac{1}{a}-1$ is redshift parameter.
However, it is to be mentioned that finding exact expression of $\rho_m$ was not possible from equation (\ref{continuity DM}) for the chosen specific form of particle creation rate $\Gamma$ and therefore we assumed $H\approx H_0$ while performing integration involved in the calculation and obtained an approximate value of $\rho_m$.

\section{Fitting the Model Parameters}
In this section, we consider the recent observational datasets such as the 580 data points from the Union-2 supernova Ia database (\cite{Suzuki2012}) and Hubble data set (\cite{Cao2018,Zhang2014}) and investigate the evolutionary
behavior of our universe.
For the SN Ia dataset the $\chi^2$ function is given by
\begin{equation}
	\chi_{SN}^2=\sum_i^{580} \frac{\left[\mu_{obs}(z_i)-\mu_{theo}(z_i)\right]^2}{\sigma_i^2},\label{36}
\end{equation}
%%%%%%%%
where the theoretical distance modulus $\mu_{theo} $ is given by
\begin{equation}
	\mu_{theo}=5log_{10}d_{L}(z_i)-\mu_0\label{37}
\end{equation}
%%%%%%%%
with $\mu_0=42.38-5log_{10}h$. The Hubble-free definition of the luminosity parameter $d_{L}$ reads as
\begin{equation}
	d_L=(1+z)\int_0^z \frac{H_0}{H(z')}dz'\label{38}
\end{equation}
$\mu_{obs}(z_i),~ \sigma_i $ and $h=H_{0}/100/[kmsec^{-1}Mpc^{-1}]$ represent observed distance modulus, the uncertainty in the distance modulus and the dimensionless Hubble parameter respectively.\\
In case of Hubble data, $\chi^2$ is given by
\begin{equation}
	\chi_{OHD}^2=\sum_i^{27}\frac{\left[H_{Obs}(z_i)-H_{theo}(z_i)\right]^2}{\sigma_i^2}\label{39}
\end{equation}
%%%%%%%%
where $H_{obs}$ and $H_{theo}$ describe observational and theoretical values of the cosmic Hubble parameter respectively.
For better result, we consider combined data of Supernova Ia and Hubble data and therefore we have
\begin{equation*}
	\chi^2=\chi_{SN}^2+\chi_{OHD}^2
\end{equation*}
and minimizing the above $\chi^2$, we obtain the best-fit values of the model parameters as follows:\\
$\Gamma_0=0.32,~ \Gamma_1=0.4,~ \omega_d=-1.11,~ \alpha=0.03,~ H_0=69.92,~ \Omega_d=0.63$. Also $
\chi^2_{min}=574.812$. So, $\chi^2/dof=0.9564(dof=580+27-6=601)$. This model is clearly consistent with the data since $\chi^2/dof \approx 1$. It is to be mentioned that for the same data set the best fit values of $\Lambda$CDM model are obtained as $H_0=70.02, \Omega_{0d}=0.73$ with $\chi^2_{min}=575.945$. However, as the interaction term $\alpha$ is estimated to be very small, so from a statistical point of view, interaction has no significant effect on the present model. It became evident when we analysed for the non-interacting case, i.e., for $\alpha=0$, and  the best fit parameters were obtained as $\Gamma_0=0.29,~ \Gamma_1=0.4,~ \omega_d=-1.11,~ H_0=69.92,~ \Omega_d=0.63$, and the minimum value of $\chi^2$ = 574.81. Hence, there is a minor difference between the interacting and non-interacting models only for the parameters $\Gamma_0$ and $\Omega_d $. The optimal value of $\chi^2$ remained the same for both models. However, to understand relations between the model parameters, a total of thirteen contour plots of $\chi^2$  considering it as the function of two parameters (considering the optimal value of other parameters in every scenario) have been drawn. It was utterly fascinating that the $\chi^2$  was invariant with $\Gamma_1$ and we obtained rectangular contours where $\Gamma_1$ was one of the two parameters on which $\chi^2$ is dependent. However, $\alpha$ has non-linear relation with DE EoS parameter, whereas with $\Gamma_0$, it has linear relationship.
Now, we use AIC and BIC (\cite{Shi2012}) methods to compare our present model with the $\Lambda$CDM model. The AIC and BIC are respectively, defined as follows:

\begin{eqnarray*}
	AIC &=& -2ln \textit{L}_{max}+2k \\
	BIC &=& -2ln \textit{L}_{max}+ k~ln N
\end{eqnarray*}
where $k,~N$ are the number of parameters and number of data points used in the fit. $\textit{L}$ is maximum likelihood and for Gaussian posterior distribution $\chi^2_{min}=-2ln \textit{L}_{max}$. The absolute values of AIC and BIC are not of any interest, only relative value between different models are useful to predict better one compared to other. The AIC and BIC values of our present model and $\Lambda$CDM model are shown in the Table \ref{tab1}.
\begin{table}
	\caption{Model Comparison}\label{tab1}
	\begin{tabular}{|l|l|l|}
		\hline
		~ & AIC & BIC \\
		\hline
		Particle creation model & 586.812 & 613.263 \\
		$\Lambda$CDM & 579.945 & 588.762 \\
		\hline
	\end{tabular}
\end{table}
%%%%%%%%%%%%
From the Table-\ref{tab1}, it is evident that $\Lambda$CDM model is still best fitted model, but our present model of particle creation is considerably good in terms of compatibility with observational data. This can be perceived more clearly if we calculate the correlation coefficient $r$ between the theoretical and observed value of Hubble parameter,
\begin{figure}
	\centering
	\subfigure[]{\includegraphics[width= 0.3\textwidth]{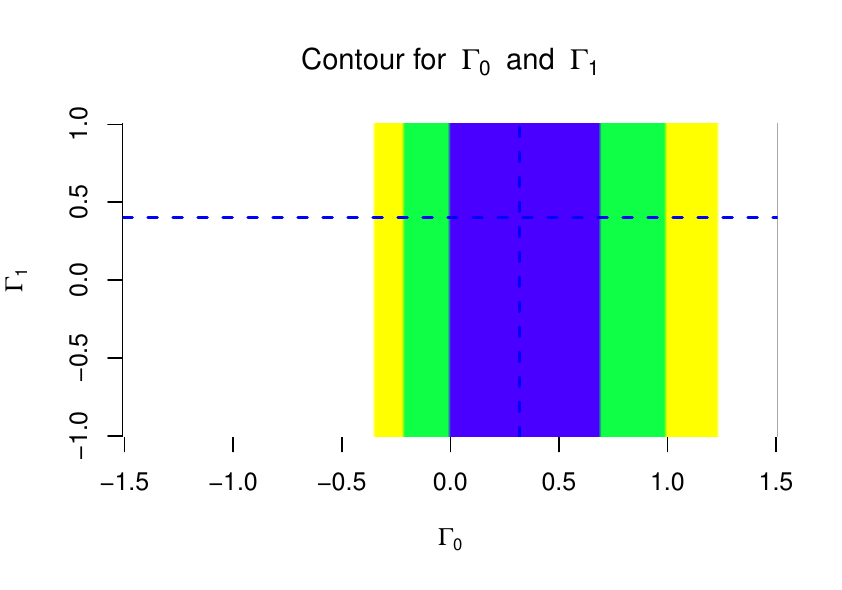}}
	\subfigure[]{\includegraphics[width= 0.3\textwidth]{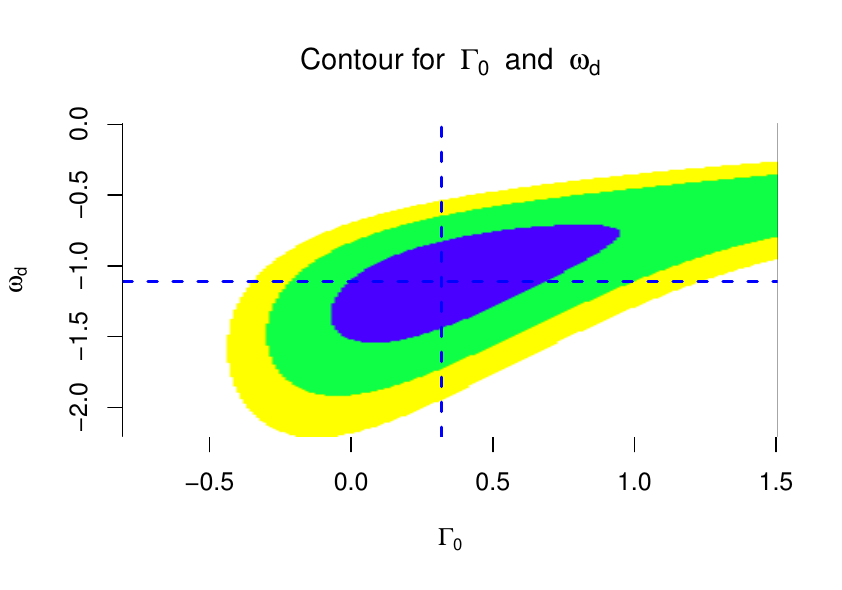}}
	\subfigure[]{\includegraphics[width= 0.3\textwidth]{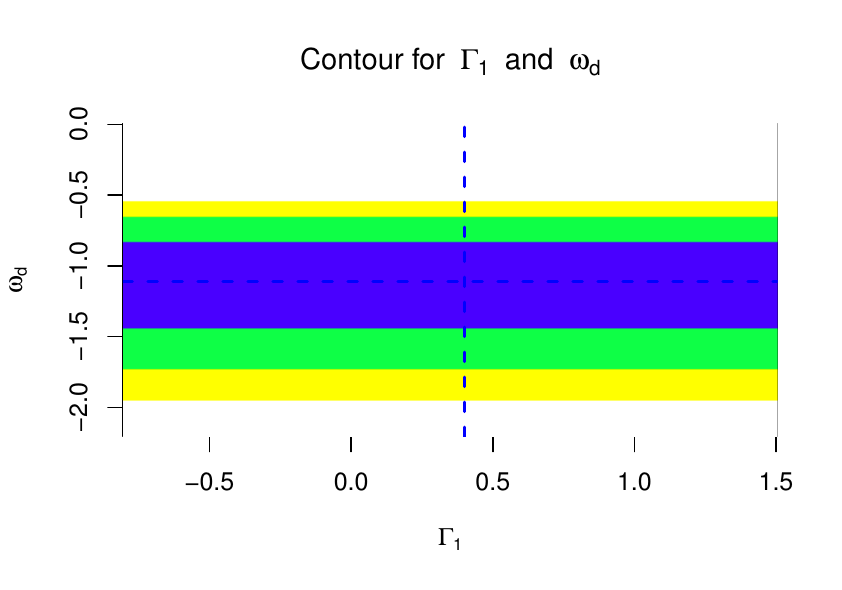}}
	\newline
	
	\centering
	\subfigure[]{\includegraphics[width= 0.3\textwidth]{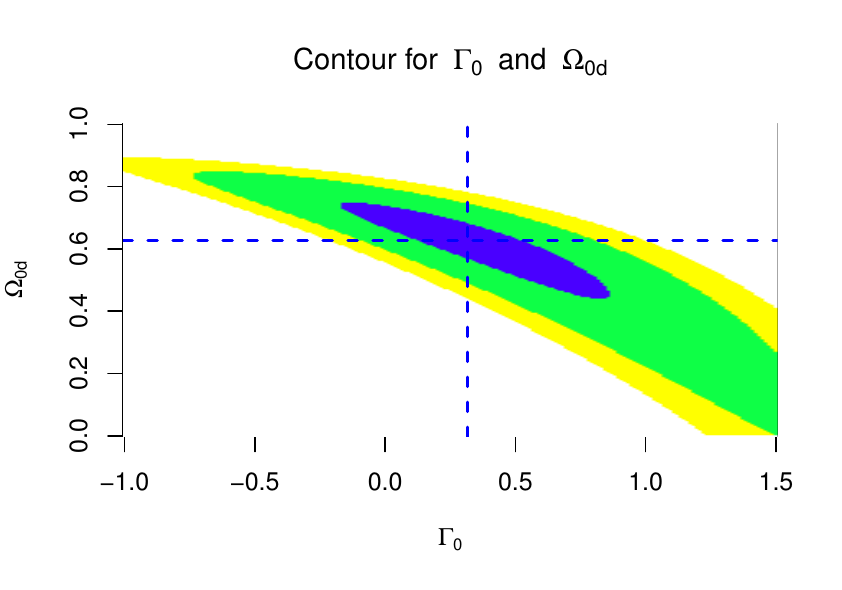}}
	\subfigure[]{\includegraphics[width= 0.3\textwidth]{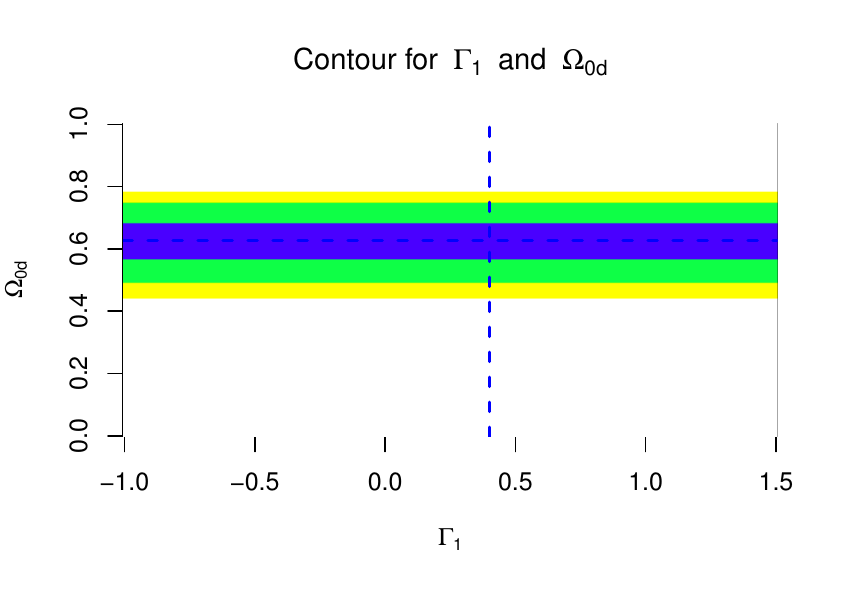}}
	\subfigure[]{\includegraphics[width= 0.3\textwidth]{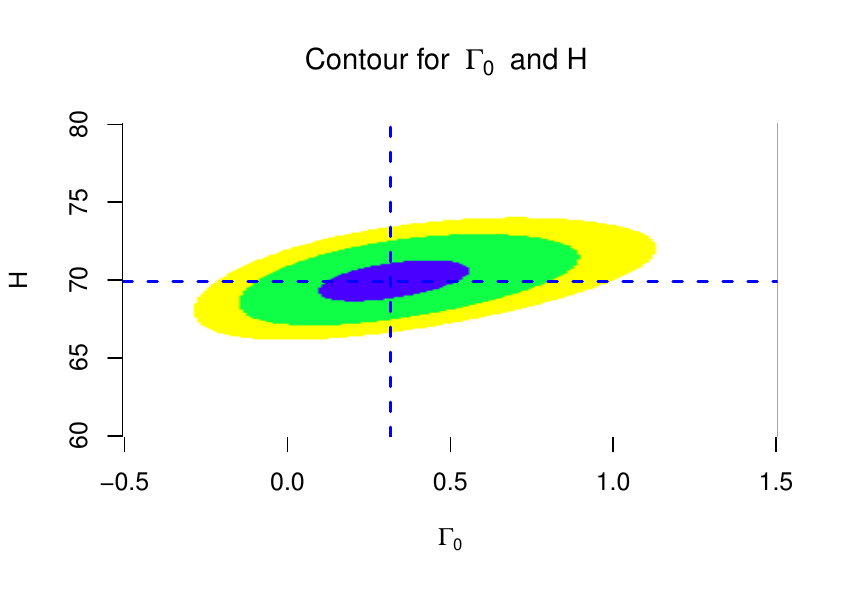}}
	\newline

	\centering
	\subfigure[]{\includegraphics[width= 0.3\textwidth]{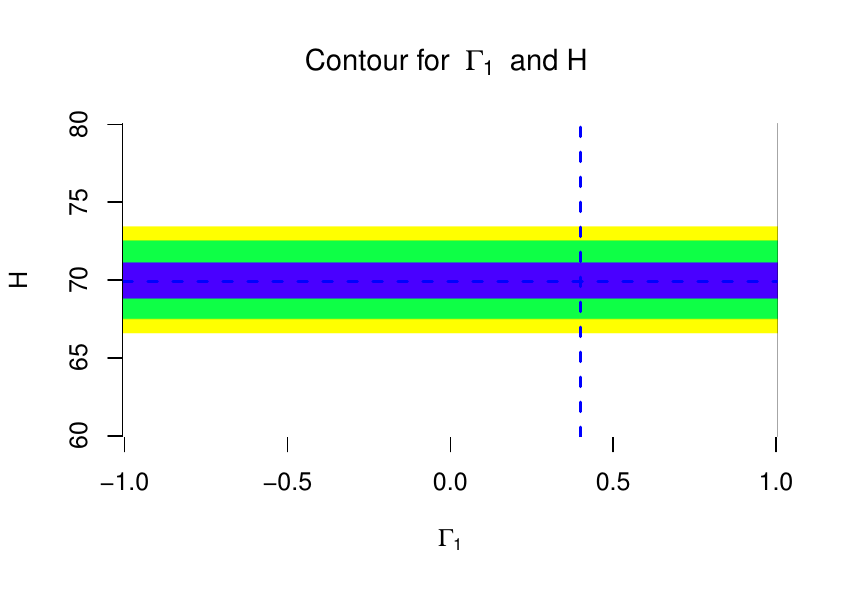}}
	\subfigure[]{\includegraphics[width= 0.3\textwidth]{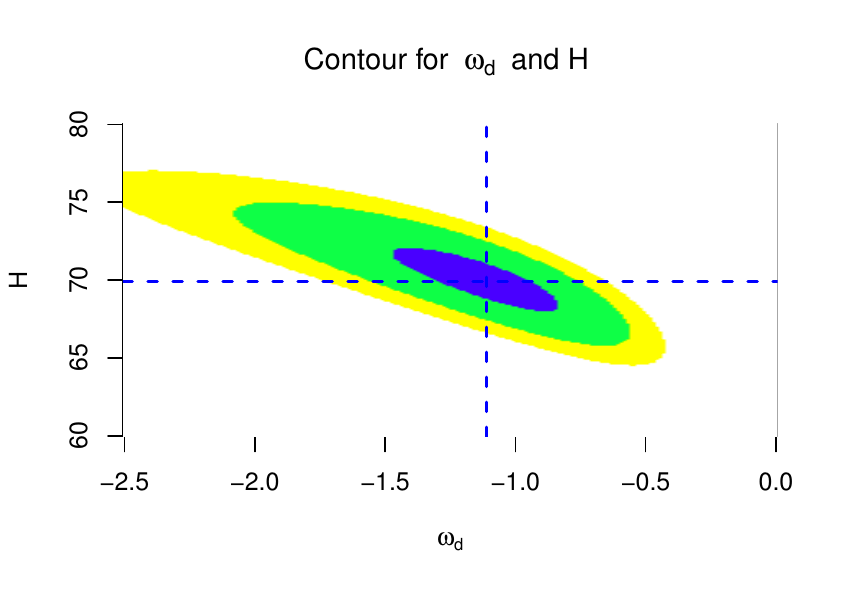}}
	\subfigure[]{\includegraphics[width= 0.3\textwidth]{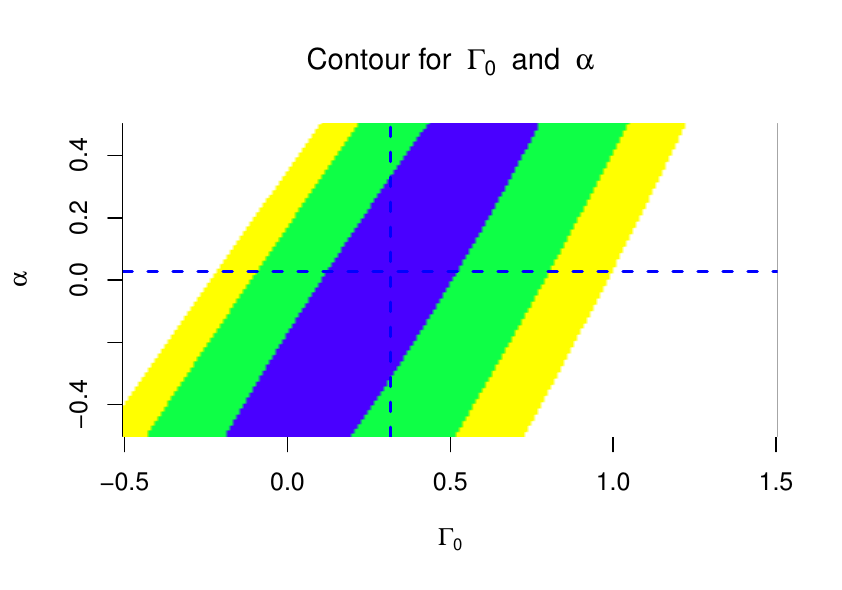}}
	\newline
	
	\centering
	\subfigure[]{\includegraphics[width= 0.3\textwidth]{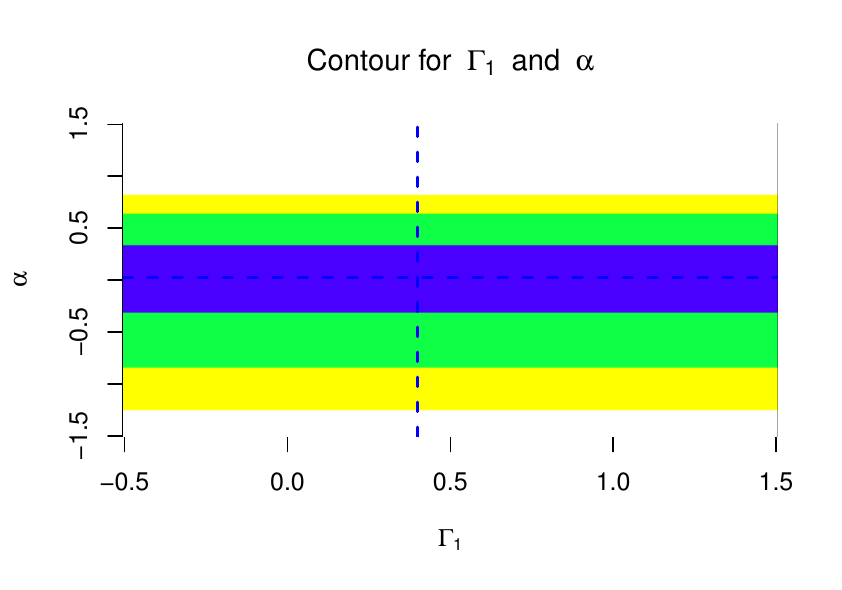}}
	\subfigure[]{\includegraphics[width= 0.3\textwidth]{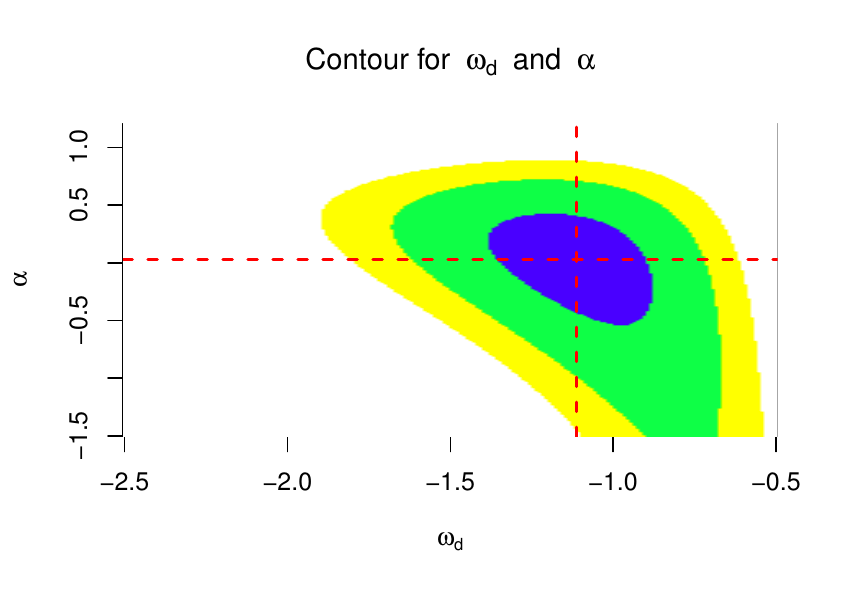}}
	\subfigure[]{\includegraphics[width= 0.3\textwidth]{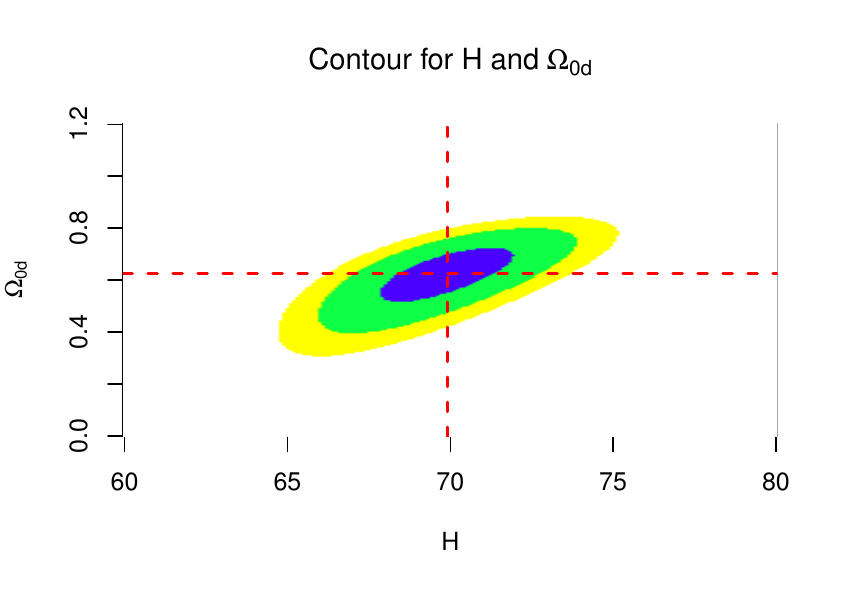}}
	\caption{$1\sigma$ (blue), $2\sigma$ (green) and $3\sigma$(yellow) confidence contours between pairs of parameters based on SN Ia+OHV datasets for present particle creation model}
	\label{fig:1}
\end{figure}
%%%%%%%%%%%%%

%%%%%%%%%%%%%

\begin{figure}[tb]
	\includegraphics[width= 0.6\columnwidth]{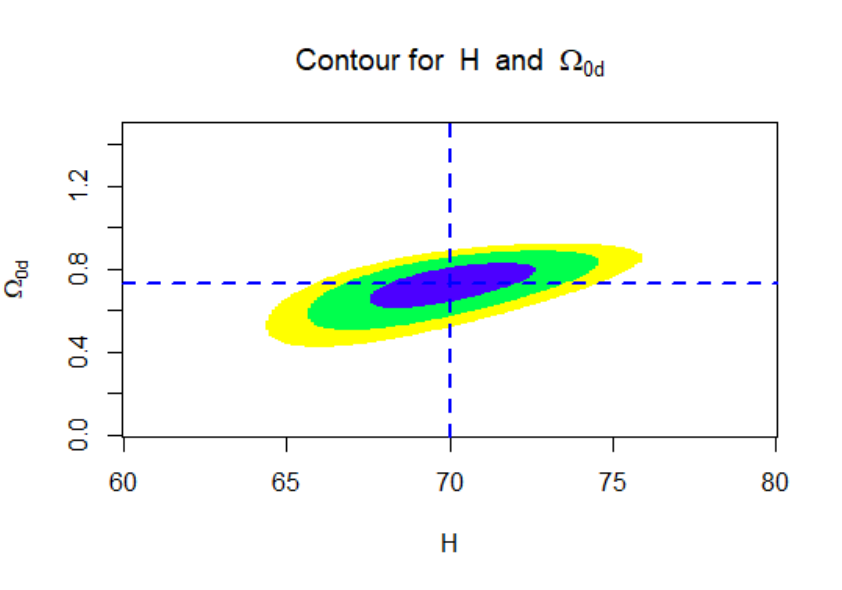}
	\caption{$1\sigma$ (blue), $2\sigma$ (green) and $3\sigma$(yellow)$\Gamma_{0d}\sim H$ confidence contour based on SN Ia+OHV datasets for $\Lambda$CDM model}
	\label{fig:2}
\end{figure}
\begin{figure}[tb]
	\includegraphics[width= 0.6\columnwidth]{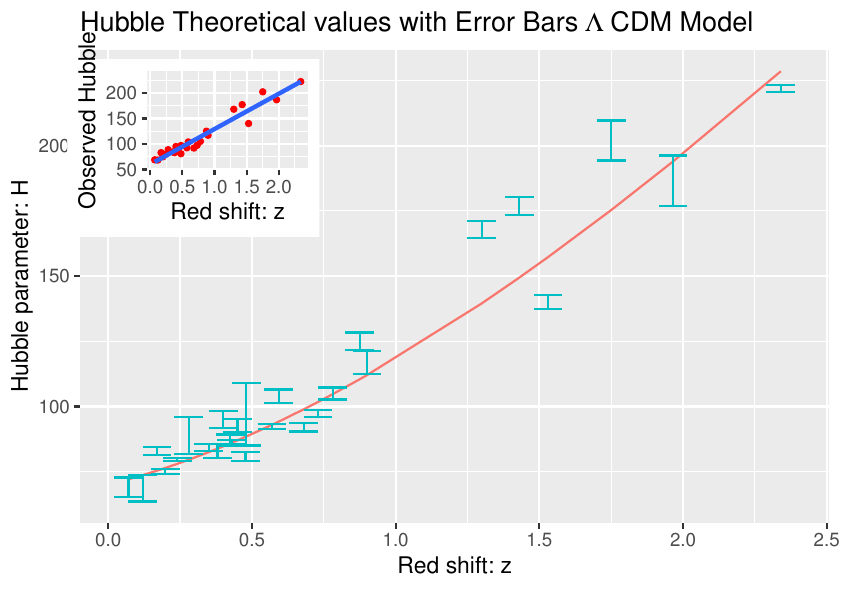}
	\caption{Graphical representation of~ $H-Z$~ relation for SN Ia+OHV datasets for $\Lambda$CDM model}
	\label{fig:3}
\end{figure}

\begin{figure}[tb]
	\includegraphics[width= 0.6\columnwidth]{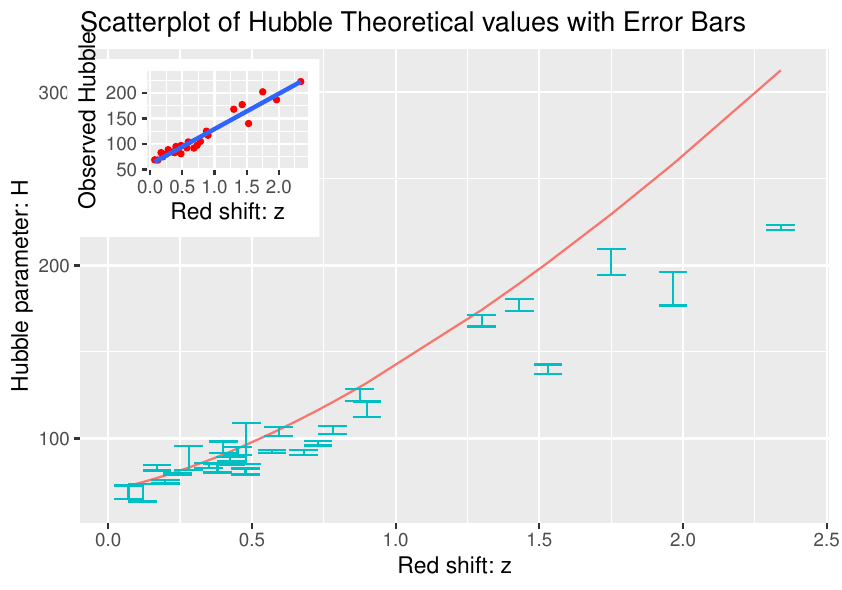}
	\caption{Graphical representation of~ $H-Z$~ relation for SN Ia+OHV datasets in the present particle creation model}
	\label{fig:4}
\end{figure}
%%%%%%%%%%%
i.e.,  correlation coefficient $r$ between $H_{theo}$ (obtained from equation (\ref{19})) and $H_{obs}$. Now, $r$ is calculated from the following equation :
\begin{equation}
	r=\frac{\sum_{i=1}^{27}\left(H_{obs}(z_i)-\bar{H}_{obs}\right)\left(H_{theo}(z_i)-\bar{H}_{theo}\right)}{(n-1)S_x S_y}\label{36}
\end{equation}
where $\bar{H}_{obs}$ and $\bar{H}_{theo}$ indicate mean values of observational and theoretical $H(z)$ datasets,respectively and $S_x,~ S_y$ are given by
\begin{eqnarray*}
	S_x &=& \sqrt{\frac{\left(H_{obs}(z_i)-\bar{H}_{obs}\right)^2}{n-1}} \\
	S_y &=&\sqrt{\frac{ \left(H_{theo}(z_i)-\bar{H}_{theo}\right)^2}{n-1}}
\end{eqnarray*}
It is to be mentioned that $r$ always lies between -1 and +1. The absolute value of the correlation
coefficient, i.e., $|r|$ indicates the relationship strength. To be specific, the larger the
absolute value of correlation coefficient, the stronger the linear relationship.  $r=0$ indicates no relationship between two variables, while $r=\pm 1$ implies that the points are on a perfect straight line with a positive (negative) slope. Now, for the present model, we have calculated $r=0.965363$ which indicates a strong relationship between $H_{obs}$ and $H_{theo}$. This is also evident from the error diagram figs.~\ref{fig:3} and \ref{fig:4}. Also, figs. \ref{fig:3} and \ref{fig:4} illustrate the evolutionary nature of the cosmic Hubble parameter according to the $\Lambda$CDM and the particle creation models, respectively, in the 1$\sigma$ confidence region. Note that the circles in figs. \ref{fig:3} and \ref{fig:4}, show the recent observable values. The two plots are pretty similar  but deviate at high redshift region. On the other hand, in figs. \ref{fig:1} and \ref{fig:2}, contour plots between different parameters in particle creation and $\Lambda$CDM model, respectively, have been shown. It is seen that contour plot for the particle creation model between $\Omega_{0d}$ and $H$ is very similar to that of $\Lambda$CDM model.

%%%%%%%%%%%%%%%%%%%%%%%%%%%%%%%%%%%%%%%%%%%%

\section{State finder diagnosis}
\begin{figure}
	\centering
	\subfigure[]{
		\includegraphics[width=0.45\textwidth]{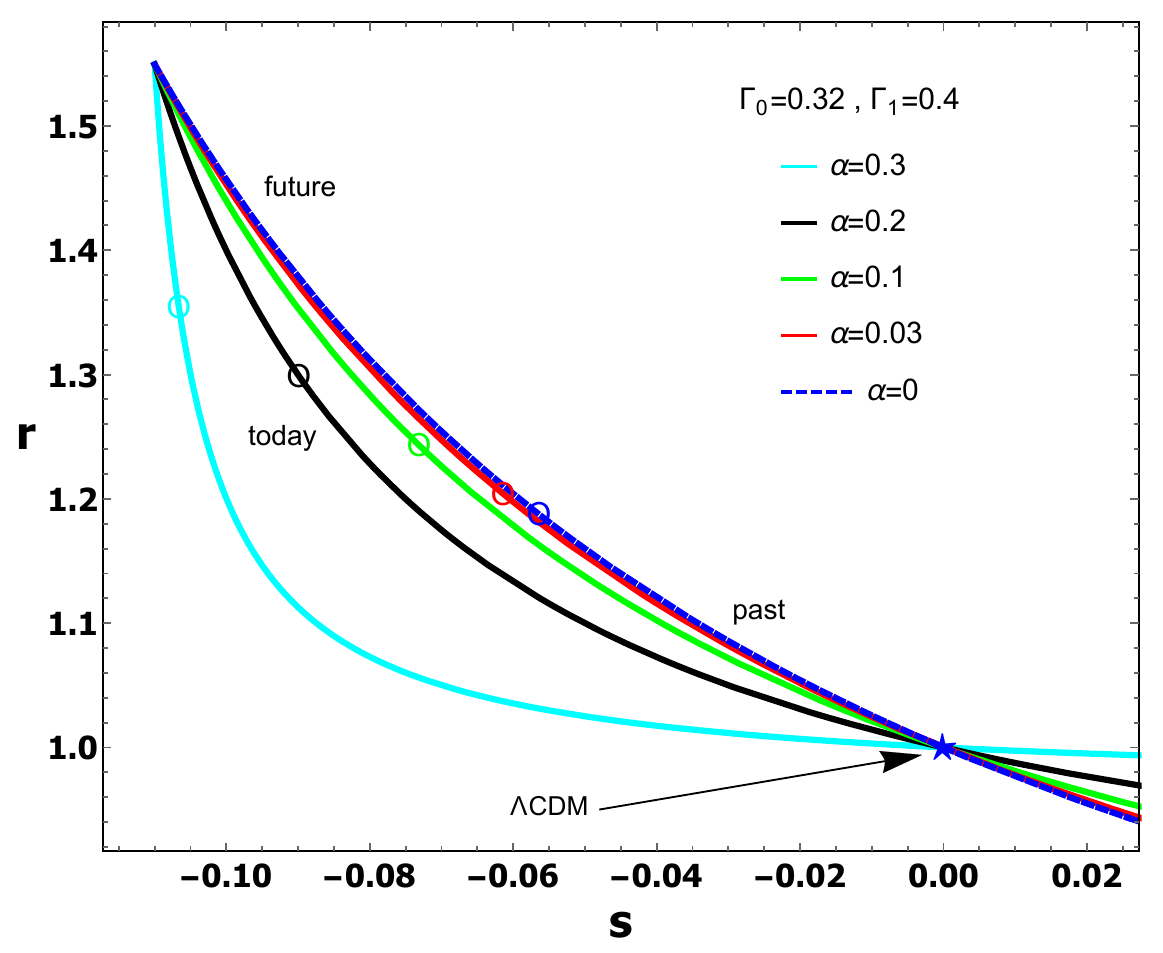}\label{sr-phantom-alpha}}
	\subfigure[]{
		\includegraphics[width=0.45\textwidth]{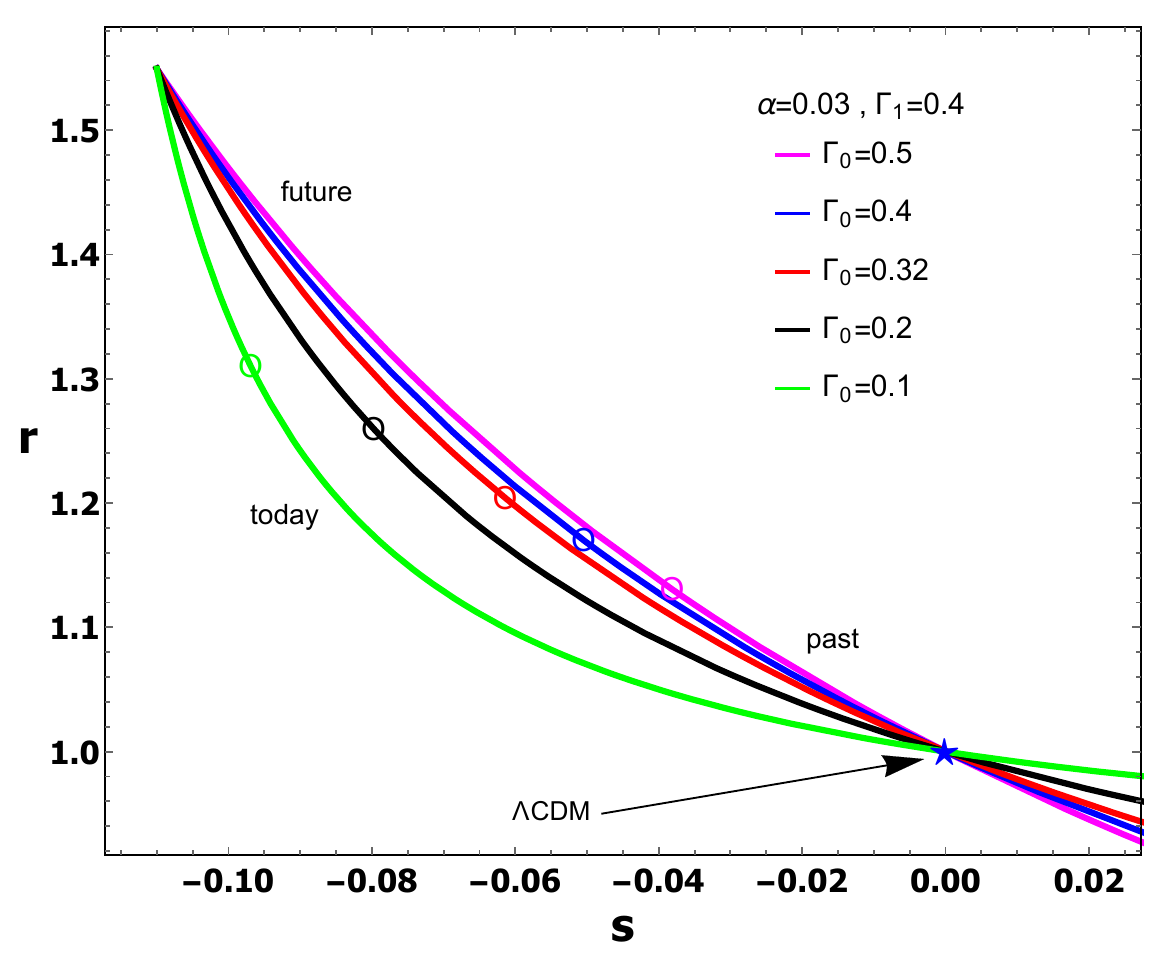}\label{sr-phantom-Gamma_{0}}}
	\subfigure[]{
		\includegraphics[width=0.46\textwidth]{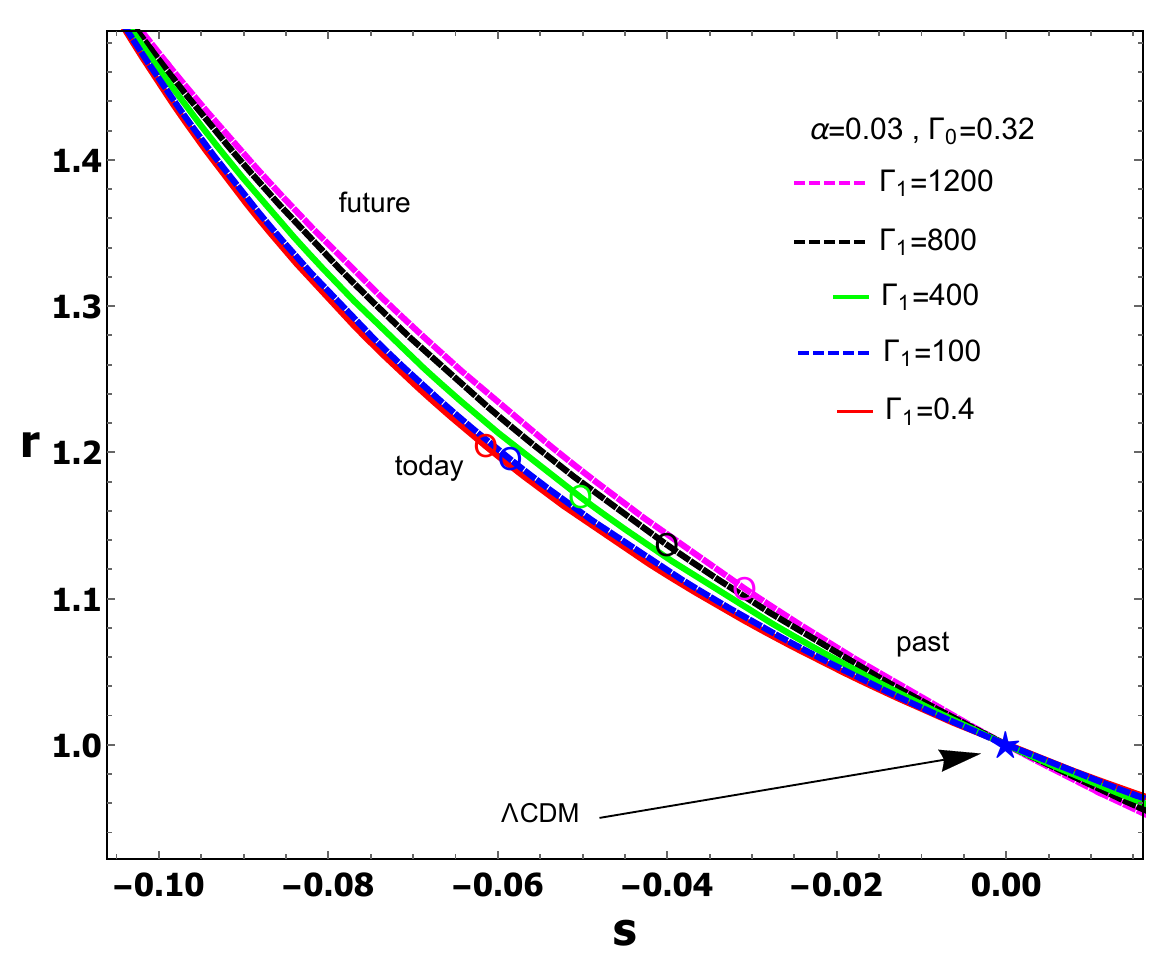}\label{sr-phantom-Gamma_{1}}}
	\caption{The figures show different time evolution trajectories of the state finder pair $(s,r)$ for
the best fit values of cosmological parameters $(H_{0},~\Omega_{0m},~\Omega_{0d})=(69.92,~0.37,~0.63)$ and $\omega_{d}=-1.11$ and for different values of free parameters as indicated in each panel.		
 Panel $(a)$ exhibits the trajectories for the values of free paramters $(\Gamma_{0},~\Gamma_{1})=(0.32,~0.4)$ and for different values of $\alpha$. Panel $(b)$ shows the trajectories for different values of $\Gamma_{0}$ when the values of free parameters are chosen to be $(\alpha,~\Gamma_{1})=(0.03,~0.4)$. In panel $(c)$, the trajectories are shown for different values of $\Gamma_{1}$ with the best fit values of free parameters $(\alpha,~\Gamma_{0})=(0.03,~0.32)$.
In each panel, the colored circles denote the present value of the state finder parameter $(s_{0},r_{0})$ and the blue star corresponds to the $\Lambda $CDM model.}
	\label{sr-phantom}
\end{figure}

\begin{figure}
	\centering
	\subfigure[]{
		\includegraphics[width=0.45\textwidth]{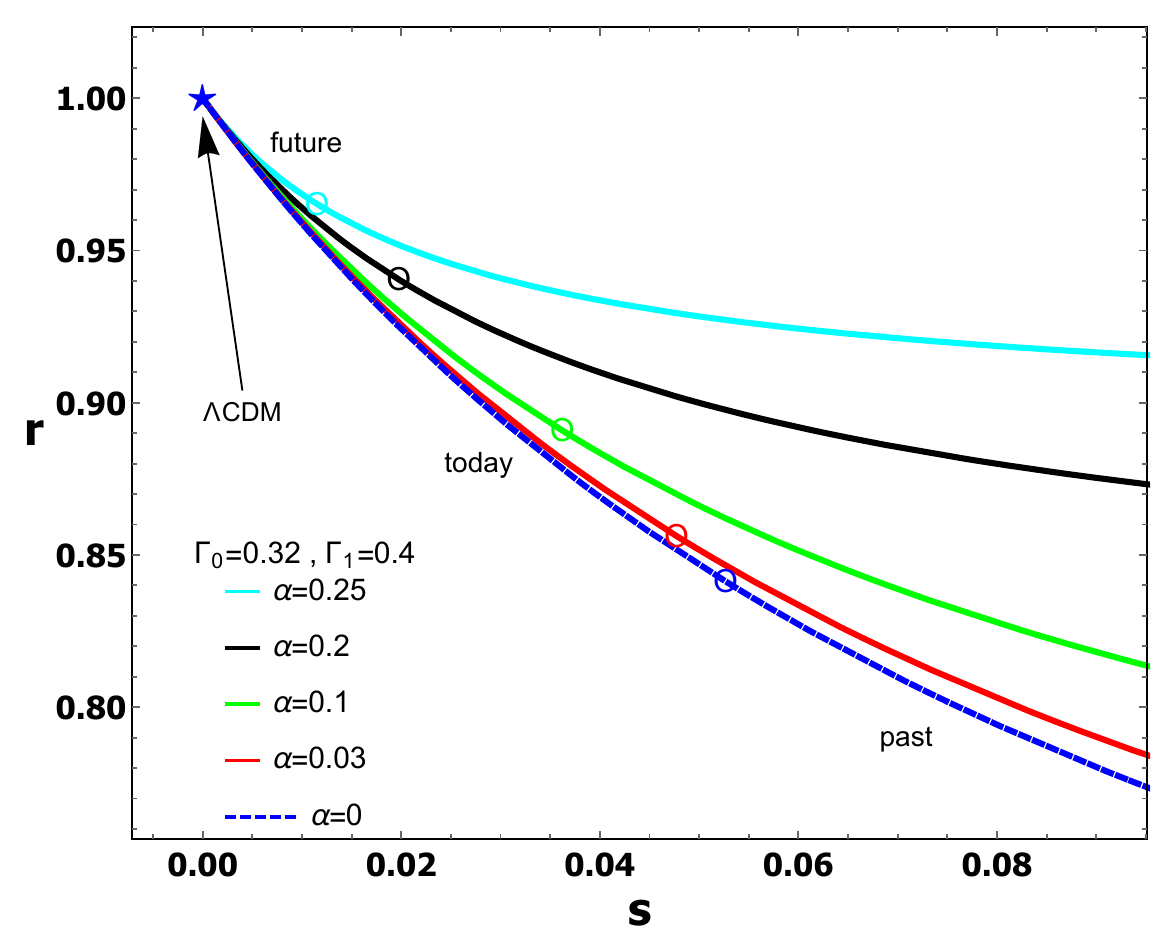}\label{sr-cosmological constant-alpha}}
	\subfigure[]{
		\includegraphics[width=0.45\textwidth]{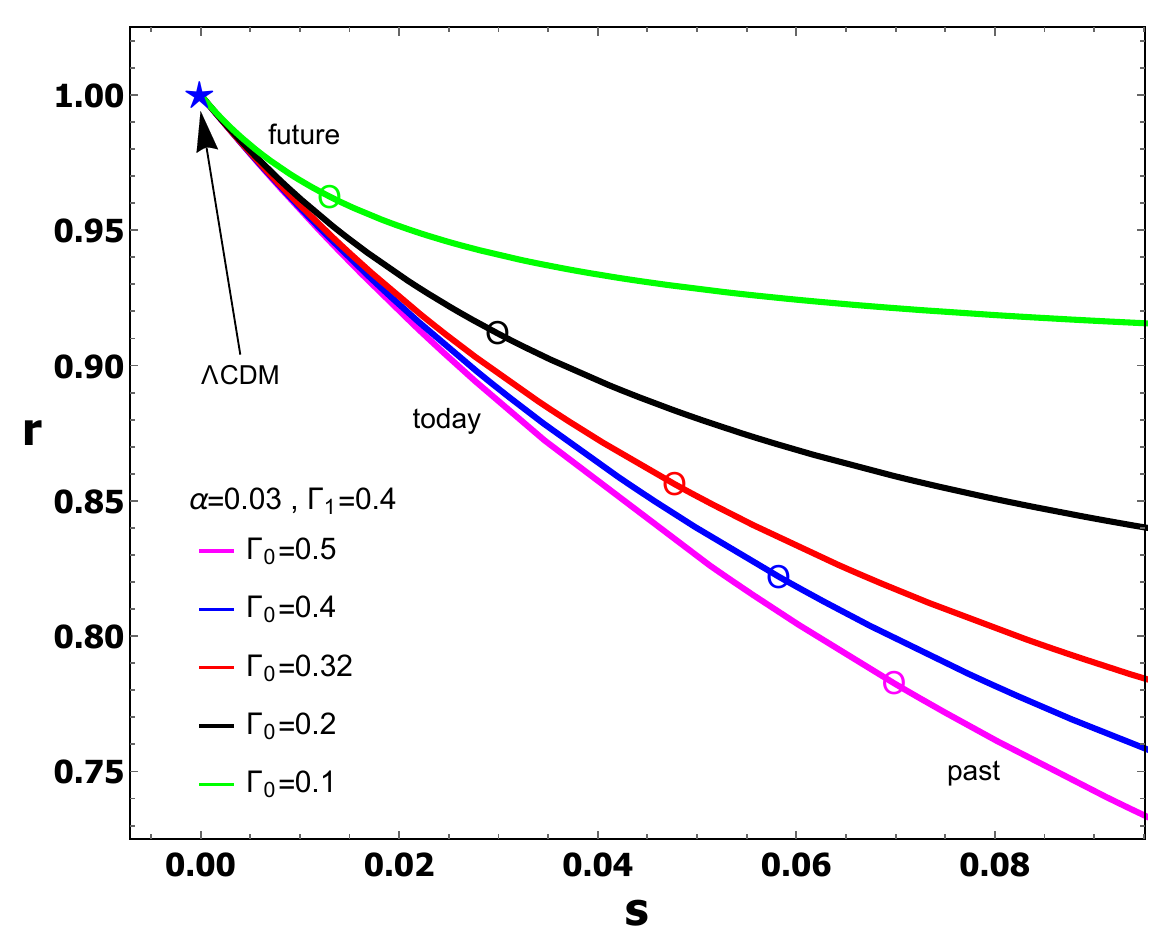}\label{sr-cosmological constant-Gamma_{0}}}
	\subfigure[]{
		\includegraphics[width=0.46\textwidth]{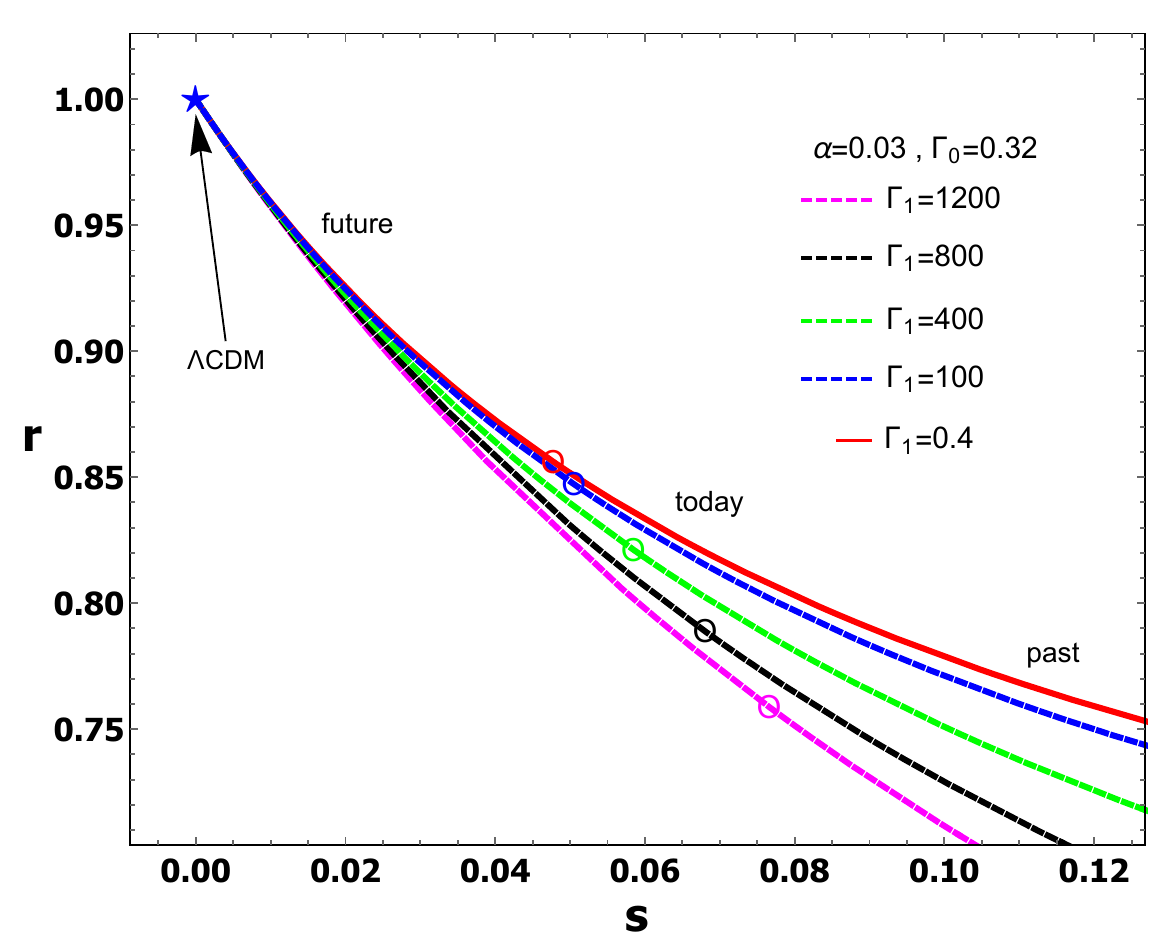}\label{sr-cosmological constant-Gamma_{1}}}
	\caption{The figures show different time evolution trajectories of the state finder pair $(s,r)$ for 
	the best fit values of cosmological parameters $(H_{0},~\Omega_{0m},~\Omega_{0d})=(69.92,~0.37,~0.63)$ and $\omega_{d}=-1$ and for different values of free parameters as indicated in each panel.		
		Panel $(a)$ shows the trajectories for the values of free paramters $(\Gamma_{0},~\Gamma_{1})=(0.32,~0.4)$ and for different values of $\alpha$. Panel $(b)$ exhibits the trajectories for different values of $\Gamma_{0}$ when the best fit values of free parameters are chosen to be $(\alpha,~\Gamma_{1})=(0.03,~0.4)$. In panel $(c)$, the trajectories are shown for different values of $\Gamma_{1}$ with the best fit values of free parameters $(\alpha,~\Gamma_{0})=(0.03,~0.32)$.
	In each panel, the colored circles denote the present value of the state finder parameter $(s_{0},r_{0})$ and the blue star corresponds to the $\Lambda$CDM model.}
	\label{sr-cosmological constant}
\end{figure}

\begin{figure}
	\centering
	\subfigure[]{
		\includegraphics[width=0.45\textwidth]{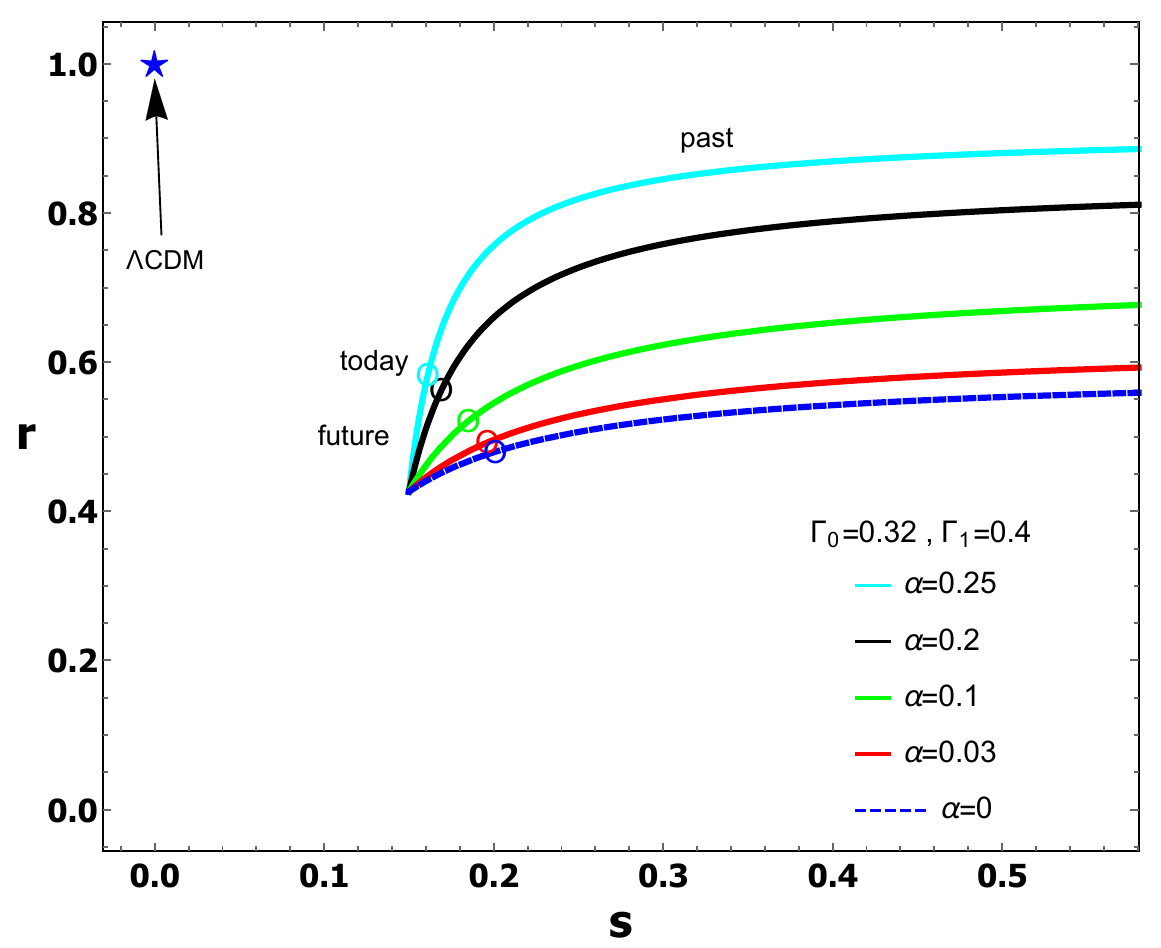}\label{sr-quintessence-alpha}}
	\subfigure[]{
		\includegraphics[width=0.45\textwidth]{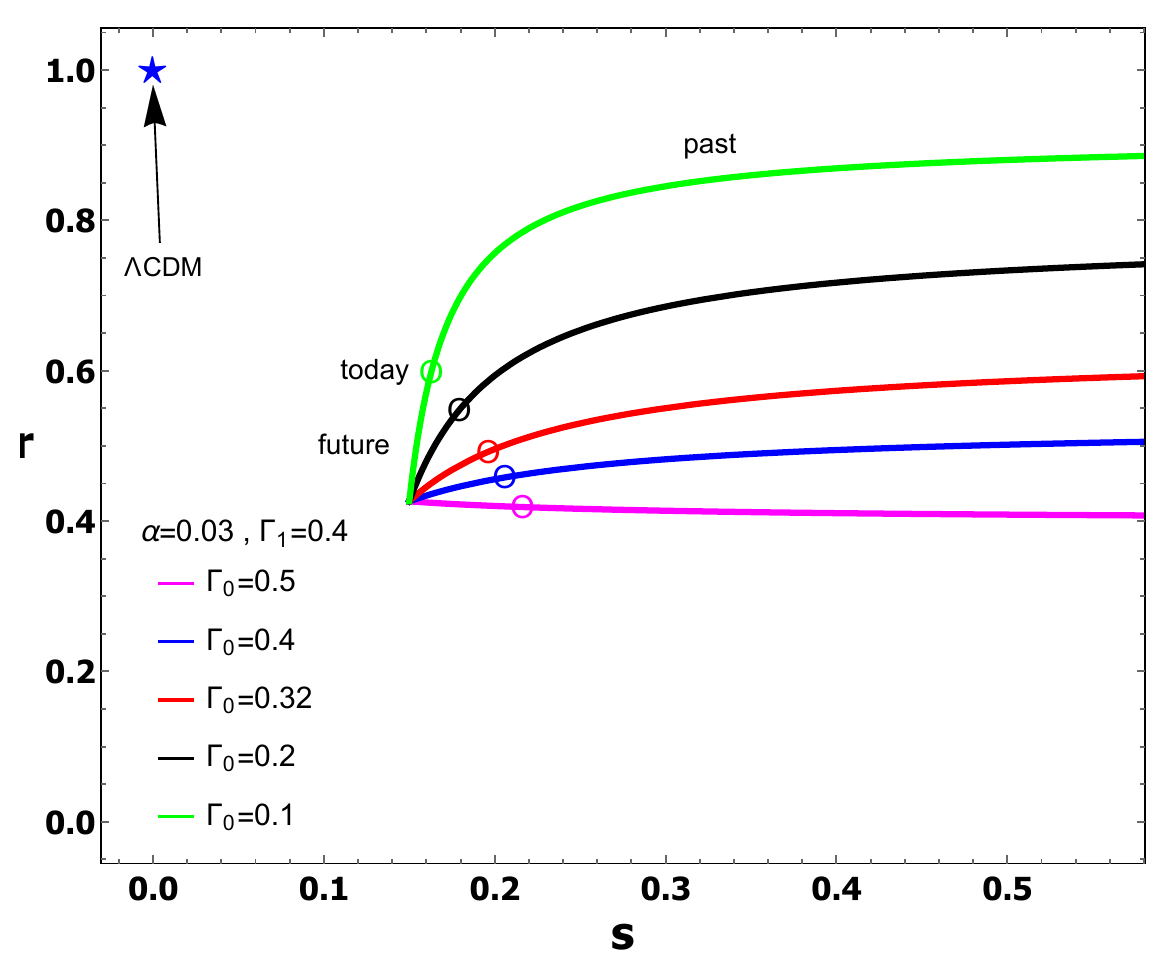}\label{sr-quintessence-Gamma_{0}}}
	\subfigure[]{
		\includegraphics[width=0.48\textwidth]{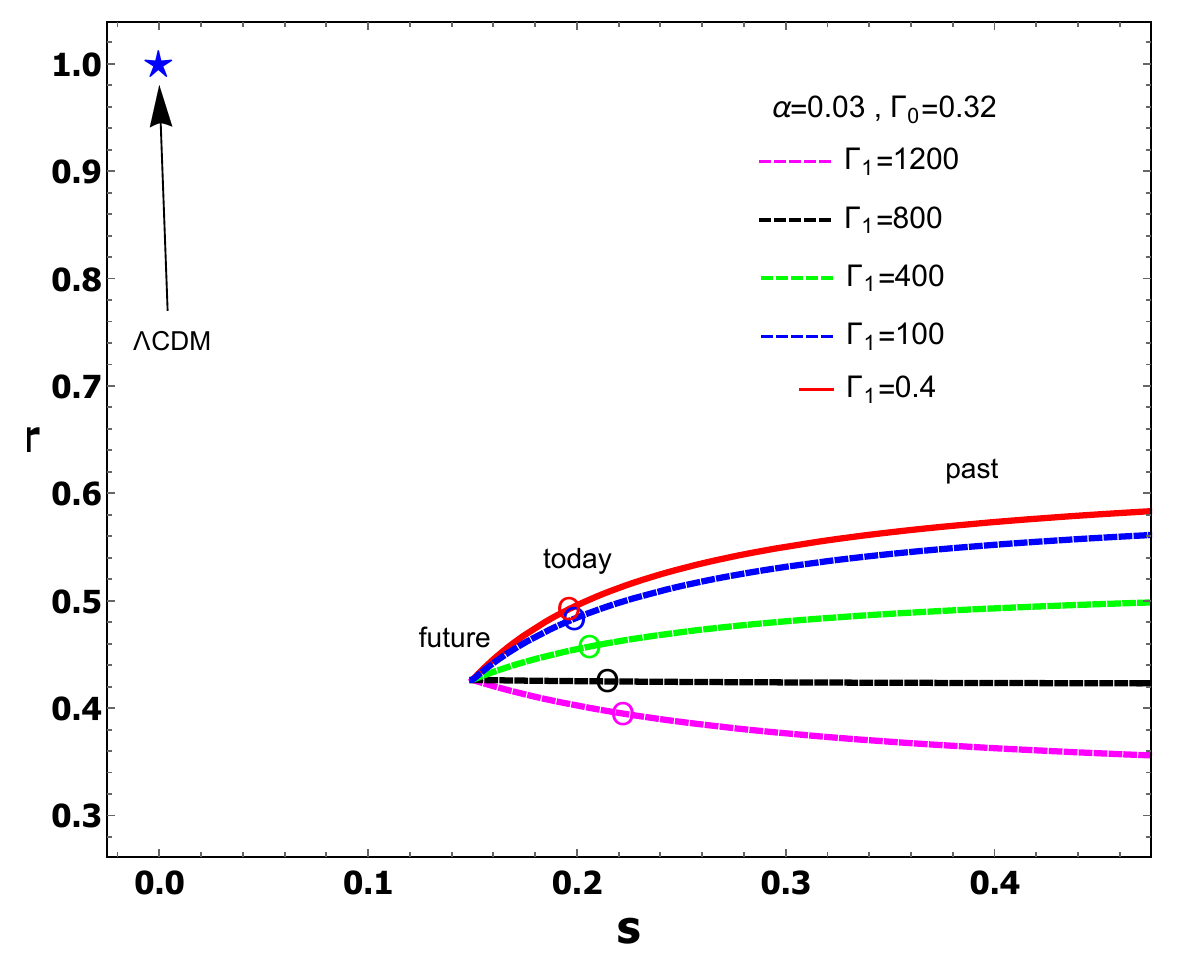}\label{sr-quintessence-Gamma_{1}}}
	\caption{The figures show different time evolution trajectories of the state finder pair $(s,r)$ for 
 the best fit values of cosmological parameters $(H_{0},~\Omega_{0m},~\Omega_{0d})=(69.92,~0.37,~0.63)$ and $\omega_{d}=-0.85$ and for different values of free parameters as indicated in each panel.		
 	Panel $(a)$ exhibits the trajectories for the values of free paramters $(\Gamma_{0},~\Gamma_{1})=(0.32,~0.4)$ and for different values of $\alpha$. Panel $(b)$ shows the trajectories for different values of $\Gamma_{0}$ when the best fit values of free parameters are chosen to be $(\alpha,~\Gamma_{1})=(0.03,~0.4)$. In panel $(c)$, the trajectories are shown for different values of $\Gamma_{1}$ with the best fit values of free parameters $(\alpha,~\Gamma_{0})=(0.03,~0.32)$.
 In each panel, the colored circles denote the present value of the state finder parameter $(s_{0},r_{0})$ and the blue star corresponds to the $\Lambda$CDM model. }
	\label{sr-quintessence}
\end{figure}

\begin{figure}
	\centering
	\subfigure[]{
		\includegraphics[width=0.45\textwidth]{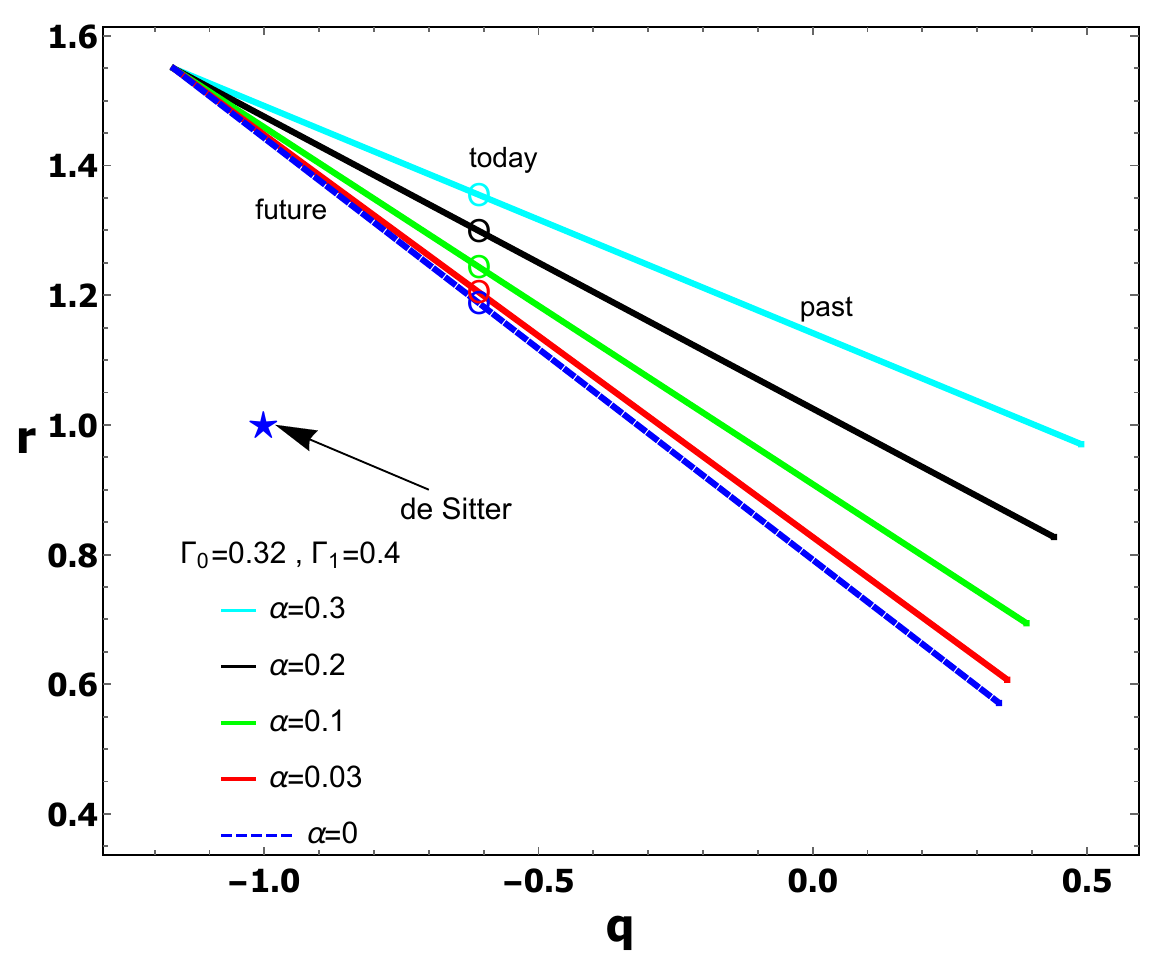}\label{qr-phantom-alpha}}
	\subfigure[]{
		\includegraphics[width=0.45\textwidth]{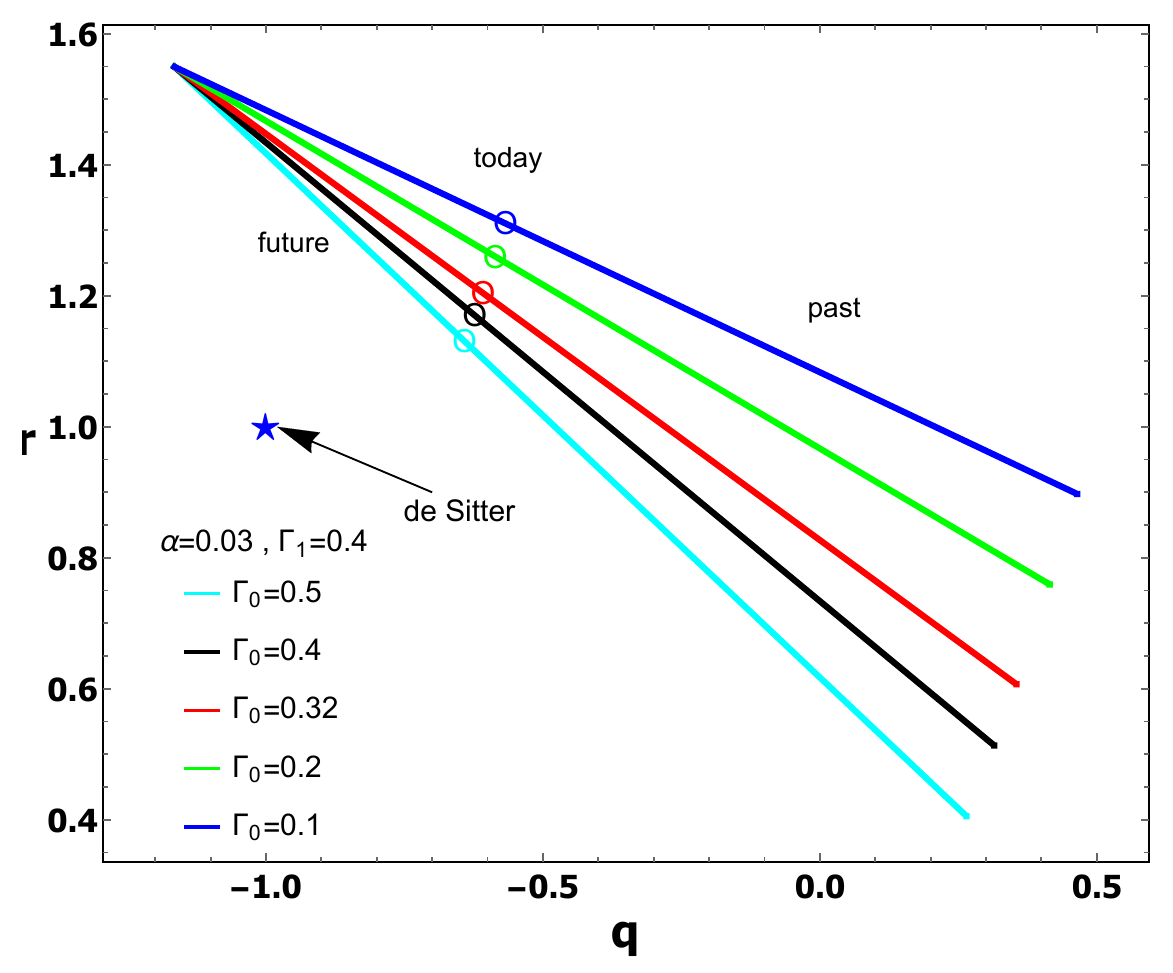}\label{qr-phantom-Gamma_{0}}}
	\subfigure[]{
		\includegraphics[width=0.45\textwidth]{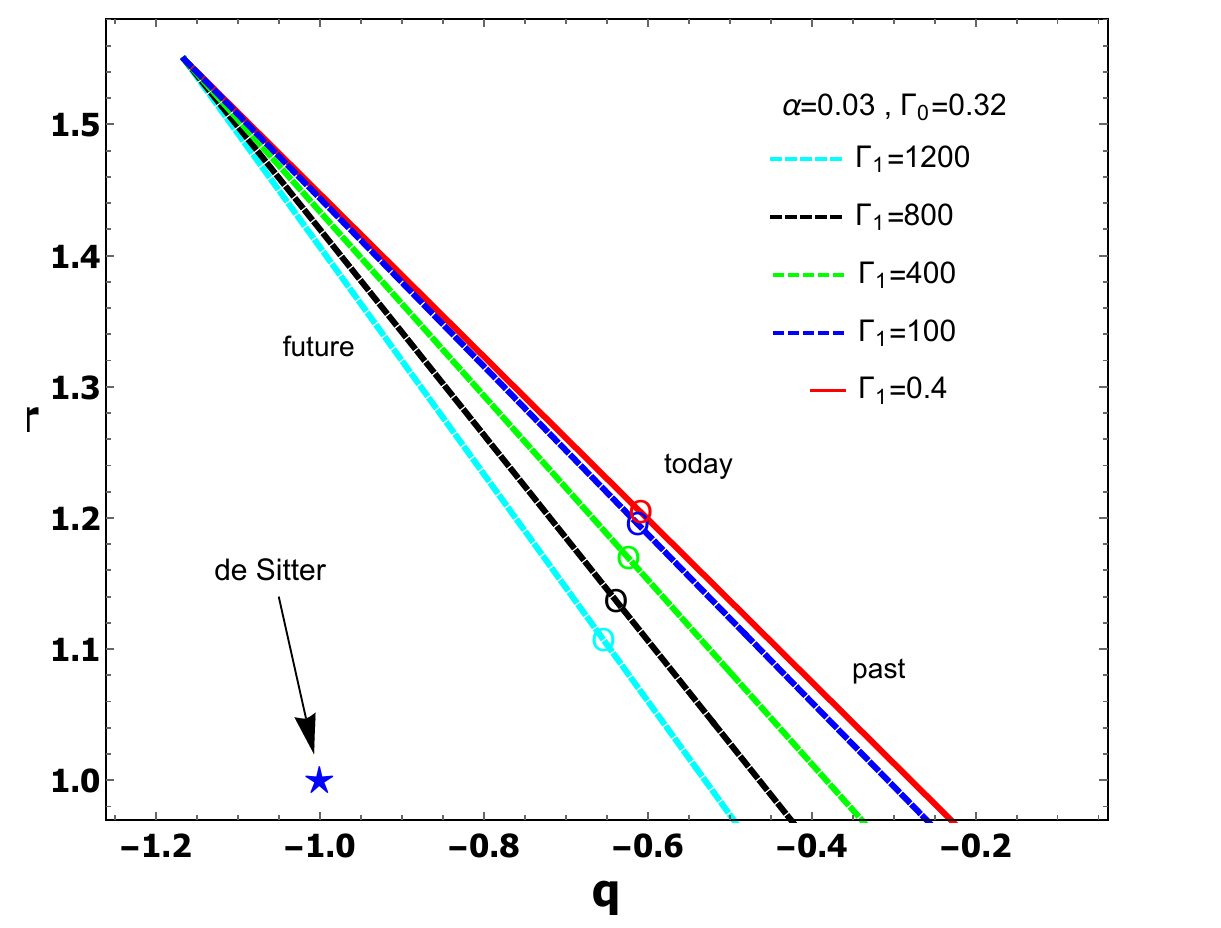}\label{qr-phantom-Gamma_{1}}}
	\caption{The figures show different time evolution trajectories of the parameter pair $(q,r)$ for 
 the best fit values of cosmological parameters $(H_{0},~\Omega_{0m},~\Omega_{0d})=(69.92,~0.37,~0.63)$ and $\omega_{d}=-1.11$ and for different values of free parameters as indicated in each panel.		
		Panel $(a)$ exhibits the trajectories for the values of free paramters $(\Gamma_{0},~\Gamma_{1})=(0.32,~0.4)$ and for different values of $\alpha$. Panel $(b)$ shows the trajectories for different values of $\Gamma_{0}$ when the best fit values of free parameters are chosen to be $(\alpha,~\Gamma_{1})=(0.03,~0.4)$. In panel $(c)$, the trajectories are shown for different values of $\Gamma_{1}$ with the best fit values of free parameters $(\alpha,~\Gamma_{0})=(0.03,~0.32)$.
	In each panel, the colored circles denote the present value of the state finder parameter $(q_{0},r_{0})$ and the blue star corresponds to the $de~Sitter $ model. }
	\label{qr-phantom}
\end{figure}

\begin{figure}
	\centering
	\subfigure[]{
		\includegraphics[width=0.45\textwidth]{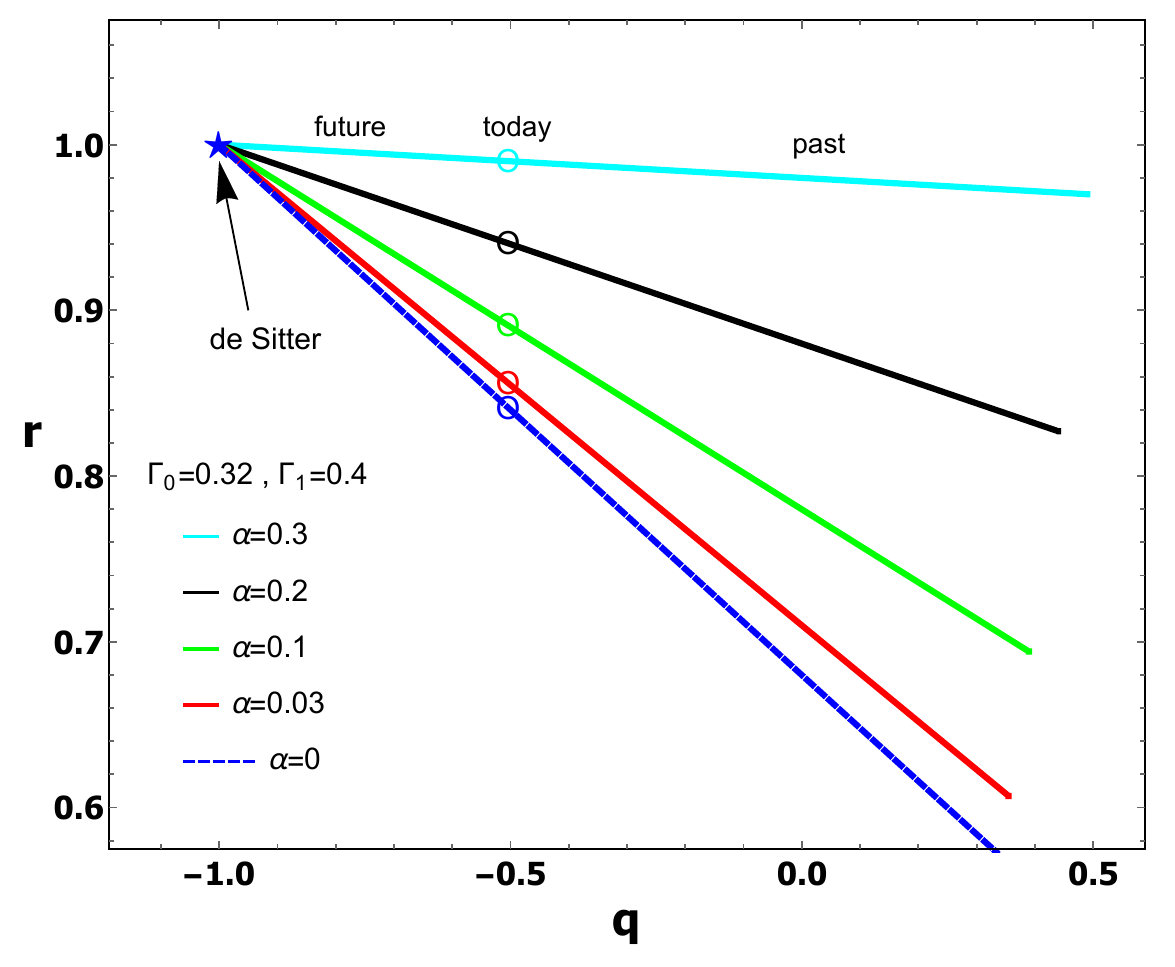}\label{qr-cosmological constant-alpha}}
	\subfigure[]{
		\includegraphics[width=0.45\textwidth]{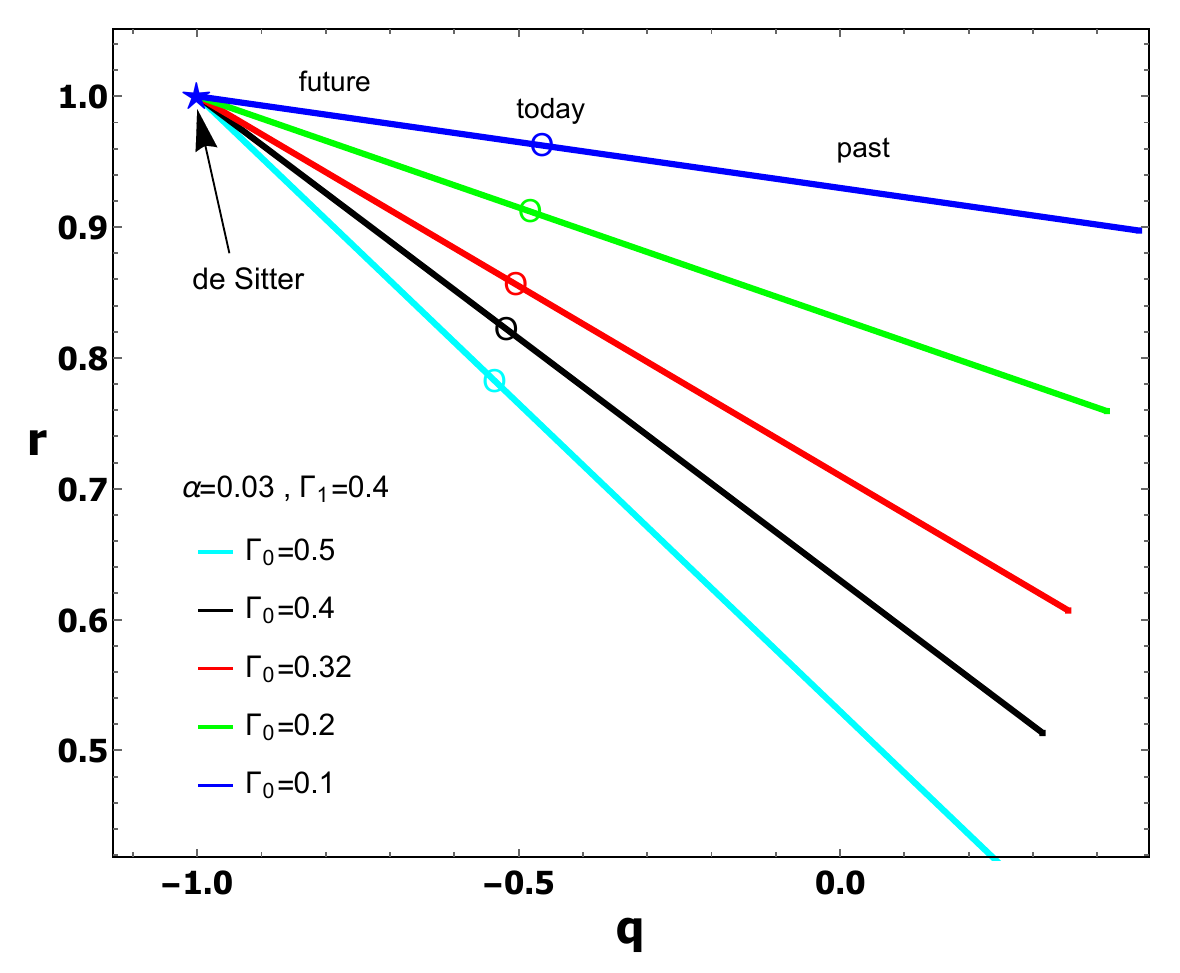}\label{qr-cosmological constant-Gamma_{0}}}
	\subfigure[]{
		\includegraphics[width=0.45\textwidth]{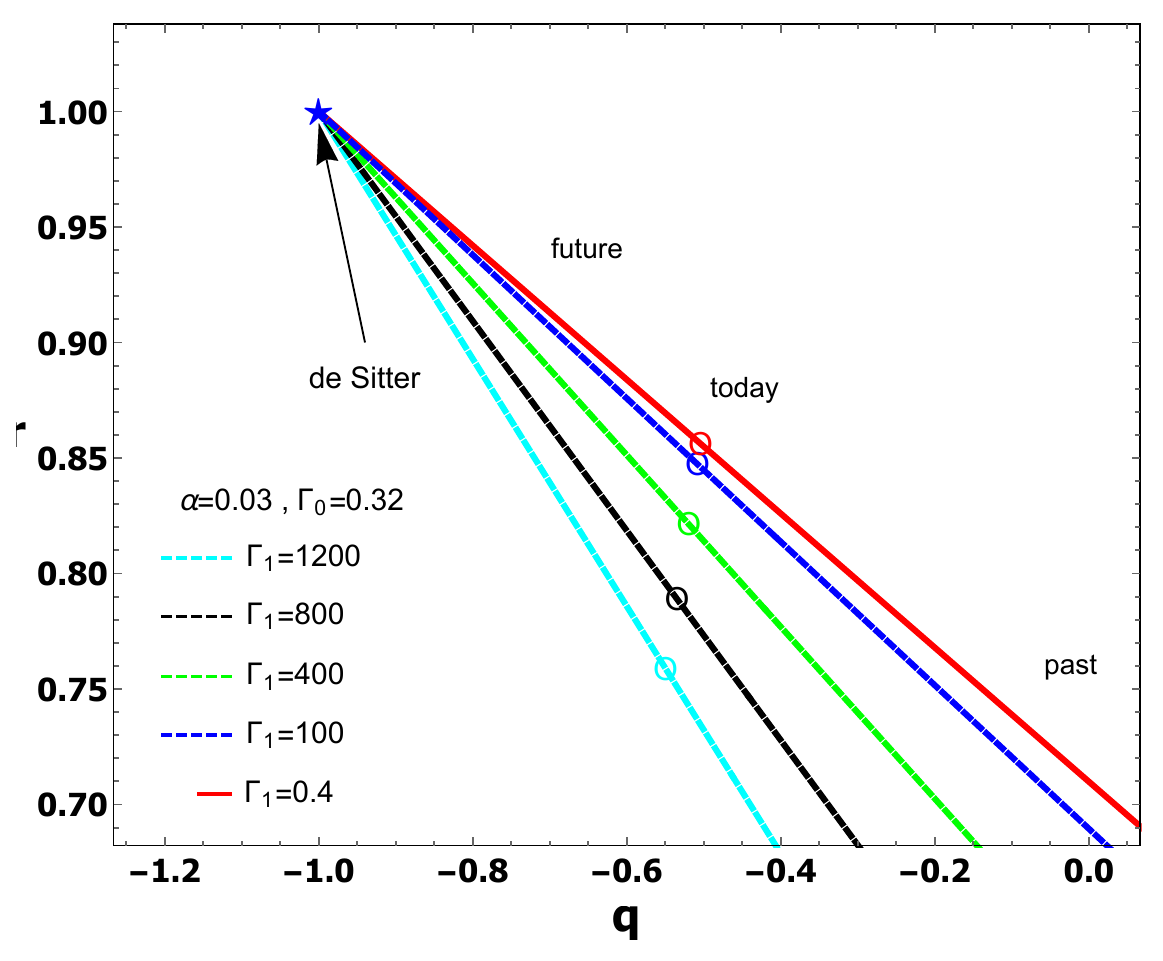}\label{qr-cosmological constant-Gamma_{1}}}
	\caption{The figures show different time evolution trajectories of the parameter pair $(q,r)$ for 
	the best fit values of cosmological parameters $(H_{0},~\Omega_{0m},~\Omega_{0d})=(69.92,~0.37,~0.63)$ and $\omega_{d}=-1$ and for different values of free parameters as indicated in each panel.		
		Panel $(a)$ exhibits the trajectories for the values of free paramters $(\Gamma_{0},~\Gamma_{1})=(0.32,~0.4)$ and for different values of $\alpha$. Panel $(b)$ shows the trajectories for different values of $\Gamma_{0}$ when the best fit values of free parameters are chosen to be $(\alpha,~\Gamma_{1})=(0.03,~0.4)$. In panel $(c)$, the trajectories are shown for different values of $\Gamma_{1}$ with the best fit values of free parameters $(\alpha,~\Gamma_{0})=(0.03,~0.32)$.
	In each panel, the colored circles denote the present value of the state finder parameter $(q_{0},r_{0})$ and the blue star corresponds to the $de~Sitter $ model. }
	\label{qr-cosmological constant}
\end{figure}

\begin{figure}
	\centering
	\subfigure[]{
		\includegraphics[width=0.45\textwidth]{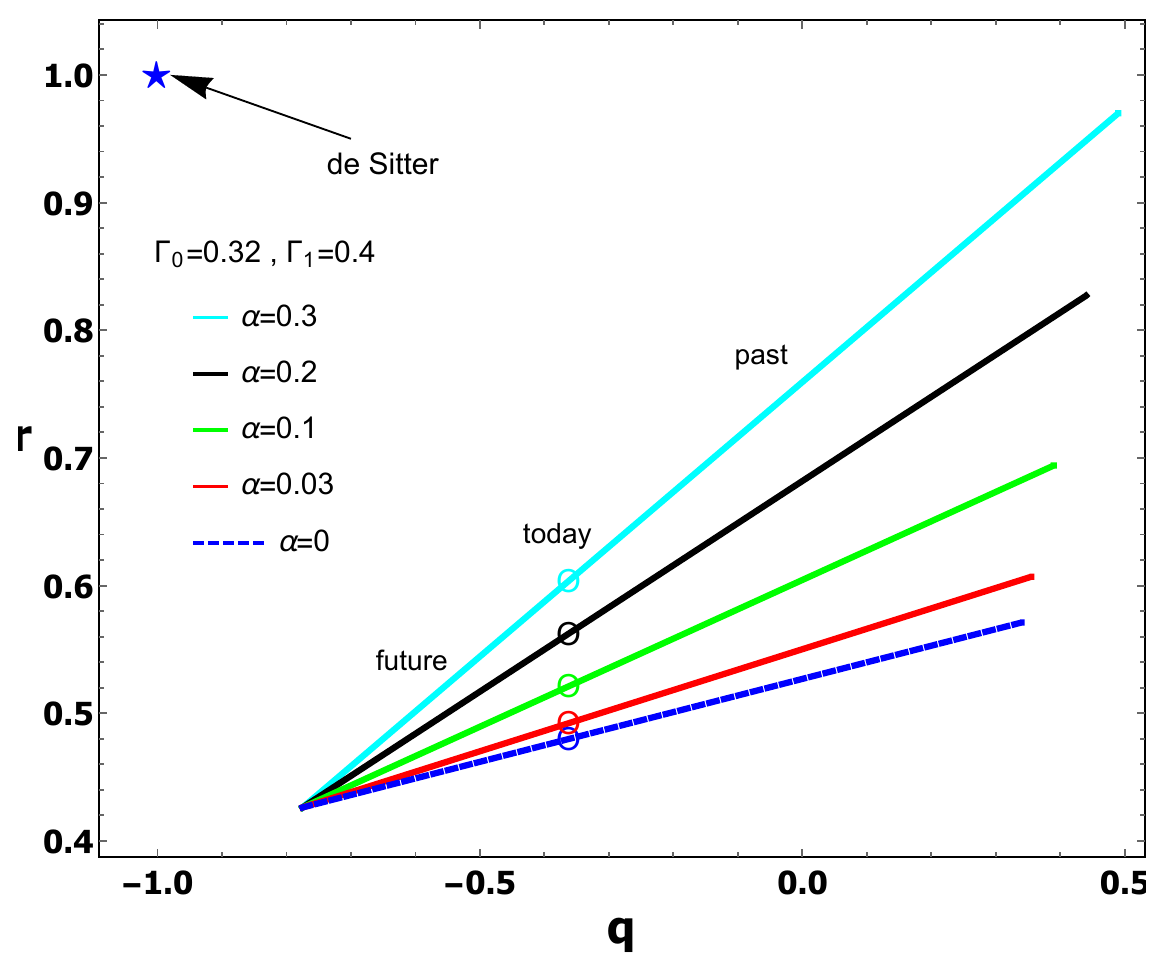}\label{qr-quintessence-alpha}}
	\subfigure[]{
		\includegraphics[width=0.45\textwidth]{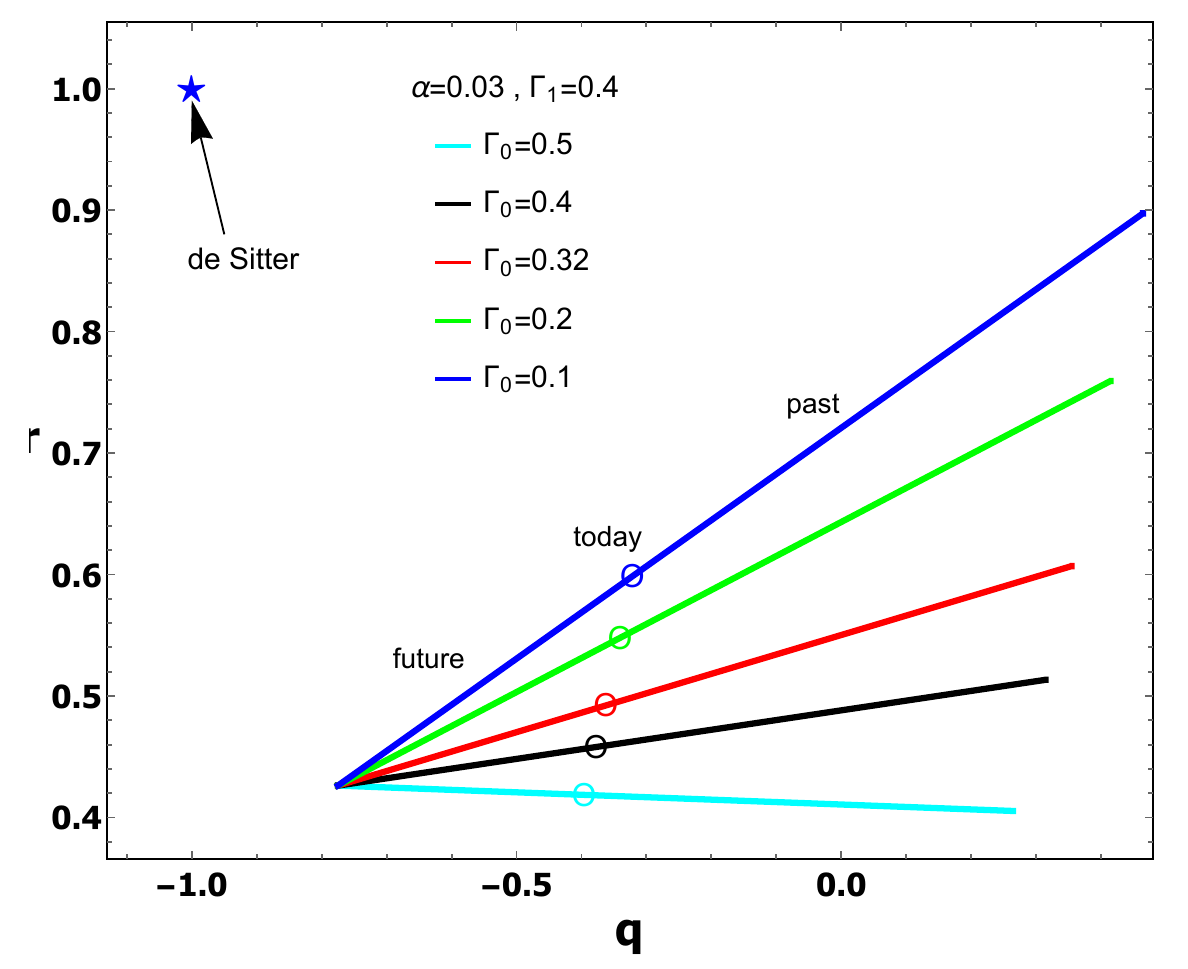}\label{qr-quintessence-Gamma_{0}}}
	\subfigure[]{
		\includegraphics[width=0.45\textwidth]{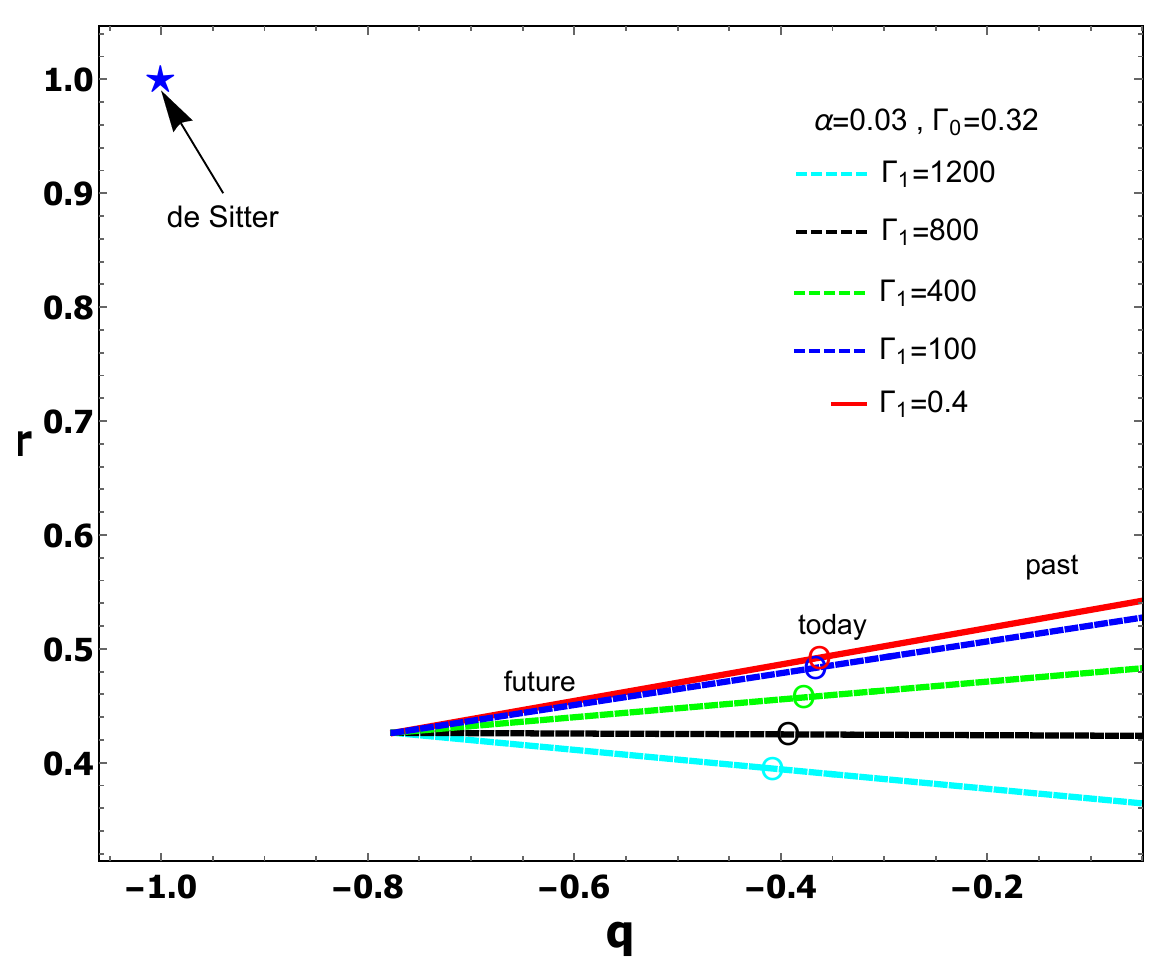}\label{qr-quintessence-Gamma_{1}}}
	\caption{The figures show different time evolution trajectories of the parameter pair $(q,r)$ for 
	the best fit values of cosmological parameters $(H_{0},~\Omega_{0m},~\Omega_{0d})=(69.92,~0.37,~0.63)$ and $\omega_{d}=-0.85$ and for different values of free parameters as indicated in each panel.		
		Panel $(a)$ exhibits the trajectories for the values of free paramters $(\Gamma_{0},~\Gamma_{1})=(0.32,~0.4)$ and for different values of $\alpha$. Panel $(b)$ shows the trajectories for different values of $\Gamma_{0}$ when the best fit values of free parameters are chosen to be $(\alpha,~\Gamma_{1})=(0.03,~0.4)$. In panel $(c)$, the trajectories are shown for different values of $\Gamma_{1}$ with the best fit values of free parameters $(\alpha,~\Gamma_{0})=(0.03,~0.32)$.
	In each panel, the colored circles denote the present value of the state finder parameter $(q_{0},r_{0})$ and the blue star corresponds to the $de~Sitter $ model. }
	\label{qr-quintessence}
\end{figure}

%%%%%%%%%%%%%%%%%
From the literature, it is evident that almost all theoretical dark energy models can predict the same late-time accelerated expansion ($H>0$ and $q<0$) of the universe. Many difficulties arise in discriminating between them. Then, it is necessary to introduce and study some new quantities (other than the known ones) which can successfully discriminate various DE models at the background level. In this context, a higher order (particularly third order) of the time derivative of the scale factor is introduced, and these would appear as a so-called statefinder parameters $r$ and $s$. The statefinder parameters are defined as the following (\cite{Sahni2003, Alam2003})
\begin{equation}
	r\equiv \frac{\dddot{a}}{a H^{3}},
\end{equation}
\begin{equation}
	s\equiv\frac{r-1}{3(q-\frac{1}{2})}\label{s}
\end{equation}
where the over dot denotes the differentiation with respect to the cosmic time $t$ and $H=\frac{\dot{a}}{a}$ is the Hubble parameter. Furthermore,
$q =-\ddot{a}/aH^{2}$ is the deceleration parameter. These parameters are purely geometrical since they depend on the scale factor. Therefore, the statefinder diagnostic is entirely geometrical diagnostic, and this allows us to characterize the properties of different DE models by finding trajectories in the $s,~r$ plane. So, it is a powerful tool that can successfully discriminate between different DE models, even if they predict similar expansion histories. It can be easily verified that for the well-known $\Lambda$CDM model, the statefinder parameters $\left\lbrace s,r \right\rbrace$  take constant values $ \left\lbrace 0,1 \right\rbrace $ in the s-r plane (\cite{Sahni2003}).
%%%%%%%%%%%%%
Now, we analyze the DE model by using statefinder diagnostic tools. For that, we first compute the statefinder parameters in terms of redshift, then compare them with that of $\Lambda$CDM and finally, we study their high and low redshift limits. This mechanism may provide very fascinating results.
%%%%%%%%%%%
For convenience, we introduce the dimensionless Hubble rate $E(z)\equiv H(z)/H_{0}$, where $H_{0}$ is the Hubble parameter at present time and it has the dimension similarly to that of $H$. Now, the parameters $\left\lbrace q,r,s\right\rbrace $ can be computed in terms of $E(z)$ as follows:
%%%%%%%%%%%%
\begin{equation}
	q(z)=-1+(1+z) \frac{E_{z}(z)}{E(z)},
\end{equation}
\begin{equation}
	r(z)=q(z)(1+2q(z))+(1+z)q_{z}(z).
\end{equation}
%%%%%%%%%%%%%%
Furthermore $s(z)$ is given by (\ref{s}), and suffix $z$ stands for derivative with respect to $z$. Using the expressions of energy densities obtained in Eqn.(\ref{17}) and Eqn. (\ref{18}), one can obtain the analytical expressions of the quantities $E(z)$, $q(z)$, $r(z)$, and $s(z)$.
%%%%%%%%%%%%%%
\begin{equation}\centering \fontsize{9pt}{20pt}
	E(z)=(1+z)^{-\frac{\frac{\Gamma_1}{H_0^2}-3-\alpha+\Gamma_0}{2}} \left[\Omega_{0m}+\Omega_{0d}(1+z)^{\frac{\Gamma_1}{H_0^2}+3\omega_d-\alpha+\Gamma_0} \left\{\frac{\alpha \Omega_{0m}}{\Omega_{0d}(\frac{\Gamma_1}{H_0^2}+3\omega_d-\alpha+\Gamma_0)}\left((1+z)^{\left( \frac{\Gamma_1}{H_0^2}+3\omega_d-\alpha+\Gamma_0\right) }-1\right)+1\right\}\right]^{\frac{1}{2}}
\end{equation}
%%%%%%%%%%%%%
and 
%%%%%%%%%%
\begin{equation}
	q(z)=\frac{3(\omega_{d}+1)H_{0}^2 \Delta +\Omega_{0m} (z+1)^{\alpha } \left\lbrace H_{0}^2 (3 \omega_{d}+\Gamma_{0} )+\Gamma_{1} \right\rbrace  \left\lbrace H_{0}^2 (\alpha -\Gamma_{0} +3)-\Gamma_{1} \right\rbrace }{2 H_{0}^2 \left[ \Delta +\Omega_{0m} (z+1)^{\alpha } \left\lbrace H_{0}^2 (3 \omega_{d}+\Gamma_{0} )+\Gamma_{1} \right\rbrace \right] }-1,
\end{equation}
%%%%%%%%%%%%%%%
\begin{equation}
	r(z)=\frac{(3 \omega_{d}+1) (3 \omega_{d}+2)H_{0}^4 \Delta +\Omega_{0m} (z+1)^{\alpha } \left\lbrace H_{0}^2 (3 \omega_{d}+\Gamma_{0} )+\Gamma_{1} \right\rbrace  \left\lbrace \Gamma_{1} +H_{0}^2 (-\alpha +\Gamma_{0} -2)\right\rbrace  \left\lbrace \Gamma_{1} +H_{0}^2 (-\alpha +\Gamma_{0} -1)\right\rbrace } { 2 H_{0}^4 \left[\Delta +\Omega_{0m} (z+1)^{\alpha } \left\lbrace H_{0}^2 (3 \omega_{d}+\Gamma_{0} )+\Gamma_{1} \right\rbrace \right]}
\end{equation}
%%%%%%%%%%%%%%%%%
\begin{equation}
	s(z)=\frac{2 \left(\frac{\left\lbrace 9 \omega_{d} (\omega_{d}+1)+2\right\rbrace  H_{0}^4 \Delta +\Omega_{0m} (z+1)^{\alpha } \left\lbrace H_{0}^2 (3 \omega_{d}+\Gamma_{0} )+\Gamma_{1} \right\rbrace \left\lbrace \Gamma_{1} +H_{0}^2 (-\alpha +\Gamma_{0} -2)\right\rbrace  \left\lbrace \Gamma_{1} +H_{0}^2 (-\alpha +\Gamma_{0} -1)\right\rbrace }{2 H_{0}^4 \left[\Delta +\Omega_{0m} (z+1)^{\alpha } \left\lbrace H_{0}^2 (3 \omega_{d}+\Gamma_{0} )+\Gamma_{1} \right\rbrace \right]}-1\right)}{3 \left(\frac{3 (\omega_{d}+1) H_{0}^2 \Delta +\Omega_{0m} (z+1)^{\alpha } \left\lbrace H_{0}^2 (3 \omega_{d}+\Gamma_{0} )+\Gamma_{1} \right\rbrace  \left\lbrace H_{0}^2 (\alpha -\Gamma_{0} +3)-\Gamma_{1} \right\rbrace }{H_{0}^2 \left[\Delta +\Omega_{0m} (z+1)^{\alpha } \left\lbrace H_{0}^2 (3 \omega_{d}+\Gamma_{0} )+\Gamma_{1} \right\rbrace \right]}-3\right)}
\end{equation}
%%%%%%%%%%%%%%%%%%%%%%
where $\Delta=(z+1)^{\left( 3 \omega_{d}+\Gamma_{0} +\frac{\Gamma_{1}}{H_{0}^2}\right) } \left[ H_{0}^2 \left\lbrace \Omega_{0d} (-\alpha +3 \omega_{d}+\Gamma_{0} )-\alpha \Omega_{0m}\right\rbrace +\Omega_{0d} \Gamma_{1} \right]$.\\
%%%%%%%%%%%%%%%%%%%%%%
\\ We shall now discuss the behavior of the parameters $q(z),r(z),s(z)$ as functions of reshift z. Numerical investigations are executed by plotting the evolutionary trajectories of the statefinder parameters in the s-r plane and q-r plane for our model.
We shall plot the figures by using the best-fit values of dimensionless free parameters $(\alpha,~\Gamma_{0},~\Gamma_{1})=(0.03,~0.32,~0.4)$ and the present values of the cosmological parameters as $(H_{0},~\Omega_{0d},~\Omega_{0m})=(69.92,~0.63,~0.37)$ which have been obtained in section III. Note that the fig. (\ref{sr-phantom}) and fig. (\ref{qr-phantom}) are plotted for  $ \omega_d=-1.11$; the figs. (\ref{sr-cosmological constant}) and (\ref{qr-cosmological constant}) are plotted for  $ \omega_d=-1$. Finally, the figs. (\ref{sr-quintessence}) and (\ref{qr-quintessence}) are plotted for $\omega_d=-0.85$.

%%%%%%%%%%%%%%%%%%%%%%%%%
%%%%%%%%%%%%%%%%%%%%%%%%%%%%%%%%%%%%%%%%%%%%%%%%%%%%%%%%%%%%%%%
%%%%%%%%%%%%%%%%%%%%%%%%%%%%%%%%%%%%%%%%%%%%%%%%%%%%%%%%%%%%%%%%%%%%%	

%%%%%%%%%%%%%%%%%
%%%%%%%%%%%%%%%%%%%%%%%%%%%%%%%%%%%%%%%%%%%%%%%%%%%%%%%%%%%%%%%%
The evolutionary trajectories are plotted in the $s-r$ plane in the figs. \ref{sr-phantom-alpha}, \ref{sr-cosmological constant-alpha}, \ref{sr-quintessence-alpha} for different values of $\alpha$, and in  figs. \ref{sr-phantom-Gamma_{0}}, \ref{sr-cosmological constant-Gamma_{0}}, \ref{sr-quintessence-Gamma_{0}} for different values of $\Gamma_{0}$. Finally, the figs. \ref{sr-phantom-Gamma_{1}}, \ref{sr-cosmological constant-Gamma_{1}}, \ref{sr-quintessence-Gamma_{1}} are drawn for different value of $\Gamma_1$. Note that for all the figures, only one parameter out of three takes different values but, remaining two are fixed. In all diagrams (\ref{sr-phantom}), (\ref{sr-cosmological constant}), and (\ref{sr-quintessence}), the blue star symbol indicates the standard $\Lambda$CDM fixed point. The present value of the statefinder pair $(s,~r)$ are shown by the putting colored circles.\\
%%%%%%%%%%%%%%%
In figs. \ref{sr-phantom-alpha}, \ref{sr-cosmological constant-alpha}, and \ref{sr-quintessence-alpha}, we obtain different trajectories in $s-r$ plane for different values of the interaction parameter $\alpha$ and figures show that $s$ decreases, and  $r$ increases when $\alpha$ will increase. The scenario shows that the present values of the statefinder pair $(s,r)$ are moving away from the standard $\Lambda $CDM fixed point at an earlier time (shown in fig. \ref{sr-phantom-alpha}), moving towards $\Lambda$CDM fixed point (exhibited in fig. \ref{sr-cosmological constant-alpha}) in the future but not even evolving around $\Lambda$CDM fixed point (shown in fig. \ref{sr-quintessence-alpha}). In conclusion, the evolutionary trajectories show that evolution starts from $\Lambda $CDM point at early times and after that, moving towards the $r$- increasing direction and $s$-decreasing (shown in fig. \ref{sr-phantom-alpha}). In fig. \ref{sr-cosmological constant-alpha}, when $\alpha$ increases, $s$ will decrease, and $r$ will increase, then the evolutionary trajectories reach at $\Lambda$CDM fixed point. But, in fig. \ref{sr-quintessence-alpha}, the scenario will not be achieved. Therefore, $\alpha$ affects on the evolutionary trajectories in the $s-r$ plane. \\
%%%%%%%%%%%%%%%%55
In figs. \ref{sr-phantom-Gamma_{0}}, \ref{sr-cosmological constant-Gamma_{0}} and \ref{sr-quintessence-Gamma_{0}}, we obtain different trajectories in the $s-r$ plane by taking different values of the parameter $\Gamma_{0}$. As $\Gamma_{0}$ decreases, the value of $s$ decreases, and $r$ increases. So, the present values of the statefinder pair $(s,r)$ are moving away from the standard $\Lambda $CDM fixed point at the earlier time (in fig. \ref{sr-phantom-Gamma_{0}}), and moving towards $\Lambda$CDM fixed point (in fig. \ref{sr-cosmological constant-Gamma_{0}}), but not evolved around $\Lambda$CDM fixed point (in fig. \ref{sr-quintessence-Gamma_{0}}). Therefore, the evolutionary trajectories show that evolution starting from $\Lambda$CDM point at early times, then moving towards the $r$ increasing-direction and the $s$ decreasing-direction (see in fig. \ref{sr-phantom-Gamma_{0}}). In fig. \ref{sr-cosmological constant-Gamma_{0}}, when $\Gamma_{0}$ decreases, then $s$ will decrease and $r$ will increase. Finally, the evolutionary trajectories reach at $\Lambda$CDM fixed point. However, \ref{sr-quintessence-Gamma_{0}} shows that the scenario does not happen. This proves that $\Gamma_{0}$ also affects the evolutionary trajectories in the $s-r$ plane.\\
%%%%%%%%%%%%%
In figs. \ref{sr-phantom-Gamma_{1}}, \ref{sr-cosmological constant-Gamma_{1}}, and \ref{sr-quintessence-Gamma_{1}}, different trajectories are obtained in the s-r plane for different values of free parameter $\Gamma_{1}$. As $\Gamma_{1}$ decreasing, the value of $s$ decreases, and $r$ increases. So, the present values of the statefinder pair $(s,r)$ are moving away from the standard $\Lambda$CDM fixed point at an earlier time (see in fig. \ref{sr-phantom-Gamma_{1}}), and moving towards to $\Lambda$CDM fixed point (in fig. \ref{sr-cosmological constant-Gamma_{1}}), but not evolved around the $\Lambda$CDM fixed point (see in fig. \ref{sr-quintessence-Gamma_{1}}) when $\Gamma_{1}$ decreases. Therefore, the evolutionary trajectories show that evolution starting from $\Lambda$CDM point at early times and going forward when $r$ increases and $s$ decreases (see in fig. \ref{sr-phantom-Gamma_{1}}). In fig. \ref{sr-cosmological constant-Gamma_{1}}, when $\Gamma_{1}$ decreases, then $s$ decreases and $r$ increases. Then finally the evolutionary trajectories reach at $\Lambda$CDM fixed point. But, the fig. \ref{sr-quintessence-Gamma_{1}} shows that the scenario does not happen. So, $\Gamma_{1}$ can also affect the evolutionary trajectories in the s-r plane. \\
%%%%%%%%%%%%%%%
Similarly, the evolutionary trajectories are plotted in the q-r plane for different values of interaction coupling parameter $\alpha$ in (\ref{qr-phantom-alpha}, \ref{qr-cosmological constant-alpha}, \ref{qr-quintessence-alpha}), for different values of parameter $\Gamma_{0}$ in (\ref{qr-phantom-Gamma_{0}}, \ref{qr-cosmological constant-Gamma_{0}}, \ref{qr-quintessence-Gamma_{0}}) and for different values of $\Gamma_{1}$  in (\ref{qr-phantom-Gamma_{1}}, \ref{qr-cosmological constant-Gamma_{1}}, \ref{qr-quintessence-Gamma_{1}}). That is, we make one parameter changes only, keeping all others unaltered. In all diagrams (\ref{qr-phantom}), (\ref{qr-cosmological constant}) and (\ref{qr-quintessence}), the blue star symbol denotes the standard $de~ Sitter$ fixed point. The present value of the pair $(q,r)$ is shown with the colored circles.\\
%%%%%%%%%%%%%%%%%
In figs. \ref{qr-phantom-alpha}, \ref{qr-cosmological constant-alpha}, and \ref{qr-quintessence-alpha}, the trajectories in the q-r plane for different values of coupling parameter $\alpha$ are obtained. As the value of $\alpha$ increases, at same time the value of $q$ will decrease, and $r$ will increase. So, the present values of the statefinder pair $(q,r)$ are moving towards $de~Sitter$ fixed point (see in fig. \ref{qr-cosmological constant-alpha})  but do not evolve around $de~Sitter$ fixed point (in figs. \ref{qr-phantom-alpha}, \ref{qr-quintessence-alpha}) when the value of $\alpha$ increases. Therefore, in fig. \ref{qr-cosmological constant-alpha}, $q$ decreases and $r$ increases with the increasing value of $\alpha$. Finally, the evolutionary trajectories reach the $de~Sitter$ fixed point. But, in figs. \ref{qr-phantom-alpha}, and \ref{qr-quintessence-alpha}, this type of scenario has not happened. So, $\alpha$ can affect the evolutionary trajectories in the q-r plane. \\
%%%%%%%%%%%%%%%%%%%
In figs. \ref{qr-phantom-Gamma_{0}}, \ref{qr-cosmological constant-Gamma_{0}} and \ref{qr-quintessence-Gamma_{0}}, we get different trajectories in q-r plane for different values of the parameter $\Gamma_{0}$. As the value of $\Gamma_{0}$ increases,  $q$ and $r$ will decrease simultaneously. So, the present values of the statefinder pair $(q,r)$ are moving towards $de~Sitter$ fixed point in fig. \ref{qr-cosmological constant-Gamma_{0}} in future but do not evolve around $de~Sitter$ fixed point in figs. \ref{qr-phantom-Gamma_{0}},  \ref{qr-quintessence-Gamma_{0}} as the value of $\Gamma_{0}$ increases. Therefore, in fig. \ref{qr-cosmological constant-Gamma_{0}}, $q$ and $r$ decrease with the increasing value of $\Gamma_{0}$, and finally, the evolutionary trajectories reach to $de~Sitter$ fixed point. But, in figs. \ref{qr-phantom-Gamma_{0}}, and \ref{qr-quintessence-Gamma_{0}}, this type of scenario can not happen. So, $\Gamma_{0}$ can affect the evolutionary trajectories in q-r plane. \\
%%%%%%%
In figs. \ref{qr-phantom-Gamma_{1}}, \ref{qr-cosmological constant-Gamma_{1}}, and \ref{qr-quintessence-Gamma_{1}}, we get different trajectories in the q-r plane for different values of the parameter $\Gamma_{1}$. As $\Gamma_{1}$ increases, $q$ and $r$ will decrease simultaneously. So, the present values of the statefinder pair $(q,r)$ are moving towards the $de~Sitter$ fixed point in fig. \ref{qr-cosmological constant-Gamma_{1}} in future but do not evolve around $de~Sitter$ fixed point in figs. \ref{qr-phantom-Gamma_{1}},  \ref{qr-quintessence-Gamma_{1}} as the value of $\Gamma_{1}$ increases. Therefore, in fig. \ref{qr-cosmological constant-Gamma_{1}}, $q$ and $r$ will decrease with the increasing value of $\Gamma_{1}$, and finally, the evolutionary trajectories reach the $de~Sitter$ fixed point. But, in figs. \ref{qr-phantom-Gamma_{1}} and \ref{qr-quintessence-Gamma_{1}}, this type of scenario can not happen. So, $\Gamma_{1}$ can affect the evolutionary trajectories in q-r plane. The present value of $(q,r)$ can be important if it can be extracted from some future experiments. \\

%%%%%%%%%%%%%%%%%%%%%%%%%%%%%%%%%
\section{Dynamical analysis}

This section will discuss the critical points analysis for the model under consideration. We first convert the cosmological evolution equations into an autonomous system of ordinary differential equations by adopting suitable transformation of variables. After extracting the critical points from the autonomous system, we find the nature of critical points by evaluating the eigenvalues of the linearized Jacobian matrix at the critical points. Further, the cosmological behavior of the critical points shall be explored via the cosmological parameters evaluated at critical points.\\
%%%%%%%%%%%%%%%%%%%%%%%%%%%%%%%
We consider the following dimensionless variables as the dynamical variables
\begin{equation}
	x=\frac{\rho_{d}}{3 H^{2}} ~~,~~ y=\frac{p_{d}}{3 H^{2}}
\end{equation}
%%%%%%%%%%%%%%%
which are normalized over Hubble scale. By using these variables, the governing equations, namely, constraint equation (\ref{Friedmann equation}), acceleration equation (\ref{acceleration equation modified}) and conservation equations (\ref{continuity DE}) and (\ref{continuity DM modified}) give the following system of first-order ordinary differential equations with the creation pressure in Eqn. (\ref{creation pressure})
\begin{eqnarray}
	\begin{split}
		\frac{dx}{dN}& =\frac{Q}{3 H^{3}}-(1-x)\left\lbrace 3 y+x\left(\Gamma_{0}+\frac{\Gamma_{1}}{H^{2}} \right) \right\rbrace   ,& \\
		\frac{dy}{dN}& =\frac{y}{x} \left[\frac{Q}{3 H^{3}}-(1-x)\left\lbrace 3 y+x\left(\Gamma_{0}+\frac{\Gamma_{1}}{H^{2}} \right) \right\rbrace  \right]   ,& \\
		%
		%
		%\frac{dz}{dN} & =3z\left\lbrace 1+y-\left( \frac{1-x}{3}\right) \left(\Gamma_{0}+\frac{\Gamma_{1}}{H^{2}} \right)  \right\rbrace ,
		%
		%
		&~~\label{ode_system}
	\end{split}
\end{eqnarray}
%%%%%%%%%%%%%%%%%%%%%
where $ N=\ln a $ is the e-folding parameter taken to be independent variable. The above system of ODEs of the dynamical variable $x(N)$ and $y(N)$ contains the terms involving the cosmological variables $H(t)$. In addition, the presence of the interaction term $Q$ will lead to a non-autonomous system of ODEs. Now, to make a compact autonomous system, a new dimensionless variable $z=\frac{H_{\star}^{2}}{H^{2}}$ has to be introduced in the above system, where $H_{\star}$ is taken to be the value of Hubble parameter at $t=t_{\star}$ and it has the dimension similar to $H$. To perform dynamical analysis of cosmological models, similar approaches have been considered several times in the literature, where the Hubble parameter is chosen as dynamical variable $H(N)$ instead of cosmological variable $H(t)$ so as to make a non-autonomous system to compact autonomous system. For details, one may follow the references (\cite{Mandal2022,Oikonomou2019,Odintsov2018}). Recently, the authors in ref. (\cite{Halder2024}) followed the same approach for their study of dynamical analysis. In our paper, 
we consider the interaction term Q as in eqn. (\ref{interaction term}) and assume $H_{\star}=1$ (for mathematical simplicity) in the above system of ODEs (\ref{ode_system}), then we obtain the following compact autonomous system of ODEs in $x(N),~y(N),~z(N)$.

\begin{eqnarray}
	\begin{split}
		\frac{dx}{dN}& =(1-x) \left\lbrace \alpha-3y -x (\Gamma_{0} +\Gamma_{1}  z)\right\rbrace   ,& \\
		\frac{dy}{dN}& =\frac{y}{x} (1-x) \left\lbrace \alpha-3y -x (\Gamma_{0} +\Gamma_{1}  z)\right\rbrace ,& \\
		\frac{dz}{dN} & =3z\left\lbrace 1+y-\frac{(1-x)(\Gamma_{0}+\Gamma_{1} z)}{3} \right\rbrace ,
		&~~\label{autonomous_system1}
	\end{split}
\end{eqnarray}
%%%%%%%%%%%%%%%%%%%%%%%%%%%%%%%%%%%%%%%%%%%%%%%%%%%%%%%%%%%%%%%
Cosmological parameters associated to this model can immediately be expressed in terms of dynamical variables as follows :\\
The density parameters for dark energy and dark matter are
%%%%%%%%%%%%%%%%%%%%%%%%%%55
\begin{equation}\label{density_parameter}
	\Omega_{d}=x,
\end{equation}
and
%%%%%%%
\begin{equation}\label{density_parameter_m}
	\Omega_{m}=1-x
\end{equation}
%%%%%%%%%%%%%%
respectively.
The equation of state parameter for dark energy is
%%%%%%%%%%%%%%%%%%%%%%%%
\begin{equation}\label{eqn_of_state_parameter}
	\omega_{d}=\frac{p_{d}}{\rho_{d}}=\frac{y}{x}
\end{equation}
%%%%%%%%%%%%%%%%%%%%%
and the effective equation of state parameter for the model is
\begin{equation}
	\omega_{eff}=y-(1-x) \left(\frac{\Gamma_{0}+\Gamma_{1} z}{3}\right).
\end{equation}
%%%%%%%
The deceleration parameter for the model takes the form
\begin{equation}\label{dec_parameter}
	q=-1+\frac{3}{2}(1+\omega_{eff})=-1+\frac{3}{2}\left\lbrace 1+y-(1-x) \left(\frac{\Gamma_{0}+\Gamma_{1} z}{3}\right)\right\rbrace
\end{equation}
%%%%%%%
which shows the condition for acceleration of the universe $q<0 \Longrightarrow \omega_{eff}<-\frac{1}{3}$
and for deceleration $q>0 \Longrightarrow\omega_{eff}>-\frac{1}{3}.$\\
%%%%%%%%%%%%%%%%%%
Moreover, the evolution equation of the Hubble expansion
function reads as:
\begin{equation}
	\frac{\dot{H}}{H^{2}}=-\frac{3}{2}\left\lbrace 1+y-(1-x) \left(\frac{\Gamma_{0}+\Gamma_{1} z}{3}\right)\right\rbrace .
\end{equation}
%%%%%%%%%%%%%%%%%
\subsection{Critical points and cosmological analysis of the autonomous system (\ref{autonomous_system1}):}
\label{phase space autonomous system}
%%%%%%%%%%%%%%%%%%%%%%%%%%%%%%%%%%%%%
We shall now discuss the set of critical points and their corresponding physical parameters and stability in the phase space of $3D$ autonomous system (\ref{autonomous_system1}).\\
%%%%%%%%%%%%%%%%%%
The critical points extracted from the system (\ref{autonomous_system1}) are the following
{\bf
	\begin{itemize}
		\item  I. Set of critical points : $ A=(1,y_{c},0)$
		
		\item  II. Set of critical points : $B=(1,-1,z_{c})$
		
		\item  III. Set of critical points : $ C=\left(x_{c},\frac{\alpha-x_{c} \Gamma_{0}}{3},0 \right)$
		
		\item  IV. Set of critical Points : $ D=\left(x_{c},\frac{\alpha-x_{c} (\alpha+3)}{3},\frac{3+\alpha-\Gamma_{0}}{\Gamma_{1}}\right)$
	\end{itemize}
	
}
%%%%%%%%%%%%%%%%%%%%%%%%%%%%%%%
Note that all the sets are non-isolated sets of points in the phase space. Set of critical points and their corresponding physical parameters are shown in the Table \ref{physical_parameters} and the eigenvalues of linearized Jacobian matrix are presented in the Table \ref{eigenvalues}.

%%%%%%%%%%%%%%%%%%%%
%TCIMACRO{\TeXButton{B}{\begin{table}[tbp] \centering}}%
	%BeginExpansion
	\begin{table}[tbp] \centering
		%EndExpansion
		\caption{The critical points and the corresponding physical parameters  are presented.}%
		\begin{tabular}
			[c]{cccccccc}\hline\hline
			\textbf{Critical Points} &~~ $\mathbf{Existence}$  &~~$\mathbf{\Omega_{m}}$&  $\mathbf{\Omega_{d}}$ & $\mathbf{\omega_{d}}$ & $\mathbf{\omega_{eff}}$ & $q$ &
			\\\hline
			$A  $ & $ \mbox{Always} $ & $0$ & $1$ & $y_{c}$ & $y_{c}$ & $\frac{1}{2}+\frac{3 y_{c}}{2}$\\
			$B  $ & $ \mbox{Always} $ & $0$ & $1$ & $-1$ & $-1$ & $-1$ \\
			$C  $ &  $ 0 \leq x_{c}\leq 1 $ & $1-x_{c}$ & $x_{c}$ & $\frac{\alpha -x_{c} \Gamma_{0} }{3 x_{c}}$ & $\frac{\alpha -\Gamma_{0} }{3}$ & $~~\frac{1}{2} (\alpha -\Gamma_{0} +1)$ \\
			$D  $ & $ 0 \leq x_{c}\leq 1 ~~\mbox{and}~~\Gamma_{1}\neq 0$ ~~~~ & $1-x_{c}$ & $x_{c}$ & $\frac{\alpha -x_{c} (\alpha +3)}{3 x_{c}}$ & $-1$ & $-1$ \\
			
			\\\hline\hline
		\end{tabular}
		\label{physical_parameters} \\
		
		%TCIMACRO{\TeXButton{E}{\end{table}}}%
	%BeginExpansion
\end{table}%
%EndExpansion
%
%%%%%%%%%%%%%%%%%%%%%%%%%%%%%%%%%%%%%%%

%TCIMACRO{\TeXButton{B}{\begin{table}[tbp] \centering}}%
	%BeginExpansion
	\begin{table}[tbp] \centering
		\caption{The eigenvalues of the linearized system (\ref{autonomous_system1}) are presented where $\Sigma=(\sqrt{(x_{c}-1)}\left\lbrace x_{c}^3 (\alpha -\Gamma_{0} +3)^2+x_{c}^2 (\alpha -\Gamma_{0} +3) (\alpha +\Gamma_{0} -3)-\alpha ^2+\alpha  x_{c} (3 \alpha -2 \Gamma_{0} +6)\right\rbrace\overline{} $ }%EndExpansion
		\begin{tabular}
			[c]{ccccccc}\hline\hline
			\textbf{Critical Points} & $\mathbf{\lambda_{1}}$ &
			$\mathbf{\lambda_{2}}$ & $\mathbf{\lambda_{3}}$\\\hline
			$ A $ & $ 0 $ & $ 3 (y_{c}+1) $ &  ~~ $ \Gamma_{0}  -\alpha +3 y_{c} $ \\
			$ B $ & $ 0 $ & $ 0 $ & $ \Gamma_{0}+z_{c} \Gamma_{1} -\alpha  -3 $   \\
			$ C $ & $ 0 $ & $ \frac{x_{c} \alpha -\alpha }{x_{c}} $ & $ \alpha -\Gamma_{0} +3 $\\
			$ D $ & $ 0 $ & $ \frac{(x_{c}-1) \left\lbrace x_{c} (\alpha -\Gamma_{0} +3)+\alpha\right\rbrace -\Sigma }{2 x_{c}} $ & ~~ $ \frac{(x_{c}-1) \left\lbrace x_{c} (\alpha -\Gamma_{0} +3)+\alpha \right\rbrace +\Sigma }{2 x_{c}} $ \\
			\\\hline\hline
		\end{tabular}
		\label{eigenvalues}
		%\label{model_Eigen}
		%%TCIMACRO{\TeXButton{E}{\end{table}}}%
	%%BeginExpansion
\end{table}%
%%EndExpansion
%%%%%%%%%%%%%%%%%%%%%%%%%%%%%%%%%%%%%%%%%%%%%%%%%%%%
%\begin{figure}
%	\centering
%	\subfigure[]{%
	%		\%includegraphics[width=7cm,height=7cm]{A_stable}\label{fig:stability_A}}
%	\qquad
%	\subfigure[]{%
	%		\includegraphics[width=7cm,height=7cm]{evolution_A}\label{fig:evolution_A}}
%	\qquad
%	\subfigure[]{%
	%\includegraphics[width=9cm,height=7cm]{points A B C.pdf}\label{fig:A B C}}
%	\caption{The figure is plotted with $\alpha=1$ and $\gamma=-3$.  Panel (a) shows the phase projection on $x-z$ plane, where the scaling solution A is a stable attractor. While panel (b) shows the time evolution of the cosmological parameters with initial conditions: $ x[0]=0.816,~ y[0]=0.001,~ z[0]=0.001$ where late time scaling attractor approaches in decelerated era.}
%	\label{phasespace-figure_A}
%\end{figure}
%%%%%%%%%%%%%%%%%%%%%%%%%%%%%%%%%%%%%%%%%%%%%%%%%%%

%%%%%%%%%%%%%%%%%%%%%%%%%%%%%%%%%
\begin{figure}
	\centering
	\subfigure[]{
		\includegraphics[width=0.32\textwidth]{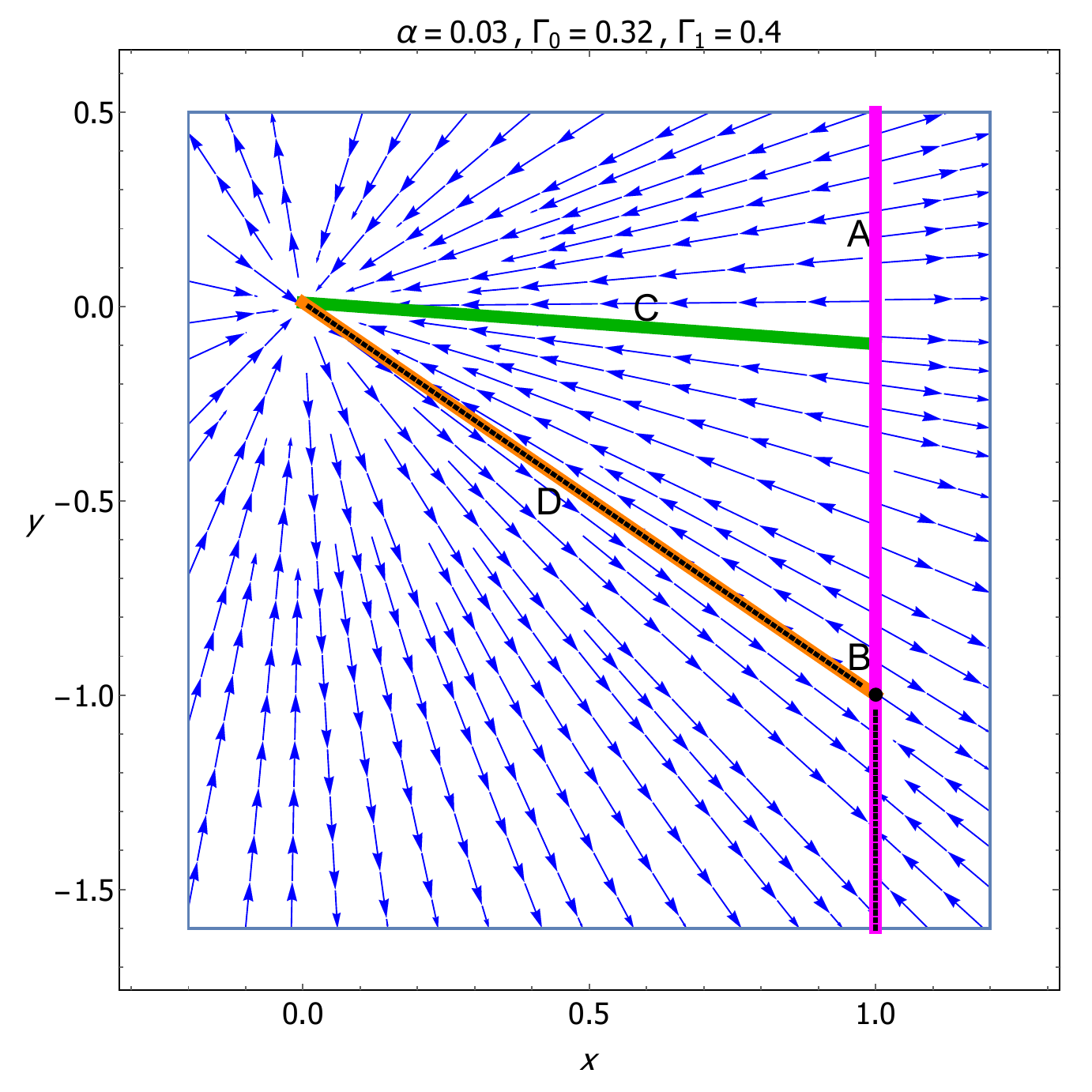}\label{fig:Stable_A}}
	\subfigure[]{
		\includegraphics[width=0.32\textwidth]{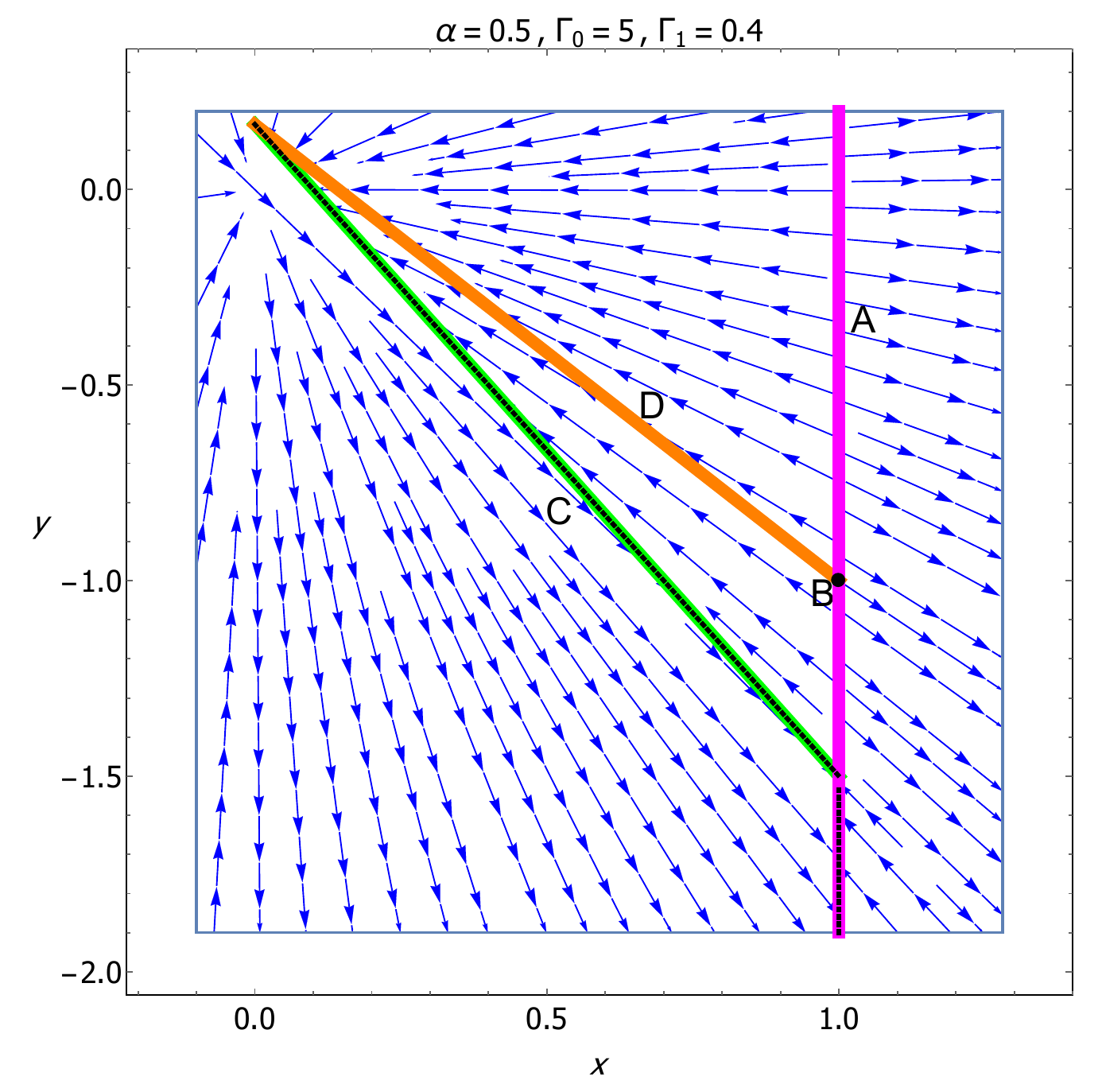}\label{fig:Stable_C}}
	\subfigure[]{
		\includegraphics[width=0.32\textwidth]{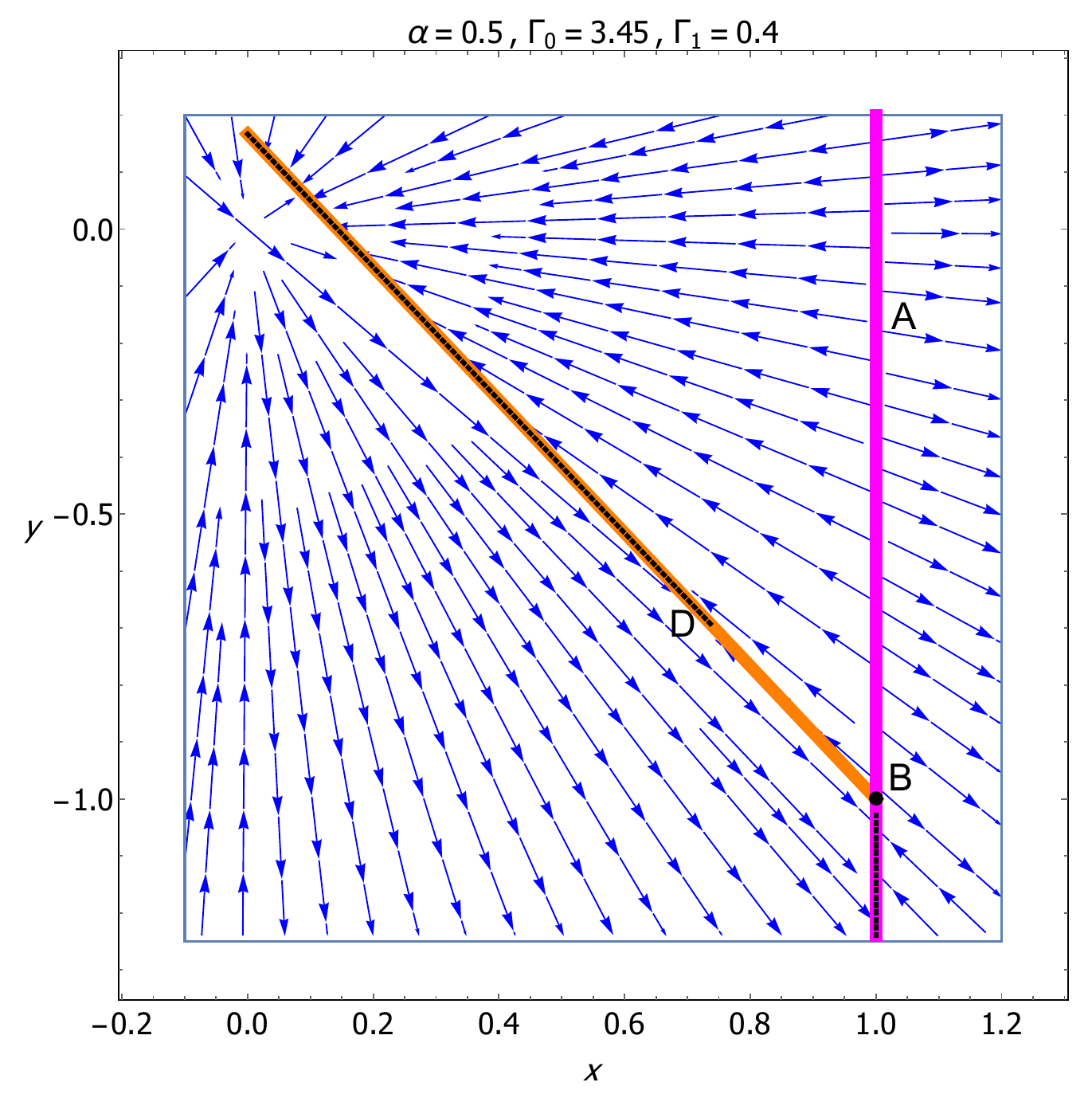}\label{fig:Stable_D}}
	\caption{The figures show the phase space projection of the autonomous system (\ref{autonomous_system1}) on the x-y plane. In panel (a) phase plot has been drawn for the free parameter values $\alpha=0.03$, $\Gamma_{0}=0.32$ and $\Gamma_{1}=0.4$. In panel (b) phase plot has been drawn for the parameter values $\alpha=0.5$, $\Gamma_{0}=5$ and $\Gamma_{1}=0.4$. In panel (c) phase plot has been drawn for the parameter values $\alpha=0.5$, $\Gamma_{0}=3.45$ and $\Gamma_{1}=0.4$. In all panels, the magenta colored line represents the set of points A, the green colored line represents the set of points C, the orange colored line represents the set of points D. The black-dashed line on a line represents the stable portion on that line. The point B is unstable in all the figures. }
	\label{phasespace-figure-ACD}
\end{figure}

%%%%%%%%%%%%%%%%%%%%%%%%%%%%%%%%%%%%%%%%%%%%%%%%%%%%%%%%
\begin{figure}
	\centering
	\subfigure[]{
		\includegraphics[width=0.317\textwidth]{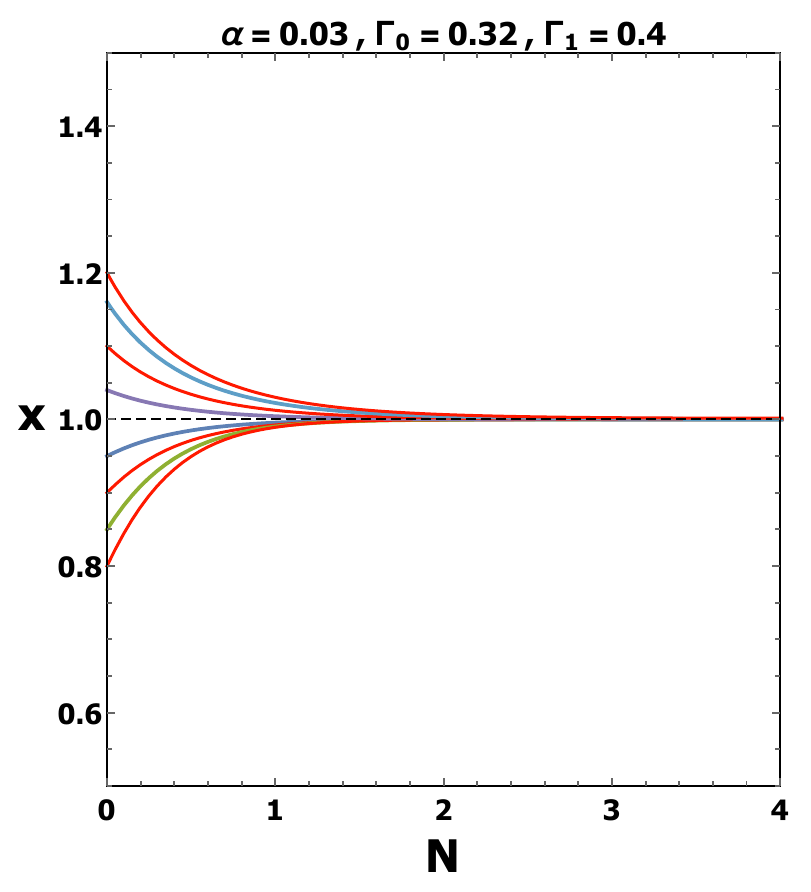}\label{fig:Stable_Nx}}
	\subfigure[]{
		\includegraphics[width=0.3\textwidth]{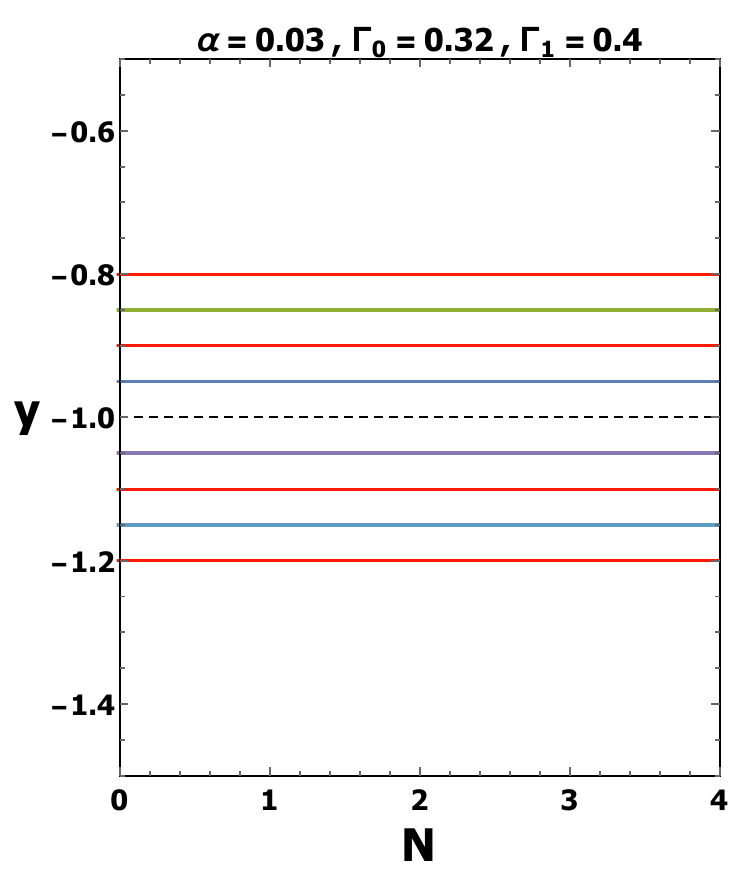}\label{fig:Stable_Ny}}
	\subfigure[]{
		\includegraphics[width=0.3\textwidth]{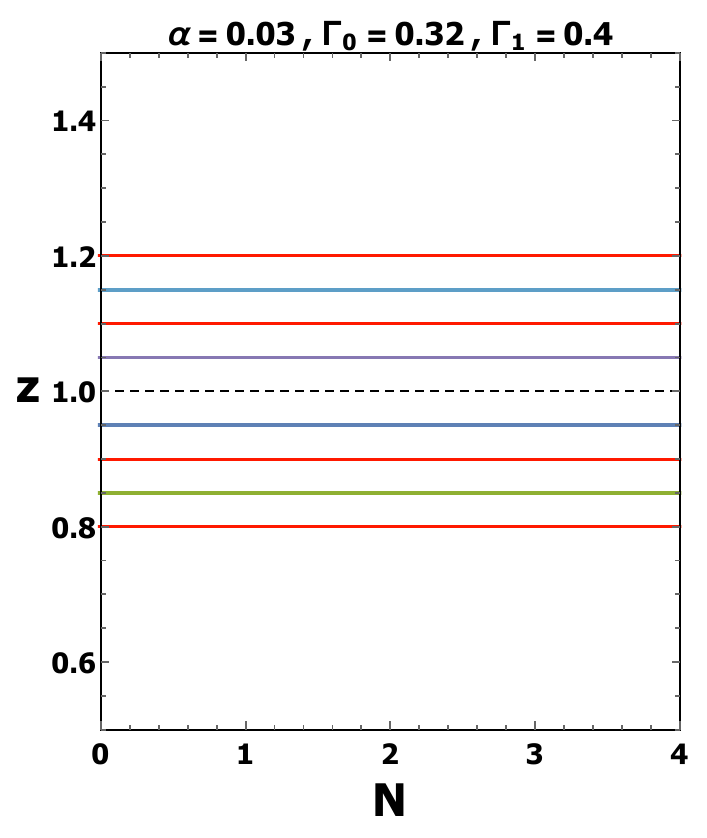}\label{fig:Stable_Nz}}
	\caption{The figures show the phase space projection of perturbation of the autonomous system (\ref{autonomous_system1}) along the x,y and z axes for the set of critical points $B$ for parameter values  $\alpha=0.03$, $\Gamma_{0}=0.32$ and $\Gamma_{1}=0.4$. In panel (a) perturbations come back but in panels (b), (c) perturbations do not come back. This indicates that the set $B$ is unstable for parameters value  $\alpha=0.03$, $\Gamma_{0}=0.32$ and $\Gamma_{1}=0.4$. }
	\label{B-perturbation}
\end{figure}

%%%%%%%%%%%%%%%%%%%%%%%%%%%%%%%%%%%%%%%%%%%%%%%%%
\begin{figure}
	\centering
	\subfigure[]{
		\includegraphics[width=0.45\textwidth]{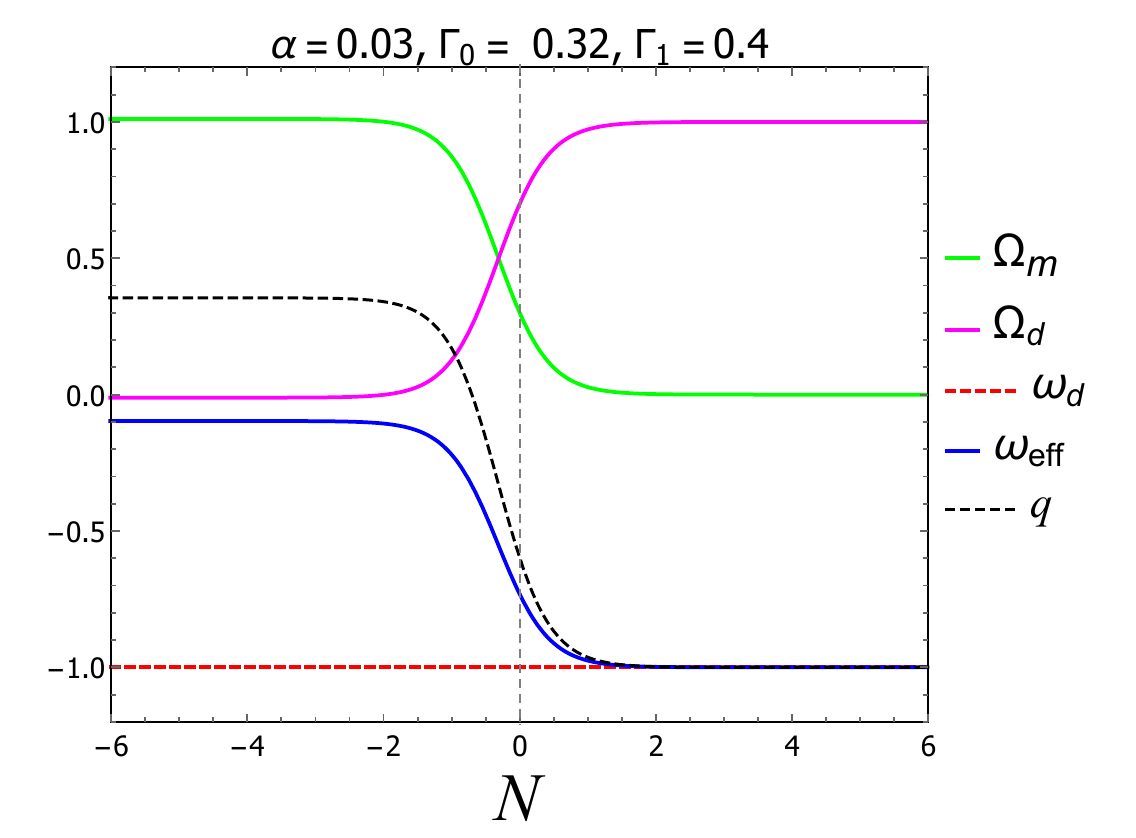}\label{fig:evolution1}}
	\subfigure[]{
		\includegraphics[width=0.45\textwidth]{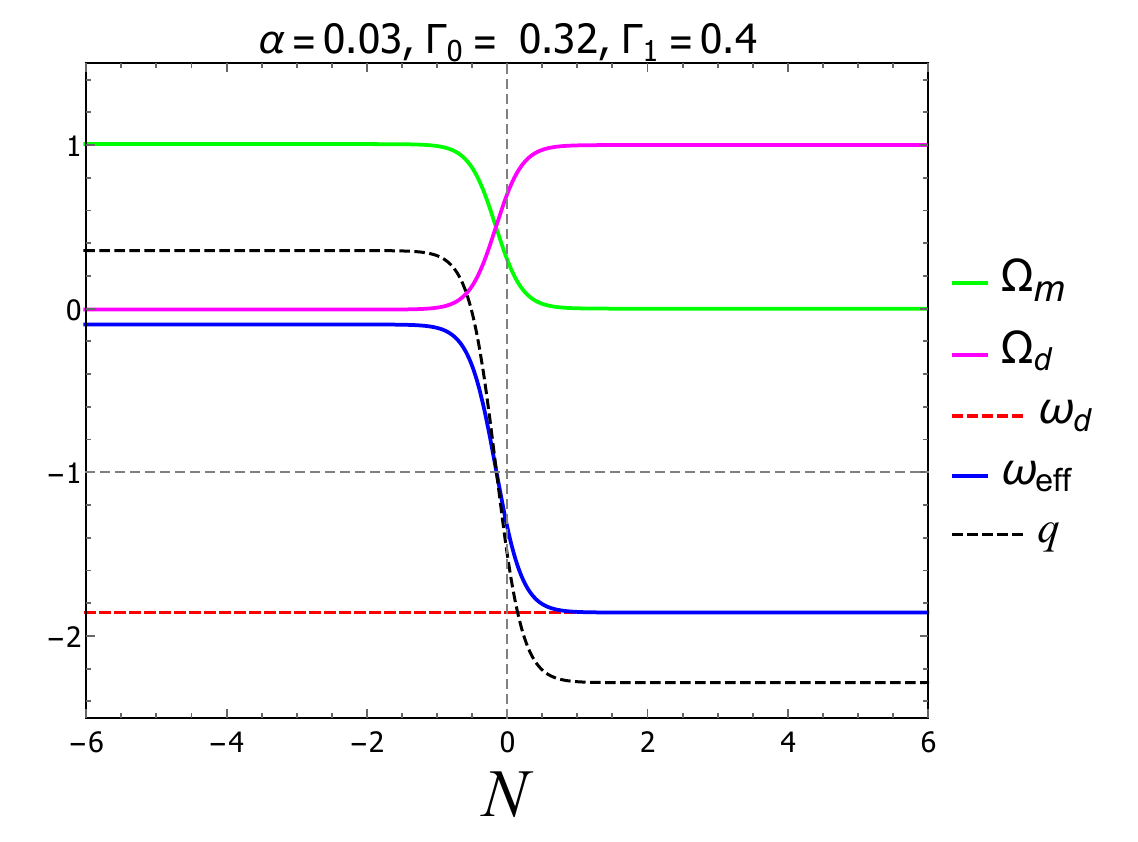}\label{fig:evolution2}}
	\caption{Figures show that the evolution of cosmological parameters like the DE density parameter $\Omega_{d}$, the density parameter for DM  $\Omega_{m}$,  the equation of state parameter $\omega_{d}$, the effective equation of state parameter $\omega_{eff}$, and the decelerating parameter $q$ with respect to $N=\ln a$ of the autonomous system (\ref{autonomous_system1}) with the parameter values  $\alpha=0.03$, $\Gamma_{0}=0.32$ and $\Gamma_{1}=0.4$. In panel (a) Numerical investigation shows that the ultimate fate of the universe is late time accelerated evolution in cosmological constant era followed by an intermediate matter dominated phase with initial values $x(0)=0.7, y(0)=-0.7, z(0)=0$ and in panel (b) Numerical investigation shows that the ultimate fate of the universe is late time accelerated evolution in phantom era followed by an intermediate matter dominated phase with initial values $x(0)=0.7, y(0)=-1.3, z(0)=0$. }
	\label{phasespace-parameters-evolution}
\end{figure}

%%%%%%%%%%%%%%%%%%%%%%%%%%%%%%%%%%%%%%%%%%%%%%%
%%%%%%%%%%%%%%%%%%%%%%%%%%%%%%%%%%%%
%%%%%%%%%%%%%%%%%%%%%%%%%%%%%%%%%%%%%%%%%%%%%%%%%%%%%%%%
\begin{itemize}
	
	\item
	The set of critical points $A$ represents a completely dark energy dominated ($\Omega_{d}=1$) solution where DE mimics any perfect fluid model since $\omega_{d}= y_c$ (see Table \ref{physical_parameters}) and $y_c$ takes any real value. Depending upon $y_c$, DE can behave as quintessence, cosmological constant, and phantom fluid, respectively, according to $-1<y_c<-\frac{1}{3}$, $y_c=-1$ and $y_{c} <-1$ in the phase space. It should be mentioned that the DE can mimic dust fluid for $y_c=0$. However, there exists an accelerating phase only for $y_{c}<-\frac{1}{3}$. From Table \ref{eigenvalues}, we observe that the set has precisely one vanishing eigenvalue, so by definition, it is a normally hyperbolic set, and the stability of this set can be found by evaluating the signature of remaining non-vanishing eigenvalues (\cite{Coley2003}). The set is stable for
	%%%%%%%%%%%%%%%%%%%
	$$y_c<\mbox{min} \left\{-1,~~ \frac{\alpha-\Gamma_{0}}{3} \right\}.$$
	%%%%%%%%%%%%%%%%%%%%%
	From the stability condition, one may conclude that the set of points $A$ is stable only in the phantom regime where DE behaves as a phantom fluid. Therefore, a DE-dominated accelerated universe is always observed by the set $A$ where the universe is evolving in phantom regime only. On the other hand, a decelerated dust-dominated universe is achieved when the DE behaves as dust. Here, the point with co-ordinate (1,0,0) on the set is an unstable saddle-like solution.	
	
	\item
	
	Set of critical points $B$ exists for all parameter values. It is completely DE dominated and DE behaves as cosmological constant. Acceleration is always possible for this set. It is non-hyperbolic set. Since two of its eigenvalues having zero real part, it is not a normally hyperbolic set. Therefore, its stability can be checked by numerical investigation and is done in the fig. \ref{B-perturbation}. Figure shows that the trajectories are attracted only in one direction.\\
	
	\item
	
	Set of critical points $C$ exists for $0\leq x_c\leq 1$. It represents a scaling solution in the phase space with the combination of both DE and DM, and the ratio is $\Omega_{d}/ \Omega_{m}=x_c /(1-x_c)$. Here, DE behaves as perfect fluid in nature. Acceleration of the universe exists for $\alpha<\Gamma_{0}-1$. The phantom behavior is obtained for $\alpha<\Gamma_{0}-3$.  It is a non-hyperbolic set with exactly one vanishing eigenvalue, so the set is normally hyperbolic set. Therefore, the set $C$ is stable for $ \left( 0<x_{c}<1~~\mbox{and}~~ 0<\alpha< \Gamma_{0}-3  \right) $. Therefore, late-time accelerated universe is obtained by the set and the universe evolves in the phantom regime.\\
	
	\item
	
	Set of points $D$ exists for all parameter values with $\Gamma_{1}\neq0$. It is also a solution with the combination of both the DE and DM with the ratio $\Omega_{d}/ \Omega_{m}=x_c /(1-x_c)$. It may be a scaling solution at late-times. DE behaves as perfect fluid in nature. Accelerating universe exists for this set. The set of critical points $D$ is also non-hyperbolic. Moreover, the set is normally hyperbolic. The set of critical points $D$ is stable for $ \left\lbrace  0<x_{c}<1~~\mbox{and}~~ \left(\frac{\left(x_{c}^{3/2}+x_{c}+\sqrt{x_{c}}-1\right) \alpha }{x_{c}^{3/2}+x_{c}}+3\leq \Gamma_{0} <\alpha +3~~\mbox{or}~~ \frac{\left(x_{c}^{3/2}-x_{c}+\sqrt{x_{c}}+1\right) \alpha }{x_{c}-x_{c}^{3/2}}+\Gamma_{0} \leq 3\right)~~\mbox{and}~~ \Gamma_{1} \neq 0~~\mbox{and}~~ \alpha >0 \right\rbrace $.
	
\end{itemize}
%%%%%%%%%%%%%%%%%%
\subsection{Cosmology of the critical points}

Late-time accelerated evolution is achieved by the set of critical points A where the universe is evolving in phantom phase $(\omega_{eff}<-1)$ and is entirely dominated by phantom fluid $(\omega_{d}<-1)$. The figs. \ref{fig:Stable_A}, \ref{fig:Stable_C}, and \ref{fig:Stable_D} exhibit that the set of points A will be stable attractor when $y_{c}<-1$, i.e. the universe evolves with a phantom-like effective equation of state $(\omega_{eff}<-1)$. The black-dashed line corresponds to the stable solution in the phase diagrams. On the other hand, for $y_{c}=0$ the set of points A describes the dust dominated decelerated phase where the universe evolves with a dust-like efffective equation of state $\omega_{eff}=0$. In this case, the DE behaves as dust ($\omega_{d}=0$) matter.  Being a saddle-like solution in phase space, the set A describes the intermediate evolution of the universe. \\

The set of critical points B is completely dominated by cosmological constant-like fluid $(\omega_{d}=-1)$ and it corresponds to a solution with de Sitter expansion of the universe (since $\Omega_{d}=1,~\Omega_{m}=0,~\omega_{eff}=-1,~q=-1$). However it fails to be a stable solution in the phase space. The numerical investigation shows the behaviour of the variables $x(N),~y(N),~z(N)$ in the fig. \ref{B-perturbation} for free parameters $(\alpha=0.03,~\Gamma_{0}=0.32,~\Gamma_{1}=0.4)$ showing that the set of points B is saddle-like solution. Therefore, transient behaviour of the accelerated de Sitter expansion of the universe is observed by this set. \\

The DE-DM scaling solution is achieved by the set of critical points C which corresponds to an accelerated evolution of the universe at late-times. For the parameter region: $0<\alpha<\Gamma_{0}-3$, the set evolves in phantom era $(\omega_{eff}<-1)$ at late-times with the similar order of energy densities: $0<\Omega_{d}<1$ and $0<\Omega_m<1$ (since $0<x_{c}<1$) providing the possible mechanism to alleviate the coincidence problem. Fig. \ref{fig:Stable_C} with parameters values: $\alpha=0.5,~\Gamma_{0}=5,~\Gamma_{1}=0.4$, shows that the scaling attractor C evolves in phantom era (i.e. $\omega_{eff}<-1$) in the region of phase space $0<x_{c}<1$, i.e. satisfying the restriction: $0<\Omega_{d}<1$. In particular, if we assume $x_{c}=0.7$, the cosmic evolution of physically relevant quantity will be: ($\Omega_{d}=0.7,~\Omega_{m}=0.3,~\omega_{eff}\approx -1.5,~q=-1.7,~\omega_{d}=-1.43$) showing to be in slightly in agreement with observations.\\

Finally, the set of critical points D corresponds to a DE (cosmological constant)-DM scaling solution in phase space. For $\alpha>\Gamma_{0}-3$ and $\Gamma_{1} \neq 0$, the set represents stable solution where the constraints  $0<\Omega_{d}<1$ and $0<\Omega_{m}<1$ are satisfied. This mechanism can alleviate the coincidence problem successfully. Note that, the scaling solutions are also obtained in Ref. (\cite{Biswas2017}) where particle creation rate was taken as $\Gamma=\Gamma_{0}H$. However, the scaling solutions were unable to solve the coincidence problem. Here, from our study, we obtained the scaling solutions C and D correspond to late-time accelerated evolution of the universe. Infact, they can solve the coincidence problem. Fig.\ref{fig:Stable_A} with values of free parameters: $\alpha=0.03,~\Gamma_{0}=0.3,~\Gamma_{1}=0.4$ exhibits that the set D represents the accelerated late phase evolution of the universe in cosmological constant era satisfying the restriction: $0<x_{c}<1$. For a particular choice of $x_{c}=0.7$ the numerical values of the physical parameters will be $\Omega_{d}=0.7,~\Omega_{m}=0.3,~\omega_{eff}=-1,~q=-1,~\omega_{d}=-0.99$. It shows that the late-time accelerated ($q=-1$) universe evolves in cosmological constant era ($\omega_{eff}=-1$) where the dark energy behaves as quintessence fluid. The fig.\ref{fig:Stable_D} for $\alpha=0.5,~\Gamma_{0}=3.5,~\Gamma_{1}=3.4$ shows the cosmic evolution of the physically relevant quantities: $\Omega_{d}=0.7$ (since we have taken $x_{c}=0.7$), $\Omega_m=0.3,~\omega_{eff}=-1, ~q=-1$ where the DE behaves as quintessence ($\omega_{d} \approx-0.93$). \\
 
The evolutions of the relevant cosmological parameters: density parameter for DE ($\Omega_{d}$) and DM ($\Omega_{m}$), effective equation of state parameter for the model ($\omega_{eff}$), equation of state parameter for DE ($\omega_{d}$) and the deceleration parameter (q) are plotted against the e-folding parameter $N=\ln(a)$ in the 
fig.\ref{phasespace-parameters-evolution} for the parameter values $\alpha=0.03,~\Gamma_{0}=0.32,~\Gamma_{1}=0.4$.
With suitable initial conditions, the fig. \ref{fig:evolution1} exhibits numerical solution describing the evolution history of the universe from early DM-dominated solution (with $\omega_{eff} \approx  0,~q \approx \frac{1}{2}$) to late-time DE-dominated scaling solution with $\omega_{eff}=-1,~q=-1$. The fig. \ref{fig:evolution2} shows the cosmological evolution where $\omega_{eff}$ evolves starting from DM(dust)-dominated intemediate phase ($\omega_{eff} \approx 0 $) to late-time DE-dominated accelerated evolution which is being attracted by scaling solution with  $\omega_{eff}<-1$. \\
 
In summary, one can conclude that our model extracts some interesting critical points representing cosmologically viable solutions: a clear trajectory is identified connecting from a dust dominated ($\omega_{eff} \approx 0$) critical point to a DE-dominated scaling solution. Also, the cosmic evolution of the physical quantities shown in 
fig.\ref{phasespace-parameters-evolution} for different initial conditions exhibiting that the universe is attracted in cosmological constant era (shown in fig.\ref{fig:evolution1}) or in phantom phase (in fig.\ref{fig:evolution2}) in near future.\\

%%%%%%%%%%%%%%%%%%%%%%%%%%%%%%%%%%%%%%%%%%%%
		
\section{Thermodynamical Analysis of Energy transfer between the dark sectors of the matter distribution due to particle creation}
The particle creation mechanism gives rise to an energy transfer between the dark species causing the two subsystems to have different temperatures, and as a result, thermodynamics of irreversible process comes into the picture. Starting from Euler's thermodynamical relation $nTs=\rho+p$ and using the energy conservation equations (\ref{continuity DM}) and (\ref{continuity DE}) along with the modified particle number conservation relations (\ref{9}) and (\ref{10}) for the two dark subsystems, the evolution equations for the temperatures of the individual dark fluid elements read as
		\begin{equation}
			\frac{\dot{T}_{m}}{T_{m}}=-\alpha H\label{20}
		\end{equation}
		and
		\begin{equation}
			\frac{\dot{T}_{d}}{T_{d}}=-3H\omega_d-\alpha H\frac{\rho_m}{\rho_d}\label{21}
		\end{equation}
		Here $T_m, T_d$ are temperatures of DM and DE respectively. Integrating the above equations we obtain
		\begin{equation}
			T_m=T_0 \left(\frac{a}{a_0}\right)^{-2-\alpha}\label{22}
		\end{equation}
		and
		\begin{equation}
			T_d=T_0 \left(\frac{a}{a_0}\right)^{-3\omega_d}\left\{\frac{\alpha r_0}{\eta}\left(\left(\frac{a}{a_0}\right)^\eta-1\right)+1\right\}\label{23}
		\end{equation}
where ~$T_{0}$~ is the common temperature of the two subsystems in equilibrium configuration. It is to be noted that in deriving equation (\ref{22}) one has to take into account of the temperature ~$T_{m_{0}}\propto a^{-2}$~ for the DM sector in the absence of interaction.\\
However, in presence of interaction, when the temperature of the system differs from that of the horizon, there will be spontaneous heat flow between  the horizon and the fluid components, and hence there will no longer be any thermal equilibrium. As we are considering the universe bounded by the apparent horizon as an isolated system, so at the thermal equilibrium, the common temperature~ $T_{0}$~ is nothing but the Hawking temperature at the horizon i.e. ~$T_{0}=\frac{1}{2\pi R_{A}}$~ where ~$R_{A}$~ is the radius of the apparent horizon for the FRW model. In case of a flat universe $R_A=\frac{1}{H}$.\\
For the present isolated system, if we denote the entropies of the two subsystems as~ $S_{m}$ ~and ~$S_{d}$~respectively and ~$S_{A}$~ as the entropy of the bounding apparent horizon, then
		\begin{equation}
			T_{m}\frac{dS_{m}}{dt}=\frac{dQ_{m}}{dt}=\frac{dE_{m}}{dt}\label{24}
		\end{equation}
		and
		\begin{equation}
			T_{d}\frac{dS_{d}}{dt}=\frac{dQ_{d}}{dt}=\frac{dE_{d}}{dt}+p_{d}\frac{dV}{dt}\label{25}
		\end{equation}
		while from the Bekenstein area formula,
		\begin{equation}
			\frac{dS_{A}}{dt}=2\pi R_{A}\dot{R_{A}}\label{26}
		\end{equation}
		Here ~$V=\frac{4}{3}\pi R_{A}^{3}$~ is the volume of the universe bounded by the apparent horizon  and ~$E_{m}=\rho_{m}V$~ and ~$E_{d}=\rho_{d}V$.\\As the overall system is isolated so the heat flow across the horizon ~$Q_{h}$~ will satisfy
		\begin{equation}
			\dot{Q_{h}}=-\left(\dot{Q_{m}}+\dot{Q_{d}}\right)\label{27}
		\end{equation}
		
In equilibrium configuration, the entropy of the whole system depends on the energy densities and volume only. Furthermore, from the extensive property, it is just the sum of the entropies, i.e. $ S_{m}+S_{d}+S_{A} $. However in non-equilibrium thermodynamics, one has to take into account the irreversible fluxes such as energy transfers in the total entropy, and hence the time variation of the total entropy is given by (\cite{Karami2010,Zhou2009})
		\begin{equation}
			\frac{dS_{T}}{dt}=\frac{dS_{m}}{dt}+\frac{dS_{d}}{dt}+\frac{dS_{A}}{dt}-A_{d}\dot{Q_{d}}\ddot{Q_{d}}-A_{h}\dot{Q_{h}}\ddot{Q_{h}}\label{28}
		\end{equation}
where $A_{d}$ and $A_{h}$ are the energy transfer constants between DE and DM within the universe and between the universe and the horizon respectively. Now using equations (\ref{24}) - (\ref{26}) the explicit form of different terms on the r.h.s of equation (\ref{28}) are given by
		
		\begin{equation}
			\frac{dS_m}{dt}=3\pi H^{-1}\Omega_m \left(\frac{a}{a_0}\right)^{\alpha+2}\left(\frac{\Gamma}{3H}-\frac{\alpha}{3}+q\right)\label{29}
		\end{equation}
		\begin{equation}
			\frac{dS_d}{dt}=3\pi H^{-1}\Omega_d \left(\frac{a}{a_0}\right)^{3\omega_d}\frac{1}{\left\{\frac{\alpha r_0}{\eta}\left(\left(\frac{a}{a_0}\right)^\eta-1\right)+1\right\}}\left[q(1+\omega_d)+\frac{\alpha}{3}\left(\frac{1}{\Omega_d}-1\right)\right]\label{30}
		\end{equation}
		\begin{equation}
			\frac{dS_{A}}{dt}=2\pi H^{-1}(1+q)\label{31}
		\end{equation}
		\begin{equation}
			A_{d}\dot{Q_{d}}\ddot{Q_{d}}=\frac{3A_d}{4}\left[\alpha+3\Omega_d\left\{q(1+\omega_d)-\frac{\alpha}{3}\right\}\right]\left[\dot{\Omega_d}\left\{q(1+\omega_d)-\frac{\alpha}{3}\right\}+\dot{q}(1+\omega_d)\Omega_d\right]\label{32}
		\end{equation}
		
		\begin{equation}
			A_{h}\dot{Q_{h}}\ddot{Q_{h}}=\frac{3A_h}{4}\left[3q(1+\omega_d \Omega_d) +\Gamma \Omega_m H^{-1})\right]\left[\dot{q}(1+\omega_d \Omega_d)+\dot{\Omega_d}\left(q\omega_d-\frac{\Gamma H^{-1}}{3}\right)+\frac{H^{-1}}{3}\left\{\Gamma+2\Gamma_1 \Omega_m(1+q)\right\}\right]\label{33}
		\end{equation}
		
where $\Omega_m, \Omega_d$ denote energy density parameters of DM and DE respectively given by \\
$\Omega_m=\frac{\rho_m}{3H^2},~~\Omega_d=\frac{\rho_d}{3H^2}$ and $q=-1-\frac{\dot{H}}{H^2}$ being the deceleration parameter. \\
$\dot{\Omega_d}$ and $\dot{q}$ are given by
		\begin{equation}
			\dot{\Omega_d}=H\left[\Omega_d(2q-3\omega_d-1-\alpha)+\alpha\right]\label{34}
		\end{equation}
		\begin{equation}
			\dot{q}=H\left[(j+3q+2)+2(1+q)^2\right]\label{35}
		\end{equation}
where $j=\frac{\ddot{H}}{H^3}-3q-2$ denotes jerk parameter.
As the expression for ~$\frac{dS_{T}}{dt}$~ is very lengthy, so to get an idea about its sign we make use of the  estimated values of different parameters at present epoch (i.e a=1) from section III as follows:
$\omega_{0d}=-1.11,~\Omega_{0d}=0.63,~ H_0=69.92,~ \Gamma_0=0.32, ~\Gamma_1=0.4,~\alpha=0.03$ and from these best fit values of the parameters we eventually obtain present estimated values of other parameters such as $q=-0.61, j=1.21$ and $r_0=\frac{37}{63}$\\
		
\begin{figure}
			\centering
			\subfigure[]{
				\includegraphics[width=0.32\textwidth]{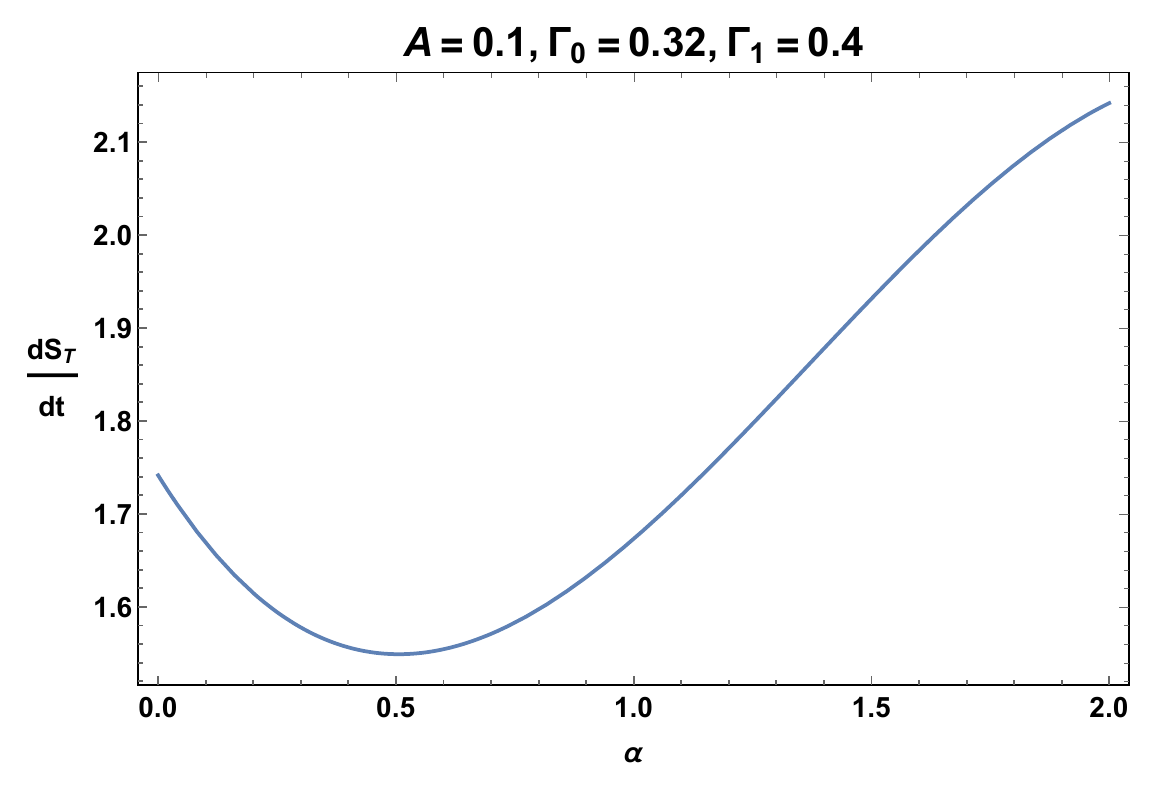}\label{fig:entropy1}}
			\subfigure[]{
				\includegraphics[width=0.32\textwidth]{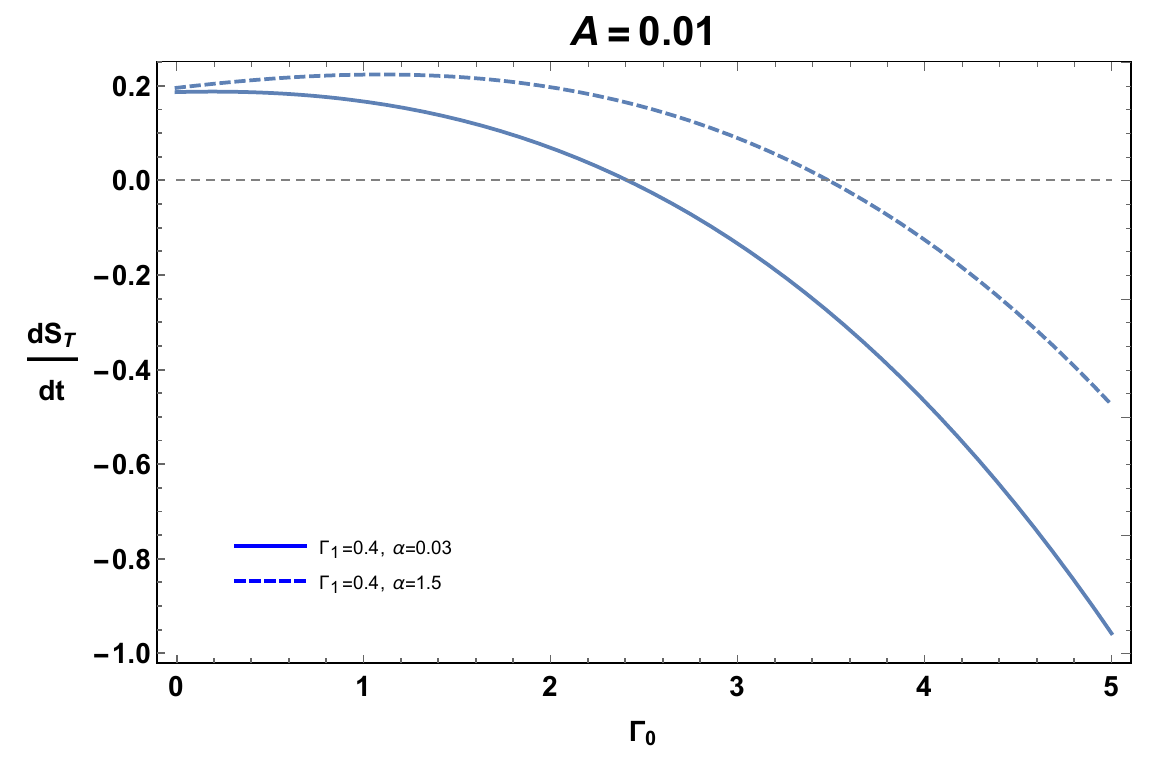}\label{fig:entropy2}}
			\subfigure[]{
				\includegraphics[width=0.32\textwidth]{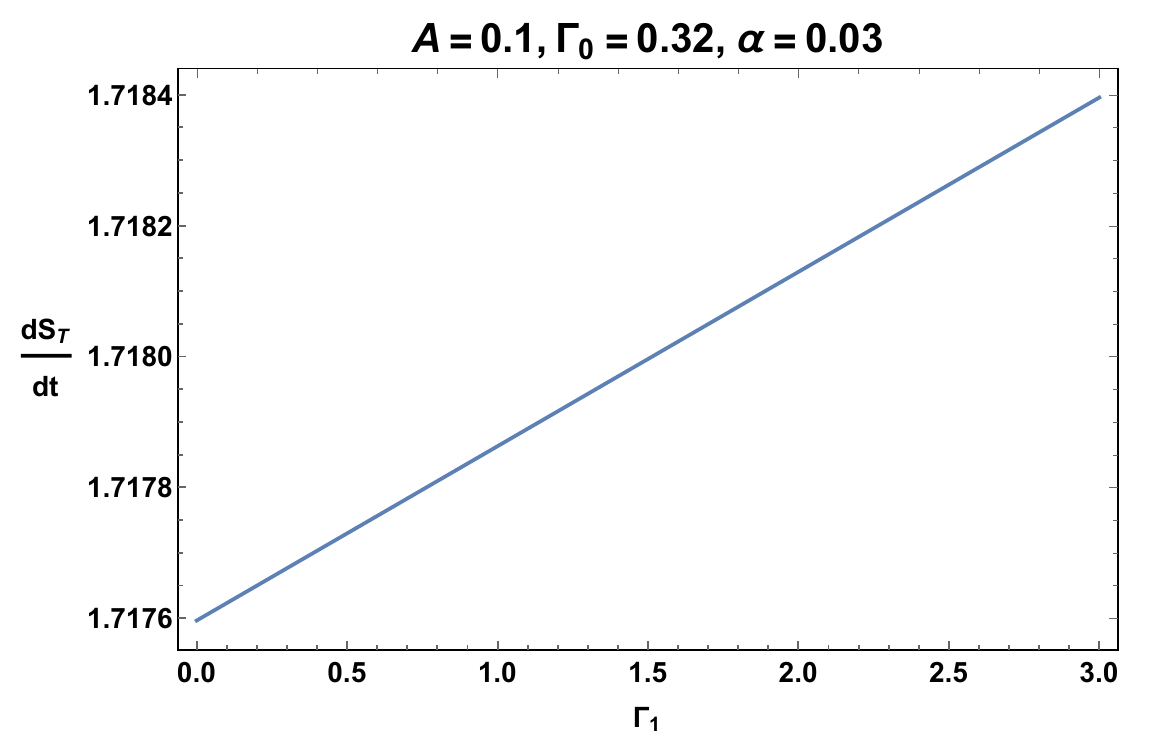}\label{fig:entropy3}}
			\caption{Figures shows change of total entropy with respect to $\alpha,\Gamma_0$ and $\Gamma_1$ at present time when other parameters takes best fit values i.e $\omega_{0d}=-1.11, \Omega_{0d}=0.63,  H_0=69.92$}
			\label{entropy}
		\end{figure}

So, if we assume, $A_d=A_h=A$, we have
		\begin{equation}
			\frac{dS_{T}}{dt}=0.017+17.0043A\label{36}
		\end{equation}
So, $\frac{dS_{T}}{dt}\geq 0$ if $A\geq -0.001$.\\
To understand the nature of $\frac{dS_T}{dt}$ with the change of particle creation rate and interaction term, we have plotted  $\frac{dS_T}{dt}$ with respect to $\alpha, \Gamma_0$ and $\Gamma_1$ in three separate graphs given in fig (\ref{entropy}) keeping other parameters fixed.
		
We see from the figs. \ref{entropy}(a) and \ref{entropy}(c) that $\frac{dS_{T}}{dt}$ remains positive for all positive values of $\alpha$ and $\Gamma_1$. It is to be noted that we have taken energy transfer constant to be 0.1 i.e., greater than -0.001, keeping parity with constraint obtained from eqn. (\ref{36}). On the other hand, from fig. \ref{entropy}(b) it is observed that for a specific range of $\Gamma_0$, depending on the value of the interaction parameter $\alpha$, $\frac{dS_{T}}{dt}$ remains positive. To be specific we see that when $\Gamma_0<\alpha+3$ i.e., when $\alpha>\Gamma_0-3$, we see that $\frac{dS_{T}}{dt}$ remains positive. Therefore the validity of GSLT constrained mainly by the value of $\Gamma_0$, interaction parameter $\alpha$, and the energy transfer constant $A$.
%%%%%%%%%%%%%%%%	
%%%%%%%%%%%%%
	
\section{Summary and Conclusion}
%%%%%%%%%%%%%%%%%%%%%%%%%%%%%%%%%%%%%%%%%%%%%%%%%%
In this work, we have studied the cosmological consequences of a particle creation mechanism from different perspectives for a particular form of particle creation rate $\Gamma = \Gamma_0 H+\Gamma_1 H^{-1}$ in an interacting two-fluid model in a spatially flat FRW universe. Here we have considered the creation of dark matter only, and energy flow from dark energy to dark matter due to interaction. We have considered the interaction term $Q=\alpha H\rho_m$ and  calculated the analytic expressions of dark fluid energy densities $\rho_m$ and $\rho_d$ given by Eqn. (\ref{17}) and Eqn. (\ref{18}), following which we have obtained the analytic expression of the Hubble parameter given in Eqn. (\ref{19}). Then we performed $\chi^2$ data analyses with combined data of Supernova-Ia (580 point), and Hubble data (27 point). We found the best fit values of the model parameters. It is also found that the present model is compatible with current observed data though it is not better than $\Lambda$CDM model. Further, the DE EoS parameter is estimated to be $\omega_{d}= -1.11$, i.e., it crossed the phantom divide line. Also, we observed that the estimated value of the interaction term is minimal, and there is no central difference between the estimated values of the model parameters in interacting and non-interacting cases. That is, from the statistical analysis viewpoint, the interaction has no significant effect on the model. Another exciting observation realized from the contour plots is that in the creation rate, one parameter, namely,  $\Gamma_1$ can not influence any other parameters. At the same time, $\Gamma_0$ is linearly or nonlinearly related to other model parameters. Therefore, creation rate $\propto H$ plays a more crucial role in the evolution dynamics of the universe than the term, which is inversely proportional to the Hubble parameter. We performed another analysis to understand the connection between the model paremeters and evolution dynamics more deeply. \\

In statefinder diagnostic analysis we obtained different trajectories depicting evolution dynamics of universe for distinct values of the model parameters $\alpha$, $\Gamma_{0}$, and $\Gamma_{1}$. It is observed that in s-r plane the distance of the present value $(s_{0},r_{0})$ from the $\Lambda$CDM fixed point $(0,1)$ becomes shorter for smaller values of interaction parameter $\alpha$, and for larger values of $\Gamma_{0}$, $\Gamma_{1}$ in phantom era though all trajectories passed through the $\Lambda$CDM fixed point at the early time. On the contrary, in cosmological constant era, the distance of the present value $(s_{0},r_{0})$ from the $\Lambda$CDM fixed point becomes shorter for larger values of $\alpha$ and for smaller values of $\Gamma_{0}$, $\Gamma_{1}$ and interestingly all the evolutionary trajectories are observed to converge at the $\Lambda$CDM fixed point in future. On the other hand, in the quintessence era, it is noticed that the distance of the present value $(s_{0},r_{0})$ from the $\Lambda$CDM fixed point becomes shorter for larger values of $\alpha$ and for smaller values of $\Gamma_{0}$, $\Gamma_{1}$ and the evolutionary trajectories never passes through the $\Lambda$CDM fixed point. Therefore, the particle creation mechanism in a two dark fluid model can mimic the standard $\Lambda$CDM model at the early time in phantom era and in the future in cosmological constant era.\\
% Later Therefore the particle creation mechanism in a two dark fluid model can mimic the standard $\Lambda CDM$ fixed point at the early time in phantom era and in the future in cosmological constant era.\\

We also performed the statefinder diagnostic in the $q-r$ plane for this model where different evolutionary trajectories have been attained  for various choices of the model parameters $\alpha$, $\Gamma_{0}$, and $\Gamma_{1}$. It clearly shows the dependency of evolutionary trajectories in the $q-r$ plane on the model parameters. We see that in the phantom era as well as in the quintessence era, the evolutionary trajectories never evolved through the $de~ Sitter$ fixed point $(q=-1,r=1)$. However, in cosmological constant era, the distance of the present value $(q_{0},r_{0})$ from the $de~ Sitter$ fixed point becomes shorter for larger values of $\alpha$ and for smaller values of $\Gamma_{0}$, $\Gamma_{1}$ and all the evolutionary trajectories are seen to converge at the $de~ Sitter$ fixed point in the future. So the current particle creation model mimics the $de~ Sitter$ universe in the future during the cosmological constant era. \\

From the dynamical study of the model, we obtain some interesting cosmological scenarios in the late times. A normally hyperbolic set of critical points $A$ represents the late-time accelerated Universe evolving in the phantom regime and dominated by phantom fluid for ($y_c<-1$). It also represents dust dominated decelerated intermediate phase of the universe for $y_c=0$ when DE behaves as dust. Another normally hyperbolic set $C$ represents the late-time scaling attractor in the phase space. For the parameter restriction: $\alpha<\Gamma_{0}-3$, the set $C$ corresponds to the accelerated universe evolving in phantom regime (see fig. \ref{fig:Stable_C}) and solving the coincidence problem.
On the other hand, the normally hyperbolic set $D$ can be a physically viable solution for the restriction:  $\alpha>\Gamma_{0}-3$ and $\Gamma_{1}\neq0$, where the set exhibits the late-time accelerated universe evolving in the cosmological constant era with the similar order of energy densities of DE and DM. As a result, it solves the coincidence problem (see figs. \ref{fig:Stable_A} and \ref{fig:Stable_D}). The authors in (\cite{Biswas2017}) showed that the scaling attractor could not alleviate the coincidence problem when particle creation was taken as  $\Gamma=\Gamma_0H$.
Therefore, one can say that the model can predict the late-time accelerated evolutionary scheme of the universe connected through a matter-dominated decelerated phase. After fine tuning the initial conditions, we observed that the ultimate fate of the evolution of the universe is attracted in the cosmological constant era or in the phantom phase see fig. (\ref{phasespace-parameters-evolution}).
		\\
		
Finally, we carried out a thermodynamic analysis of the current model. As there are temperature difference between the system and horizon and also between the fluid components of the universe, thermal equilibrium can not hold, and therefore, non-equilibrium thermodynamic treatment is considered. Using Euler's equation, we have calculated temperature evolution of the individual components DE and DM and finally calculated time derivative of the total entropy. As we have considered non-equilibrium thermodynamics, two extra terms corresponding to energy transfer between DE and DM within the universe and between the universe and the horizon, respectively, have been considered in the expression of total entropy. It is shown that if we use the best fit values of the model parameters obtained in statistical analysis, the generalized second law of thermodynamics is satisfied if the energy transfer constants are more significant than a particular value -0.001. Also, graphically, it has been shown that the validity of GSLT is crucially dependent on the $\Gamma_{0}$. In particular, when the energy transfer constant is greater than -0.001, GSLT is satisfied only when $\Gamma_0<\alpha+3$ i.e., when $\alpha>\Gamma_0-3$  \\
		
From the overall analysis obtained from different sections, we conjecture that the evolution dynamics of the universe are deeply connected with the particle creation rate and the interaction term. On a special note it is observed that $\Gamma_0$ plays a crucial role over $\Gamma_1$ in particle creation rate $\Gamma=\Gamma_0H+\frac{\Gamma_1}{H}$. Also, when  $\alpha>\Gamma_{0}-3$, the results obtained from statefinder diagnosis, dynamical system analysis, and thermodynamic analysis also agree with current observed data. From this point of view, we need to do more research on this particular restriction on the model parameters. However, it may be then concluded that our present cosmological model with particle creation mechanism solves the coincidence problem and explains late time acceleration of the universe and also satisfies the generalized second law of thermodynamics. Also, it can be predicted that the universe will eventually evolve in the cosmological constant era.
		
		%%%%%%%%%%%%%%%%%%%%%
\section*{Acknowledgments}
We are extremely grateful to the respected reviewers whose valuable suggestions helped us a lot in a significant way to improve our current work. We also express our deepest gratitude to Dr. Supriya Pan for providing us his support and valuable insights while modifying the previous version of our work. The author Goutam Mandal is supported by UGC, Govt. of India through Senior Research Fellowship [Award Letter No. F.82-1/2018(SA-III)] for Ph.D. and the author Sujay Kr. Biswas acknowledges University of North Bengal for financial assistance/support for the research project with Ref. No.2303/R-2022.   
		%%%%%%%%%%%%%%%%%%%%%%%%%%%%%%%
		
			\section*{DATA AVAILABILITY}
		All the data sets have been used in this paper are publicly available.
		
		\bibliography{PCM14}
	\end{document}